\documentclass[aps,prd,reprint,amsmath,amssymb,longbibliography,nofootinbib,nobibnotes,floatfix]{revtex4-2}
\usepackage[T1]{fontenc}
\usepackage{graphicx,booktabs,bm}
\usepackage[colorlinks=true,linkcolor=blue,citecolor=blue,urlcolor=blue]{hyperref}
\usepackage{tikz,pgfplots}
\usepgfplotslibrary{groupplots}
\pgfplotsset{compat=1.17}
\definecolor{plotblue}{RGB}{35,107,156}
\definecolor{plotred}{RGB}{179,78,40}
\definecolor{plotgreen}{RGB}{79,124,77}
\pgfplotsset{reviewaxis/.style={font=\footnotesize,tick label style={font=\scriptsize},label style={font=\footnotesize},legend style={font=\scriptsize,draw=none,fill=white},grid=major,grid style={gray!15},axis line style={black!75},scaled ticks=false}}
\newcommand{\DM}{D_{\mathrm M}}
\newcommand{\DA}{D_{\mathrm A}}
\newcommand{\DL}{D_{\mathrm L}}
\renewcommand{\DH}{D_{\mathrm H}}
\newcommand{\rd}{r_{\mathrm d}}
\newcommand{\Hp}{H_\perp}
\newcommand{\Hl}{H_\parallel}
\newcommand{\Lp}{\mathcal L_\perp}
\newcommand{\Ll}{\mathcal L_\parallel}
\newcommand{\fid}{\mathrm{fid}}

\usepackage{xcolor}

\begin{document}
\title{BAO and Type Ia Supernovae in LTB Cosmology: Constraints on Radial Inhomogeneity and Acoustic-Ruler Evolution}
\author{Myrto Magklara}
\affiliation{Department of Physics, University of Ioannina, GR-45110 Ioannina, Greece}
\author{Leandros Perivolaropoulos}
\email[Contact author: ]{leandros@uoi.gr}
\affiliation{Department of Physics, University of Ioannina, GR-45110 Ioannina, Greece}
\date{19 November 2026}
\begin{abstract}
We investigate how baryon acoustic oscillations (BAO) and Type Ia supernovae constrain radial density profiles and acoustic-ruler evolution in Lema\^itre--Tolman--Bondi (LTB) cosmology with a cosmological constant. Extending a homogeneous Friedmann--Robertson--Walker analysis requires both modified distances and unequal radial and transverse stretching of the BAO ruler. We assume one initial sound horizon and follow the material separation carrying this feature: spatially varying expansion makes its observed length depend on redshift and direction. Transverse BAO and supernovae isolate the evolution of the transverse ruler without fixing its absolute length; radial BAO provides complementary expansion and ruler information.
We apply this calibration-free interpretation to exact nonlinear models with a low-density region compensated by surrounding excess matter. Its depth, width, radial location, and exterior matter density are varied, with the observer always at the symmetry centre. We fit DESI DR2 BAO and DES-Dovekie supernovae using full covariances and matched calibration freedom in LTB and flat $\Lambda$CDM. Conditional on evaluating the BAO predictions at their effective redshifts, the full-sample LTB fit improves $\chi^2$ by $8.5$ with three additional parameters. DESI rejects the specific deep void preferred without it, but nearby narrow and more distant broad density depressions remain nearly degenerate.
Selecting DES supernovae alone gives a shallower shell and an improvement of $7.44$, almost entirely from BAO and unchanged by the additional magnitude-uncertainty cut. Two illustrative redshift-window tests retain a DES-only gain close to 7 at this geometry. The preferred profile and ruler history depend on the supernova selection. The full-sample gain occurs in at least $7\%$ of simulated flat-$\Lambda$CDM data sets; it does not establish an inhomogeneity detection, and this probability is not transferred to the DES-only fits. Finite-ruler corrections are small at DESI's effective redshifts, but do not validate averaging across a survey bin. Finally, in the local approximation, common LTB ruler evolution shifts angular and transverse three-dimensional BAO together and cannot by itself explain their discrepancy.
\end{abstract}
\maketitle

\section{Introduction}\label{sec:intro}
Baryon acoustic oscillations (BAO) provide a ruler imprinted by sound waves in the early universe. In a Friedmann--Robertson--Walker (FRW) spacetime, all locations share one expansion history, so the late-time comoving ruler is conventionally identified with the drag-epoch sound horizon $\rd$. Its angular size measures $\rd/\DM$, while its redshift extent measures $H\rd/c$. In a radially inhomogeneous universe, neighbouring matter shells need not expand at the same rate. Compared with FRW, this alters both the distance to the BAO feature and the length of the ruler itself. Since BAO measures their ratio, a consistent analysis must calculate both.

Our primary objective is to analyze BAO and Type Ia supernovae (SNe Ia) in a central Lema\^itre--Tolman--Bondi spacetime with a cosmological constant, denoted $\Lambda$LTB. We ask what these data constrain about the depth, width, and location of a radial density depression, and about the apparent evolution of the acoustic ruler with redshift. Throughout, the observer is at the symmetry centre. A displaced density minimum is an observer-centred spherical shell, not an off-centre observer in a conventional spherical void.

Supernovae and BAO supply complementary information. If the intrinsic SN luminosity is not externally calibrated, supernovae measure only relative distances: how the luminosity distance changes with redshift, but not its absolute value. Transverse BAO measures distance divided by the transverse acoustic-ruler length. Combining the two cancels their common distance dependence, so a ratio at two redshifts measures the change in ruler length without knowing its absolute value. We call this a calibration-free ruler test. Removing an overall normalization does not erase the SN constraint on the shape of the distance--redshift relation. Pedrotti et al.~\cite{PedrottiMiscalibration} emphasize this distinction when testing a uniform BAO recalibration in homogeneous late-time cosmologies. Here an overall ruler calibration is also free, but the LTB ruler can additionally evolve with redshift and stretch differently in the two directions. 

We also fit the original SN and BAO measurements directly to an LTB model. This second calculation retains the distance information that the ratio removes, and it also uses radial BAO, which measures the expansion rate along the line of sight times the radial ruler. Both directional rulers are then tied to the same density profile. The two calculations therefore answer different questions: whether the data require a change in ruler length, and which physical density profiles can reproduce the full set of measurements.

Joint BAO and supernova constraints in $\Lambda$LTB predate the present data: Marra and P\"a\"akk\"onen combined them with CMB and local expansion information~\cite{MarraPaakkonen}. Directional acoustic-ruler transport is likewise established~\cite{Zumalacarregui,GarciaBellido,ValkenburgMarraClarkson,CamarenaCopernican}. The observational analyses of Refs.~\cite{ValkenburgMarraClarkson,CamarenaCopernican} use the same underlying radial and transverse stretching prescription adopted below. Our novelty lies in the combination of diagnostics and flexible compensated profiles, in which the deepest point of the underdensity can lie at a freely varied radius from the observer, together with robustness tests on current data, rather than in introducing that prescription. Calibration-independent BAO--SN comparisons also have an existing literature~\cite{Dinda}. Zhang and Zhang~\cite{ZhangZhang} use first-order $\Lambda$LTB distances, validated against exact background solutions, and directional ruler corrections. They combine DESI DR2 with separate SN compilations, compressed CMB information, and an extension using the kinematic Sunyaev--Zel'dovich (kSZ) effect. They parametrise the enclosed density contrast with a smooth step whose width is fixed to $40\%$ of its radius in their baseline analysis, and their fits favour a mild overdensity around the observer. Our profile instead parametrises the local density contrast at each radius, rather than its enclosed mean, and allows its minimum to lie away from the central observer. We also fit the initial ruler length without a CMB calibration. The fitted density amplitudes therefore describe different models and cannot be compared as the same parameter.

Camarena et al.~\cite{CamarenaDESI2025} recently combined DESI DR2 with DES-Y5 or Pantheon+ SNe and compressed CMB information in compensated $\Lambda$LTB models specified by a central density contrast and a boundary redshift. With SNe included, their fits are comparable in quality to those of a homogeneous model with an evolving dark-energy equation of state, $w(a)=w_0+w_a(1-a)$. Their separate comparison of transverse and radial BAO distances provides a geometric perspective on BAO--SN consistency. This is a close antecedent of our analysis, including its use of directional acoustic-ruler transport.

Our contribution is to connect the calibration-free ruler test to exact nonlinear fits in which the depth, width, and radial location of the density depression are all allowed to vary. A matched flat-$\Lambda$CDM fit shows how much improvement this additional freedom provides. We then test whether the same conclusions survive changes in the SN sample and in the averaging of BAO predictions over redshift. This combination identifies which particular geometries the data reject and which different profiles remain comparably successful. It also distinguishes what is computed exactly, the LTB spacetime and the light cone, from the approximations needed to compare it with the published BAO measurements. No CMB sound-horizon calibration or kSZ likelihood is imposed here.

Local underdensities have also been studied as a possible source of the Hubble tension, including the Milgromian scenario of Haslbauer, Banik, and Kroupa~\cite{Haslbauer}. Banik, N\'ajera, and Desmond~\cite{BanikPotential} test the high-redshift gravitational-potential imprint of a local void using an effective redshift-offset description and additional early-universe and expansion data. These studies motivate complementary tests, but differ from our general-relativistic, exact-background $\Lambda$LTB calculation with no external absolute-distance calibration. Our fits therefore neither measure an absolute $H_0$ nor provide a general exclusion of local-void solutions to the Hubble tension.

We use DESI+SNe as the primary direct comparison, with the legacy angular BAO sample added as a secondary test because its cross-covariance with DESI is unavailable, and therefore we have to treat them as independent, although they may not be. As a secondary application, we consider the reported discrepancy between angular and three-dimensional BAO measurements of $\DM/\rd$ near $z\simeq0.5$~\cite{Pantos}. Under matching local weights, both transverse estimators measure $\DM/\Lp$, so common ruler transport cannot shift them relative to one another. Survey-window and estimator effects require a separate calculation.

Section~\ref{sec:geometry} starts with the FRW acoustic ruler, explains its directional evolution in LTB, and follows the ruler through angular and three-dimensional measurements. Section~\ref{sec:data} introduces the density profile and two teaching figures before specifying the data and likelihoods. Section~\ref{sec:shape} explains the ruler-shape test. Sections~\ref{sec:desi_hsw} and \ref{sec:robustness} present the main DESI fits and robustness checks. Section~\ref{sec:discrepancy} treats the estimator comparison, followed by the conclusions. The appendices give the construction, validation, and data conventions required for these results. Six-family scans, DESI-free controls, and auxiliary response derivations are retained in the accompanying Supplemental Material~\cite{SupplementalMaterial}.

\section{From FRW distances to LTB acoustic observables}\label{sec:geometry}
The interpretation of BAO involves three questions: what sets the initial ruler, how expansion carries it to the emission time, and how an observation measures it. We begin with FRW, where the answers are familiar, and then identify the changes required in LTB. The symbols $\perp$ and $\parallel$ refer to directions across and along the central observer's line of sight.

\subsection{The acoustic ruler in FRW}
Before the baryon drag epoch, pressure waves in the photon--baryon fluid propagated away from initial overdensities. When baryons ceased to be dragged by the photons, this process left a preferred comoving separation $\rd$ in the matter distribution. The later galaxy correlation function retains the associated acoustic feature. The initial sound horizon $\rd$ is the common calibration parameter in this paper. We fit a single value for all redshifts, rather than letting it vary from one measurement to another; any change of the ruler with redshift must come from the LTB expansion.

In FRW, matter expands by the same scale factor $a(t)$ at every location and in every direction. Under passive transport, an initially spherical feature therefore remains spherical: its physical radius grows as $a(t)\rd$, while its comoving radius remains $\rd$. Its angular size is $\theta_{\rm BAO}=\rd/\DM$, and its radial redshift extent is $\Delta z_{\rm BAO}=H\rd/c$. These familiar relations connect a measured angle or redshift interval to a distance or expansion rate.

An inhomogeneity changes both sides of this comparison. It changes the distance to the feature, and it changes the expansion experienced by the matter carrying the feature. A consistent LTB prediction must therefore evolve the ruler as well as solve the light propagation. We first describe the spacetime, then the ruler, and finally the measurements.

\subsection{The LTB geometry and its two expansion rates}
With physical time and explicit $c$, the dust metric and shell dynamics are
\begin{align}
 ds^2&=-c^2dt^2+\frac{R'^2(t,r)}{1+2E(r)}dr^2+R^2(t,r)d\Omega^2,\\
 \dot R^2&=\frac{2GM(r)}{R}+2E(r)c^2+\frac{\Lambda c^2}{3}R^2,\label{eq:dynamics}\\
 M'&=4\pi\rho R^2R',\qquad
 \Hp=\frac{\dot R}{R},\qquad \Hl=\frac{\dot R'}{R'}.
\end{align}
Here $R(t,r)$ is an areal radius: a sphere labelled by $r$ has area $4\pi R^2$. The mass $M(r)$ is enclosed within that shell, and $E(r)$ describes its spatial curvature. A prime denotes a radial derivative at fixed time and a dot a time derivative at fixed shell label. The two expansion rates have a simple meaning: $\Hp$ describes growth around a sphere, whereas $\Hl$ describes separation of neighboring spheres. They coincide in FRW.

A density profile alone does not specify the spacetime. We choose $R(t_0,r)=r$, so $r$ is the present areal radius, and a constant bang time, so all matter shells have the same age. Far from the density feature the model approaches an expanding, spatially flat universe containing pressureless matter and a cosmological constant. Regularity at the origin requires $M=O(r^3)$ and $E=O(r^2)$. We require positive density, $R'>0$ so neighboring shells remain ordered, and $1+2E>0$ so the radial metric has the correct sign over the domain used.

For a central observer, the past radial null cone satisfies
\begin{align}
 \frac{dt}{dz}&=-\frac{1}{(1+z)\Hl}, &
 \frac{dr}{dz}&=\frac{c\sqrt{1+2E}}{(1+z)\dot R'}.
 \label{eq:null}
\end{align}
For a central observer, and for the regular models considered here, $\DA=R(t(z),r(z))$ and $\DM=(1+z)R$.
Consequently,
\begin{equation}
 \frac{d\DM}{dz}
 =\frac{c\sqrt{1+2E}+(\Hl-\Hp)R}{\Hl}.
 \label{eq:distance}
\end{equation}
Both curvature and shear enter this derivative. Thus $c\int dz/\Hl$ is not the general LTB transverse distance. The familiar integral is recovered in the spatially flat FRW limit. With metric propagation and photon-number conservation, reciprocity gives $\DL=(1+z)^2\DA=(1+z)\DM$ for cosmological redshift.

\subsection{How the ruler changes in an inhomogeneous universe}
\paragraph*{Stretching the feature.} Assume one initial comoving sound horizon $\rd$ and an early reference time $t_i$ at which the physical acoustic ruler is isotropic, with length $a_i\rd$. Here $a_i=\bar a(t_i)$ is the early exterior scale factor. By passive transport we mean that the two matter elements defining a local separation move with the background matter flow. The expansion then gives~\cite{Zumalacarregui,ValkenburgMarraClarkson,CamarenaCopernican,CamarenaDESI2025}
\begin{align}
 \ell_\perp(t,r)&=a_i\rd\frac{R(t,r)}{R(t_i,r)},&
 \ell_\parallel(t,r)&=a_i\rd\frac{R'(t,r)}{R'(t_i,r)}.
 \label{eq:transport}
\end{align}
The proper radial length between neighbouring shells is $R'dr/\sqrt{1+2E}$; since $E$ depends only on $r$, the factor $\sqrt{1+2E}$ is the same at $t_i$ and $t$ and cancels in the ratio. Equations~(\ref{eq:transport}) hold for an infinitesimal separation. Applying them to the BAO feature, about $150\,$Mpc long, assumes that the background varies slowly over the length of the ruler, and neglects changes in the shape of the correlation function caused by inhomogeneous growth and by reconstruction.

An initially spherical acoustic feature can consequently become an ellipsoid: the expansion around a shell and the expansion between shells stretch its two axes differently. This is the reason that one must track two late-time rulers even when there is only one initial $\rd$.

\paragraph*{Observing it at a different emission time.} For either direction $i\in\{\perp,\parallel\}$, the physical ruler $\ell_i$ is the separation at the time of emission. Dividing by the exterior scale factor $\bar a(t)$, normalized to unity today, defines the background-comoving ruler $L_i=\ell_i/\bar a(t)$. A different, directly useful conversion multiplies the physical length by the measured factor $1+z$. We call the resulting quantity the observed-comoving ruler:
\begin{equation}
 \mathcal L_i=(1+z)\ell_i=Q(z)L_i,\qquad
 Q(z)=(1+z)\bar a(t(z)).\label{eq:Q}
\end{equation}
In FRW, $1+z=1/\bar a(t)$, so $Q=1$ and $\mathcal L_\perp=\mathcal L_\parallel=\rd$ under passive acoustic transport. In LTB the redshift accumulated along the light ray need not equal the inverse exterior scale factor at emission. Thus a source observed at the same redshift as an FRW source need not have emitted its light at the same exterior scale factor. The factor $Q$ records this emission-time effect and multiplies both axes equally; the unequal stretching in Eq.~\eqref{eq:transport} changes their ratio. Define $\mathcal R_i(z)=\mathcal L_i(z)/\rd$ to describe departures from a constant observed-comoving ruler.

Throughout this paper, ruler evolution means the variation of $\mathcal R_i(z)$ along the observer's past light cone. Rulers observed at different redshifts lie at different distances and are seen at different times, so this variation combines spatial inhomogeneity and time evolution; it does not describe a single ruler changing with time.

\subsection{What a thin-shell angular measurement sees}
An angular BAO analysis selects galaxies in a thin redshift shell and finds the acoustic feature in their angular correlation. In the limit that the ruler and geometry vary little across the selection, it measures the ruler at one redshift. The apparent angle is its transverse physical size divided by the angular-diameter distance. The corresponding local radial extent follows from converting a physical radial separation to a redshift interval:

\begin{equation}
 \theta_{\rm BAO}=\frac{\ell_\perp}{\DA}=\frac{\Lp}{\DM},\qquad
 \Delta z_{\rm BAO}=\frac{\Hl\Ll}{c}.\label{eq:obs}
\end{equation}
Angles in this equation are in radians. The inverse angular measurement is therefore $1/\theta_{\rm BAO}=\DM/\Lp$. An analysis that assumes FRW interprets $1/\theta_{\rm BAO}$ as $\DM/\rd$. In LTB it measures $\DM/\Lp$, so a ruler stretched beyond $\rd$ makes the inferred distance too small. A physical void also changes $\DM$, so the final prediction must include both changes. Thin-shell analyses still require finite-width projection and feature-extraction corrections; the one-redshift expression is their local interpretation, not an assumption that these corrections vanish.

\subsection{What a three-dimensional measurement sees}
A three-dimensional analysis first uses a reference (fiducial) cosmology to turn angles and redshifts into galaxy positions. It then measures the galaxy correlation function and fits it with a template in which the position of the BAO feature is rescaled separately across and along the line of sight; the two rescalings are called dilations. For a locally uniform acoustic ellipse, a template with both dilations free returns
\begin{align}
 \alpha_\perp&=\frac{\DM/\Lp}{\DM^{\fid}/\rd^{\fid}},&
 \alpha_\parallel&=\frac{c/(\Hl\Ll)}{\DH^{\fid}/\rd^{\fid}}.
 \label{eq:alpha}
\end{align}
Multiplying a fitted dilation by its fiducial distance-to-ruler ratio removes the reference conversion. The transverse result is the same $\DM/\Lp$ measured by the ideal angular method; the radial result also depends on $\Hl$ and $\Ll$. The local quantities to compare with the released summaries are therefore
\begin{equation}
 X_\perp=\frac{\DM}{\Lp},\qquad
 X_\parallel=\frac{c}{\Hl\Ll}.\label{eq:X}
\end{equation}
For an isotropic compressed measurement we use the corresponding local prescription $X_V=(zX_\perp^2X_\parallel)^{1/3}$. This extension of the FRW compression is conditional on the template interpretation; it is not a general derivation of the finite-survey isotropic likelihood in LTB.

The dilation ratio contains the ruler anisotropy:
\begin{equation}
 \frac{\alpha_\perp}{\alpha_\parallel}
 =\frac{\DM\Hl}{\DM^{\fid}H^{\fid}}
 \frac{\Ll}{\Lp}.\label{eq:AP}
\end{equation}
It cannot be interpreted as a purely geometric Alcock--Paczy\'nski ratio unless the directional ruler ratio is fixed. Table~\ref{tab:dictionary} summarizes the replacements needed in an LTB analysis.

\begin{table*}[tbp]
\caption{Observable dictionary. FRW entries assume standard passive transport of a constant comoving acoustic ruler; the distance integral shown is for spatially flat FRW. LTB entries are local central-observer predictions, before survey-window and estimator corrections.}\label{tab:dictionary}
\begin{ruledtabular}
\begin{tabular}{lll}
Quantity & Flat FRW interpretation & Central LTB interpretation\\
Transverse distance & $c\int_0^z du/H(u)$ & $(1+z)R(t(z),r(z))$\\
Observed-comoving ruler & $\rd$ & $\Lp=(1+z)\ell_\perp$, $\Ll=(1+z)\ell_\parallel$\\
Inverse angular BAO scale & $\DM/\rd$ & $X_\perp=\DM/\Lp$\\
Inverse radial BAO extent & $c/(H\rd)$ & $X_\parallel=c/(\Hl\Ll)$\\
Transverse/radial ratio & $\DM H/c$ & $X_\perp/X_\parallel=(\DM\Hl/c)(\Ll/\Lp)$\\
SN luminosity distance & $(1+z)\DM$ & $(1+z)^2R$\\
SN / transverse BAO & Constant normalization & Normalization $\times\Lp(z)$\\
\end{tabular}
\end{ruledtabular}
\end{table*}

\subsection{From one ruler to a survey measurement}\label{sec:survey}
The two measurement procedures sample different parts of the light cone. An angular analysis selects galaxies with a normalized redshift selection $\phi(z)$ and measures
\begin{equation}
 w(\theta)=\int dz_1dz_2\,\phi(z_1)\phi(z_2)\,
 \xi(z_1,z_2,\theta),\label{eq:window}
\end{equation}
where $\xi$ is their correlation function. If the geometry and ruler vary little across this selection, the recovered angular scale has the local interpretation $\DM/\Lp$. A three-dimensional analysis assigns each galaxy a fiducial comoving distance $\chi$ and fits dilations to pairs spread across a wider redshift interval. The survey's fiducial cosmology and $\rd^{\fid}$ remain fixed when the trial exterior density or physical ruler is varied; Eq.~\eqref{eq:alpha} retains both reference conversions.

Two distinct variations matter. First, the local dilations can change along a pair. In a passive-ruler construction with small-angle, plane-parallel geometry, endpoints at $\chi_1$ and $\chi_2$ give
\begin{equation}
\begin{gathered}
 \alpha_\parallel^{\rm pair}=\frac{1}{\chi_2-\chi_1}\int_{\chi_1}^{\chi_2}\alpha_\parallel(\chi)\,d\chi,\\
 \alpha_\perp^{\rm pair}=\frac{\chi_1\alpha_\perp(\chi_1)+\chi_2\alpha_\perp(\chi_2)}{\chi_1+\chi_2}.
\end{gathered}\label{eq:pair}
\end{equation}
The radial integral is exact for a purely radial passive pair in the assumed initial acoustic metric. Its use as a component of an oblique pair, together with the transverse endpoint expression, is a geometric approximation. The transverse component is not a line average of the transverse dilation. The Supplemental Material~\cite{SupplementalMaterial} derives these expressions, their short-pair expansions, and the conversion to a fixed survey fiducial convention.

Second, pairs with the same separation $s$ and orientation $\mu$ have different midpoints within the bin. Let $p(\chi_c|s,\mu)$ be their normalized, survey-weighted midpoint distribution. For an illustrative template, the leading geometric contribution to the model correlation-function residual relative to the unit-dilation template is
\begin{equation}
 v_{\rm geom}(s,\mu)=\int d\chi_c\,p(\chi_c|s,\mu)
 \sum_i h_i(s,\mu;\chi_c)\,\delta\alpha_i^{\rm pair},
 \label{eq:xiobs}
\end{equation}
where $i=\perp,\parallel$, $\delta\alpha_i^{\rm pair}=\alpha_i^{\rm pair}-1$, and $h_i$ is the template derivative with respect to that dilation. Only when the response is independent of midpoint can $h_i$ be taken outside the average. This is a first-order response in the dilation perturbations, not a complete calculation of LTB clustering, redshift-space distortions, or reconstruction. Pair extent and variation across a bin are distinct effects; symmetry and survey weights can suppress either, so their relative size must be tested.

After projection onto the fitted multipoles and profiling over the nuisance parameters, the fit maps the residual data vector into a dilation shift. This map depends on the template, its nuisance treatment, and the covariance of the fitted two-point-function data. That covariance is distinct from the released $13\times13$ covariance of the BAO distance summaries. A constant dilation shared by every pair is sufficient for the fit to recover that value; it is not necessary, because weighted cancellations can give the same result. For a linear fit with both dilations free, correlation of their uncertainties alone does not turn a pure radial dilation into a transverse bias.

Our exact nonlinear calculation supplies the spacetime, light cone, and local rulers without expanding in density contrast. The numerical likelihood evaluates the corresponding observables at each published effective redshift and uses the full released summary covariance. It does not implement the survey-level response of Eq.~\eqref{eq:xiobs}. Slowly varying dilations and a suitable local template motivate this prescription, but do not by themselves validate the full estimator. The finite-ruler and window tests in Sec.~\ref{sec:robustness} examine selected departures from it. DESI's fiducial-cosmology tests do not establish the response to these radially inhomogeneous profiles~\cite{DESI,Perez,DESIValidation}. In particular, its BGS isotropic entry is interpreted here through $X_V=(zX_\perp^2X_\parallel)^{1/3}$; this is a conditional extension of the FRW compression, not a two-dilation BGS fit.

\section{Density profiles, data, and likelihoods}\label{sec:data}
Our primary fit combines DESI DR2 BAO with DES-Dovekie supernovae; we abbreviate it D+S. A secondary fit, D+S+A, adds the earlier angular BAO measurements. We first define the physical profile and illustrate what its parameters do to the ruler and the angular measurement. We then specify the data, unknown absolute scales, and joint likelihood. The illustrative curves use first-order perturbations; the fitted profiles use exact nonlinear shell dynamics.

\subsection{The density profile and its physical parameters}\label{sec:model}

We describe the present density relative to the homogeneous exterior by $\delta_m(t_0,r)=[\rho(t_0,r)-\rho_{\rm ext}(t_0)]/\rho_{\rm ext}(t_0)$. A negative value is an underdensity. The observer remains at the symmetry centre $r=0$; the location of the density minimum is allowed to move away from that centre. A depression at positive radius is therefore an observer-centred spherical shell. The parameter $r_0$ sets the present-day areal radius of the density minimum in the profile template. For $r_0=0$ the minimum lies at the observer, whereas for $r_0>0$ it lies on a spherical shell of radius $r_0$ around the observer. An overdensity at the observer can coexist with an underdense shell farther out.

The empirical profile of Hamaus, Sutter, and Wandelt motivates our radial function~\cite{Hamaus}; we refer to the resulting family as HSW-shaped. Its displaced local-density contrast is
\begin{equation}
\begin{aligned}
 \delta_m(t_0,r)&=-d\,U(r),\qquad x=\frac{r-r_0}{r_w},\\
 U(r)&=\frac{1}{\mathcal N}\frac{1-x^2}{1+(|x|/q_s)^{9.5}}.
\end{aligned}\label{eq:main_hsw}
\end{equation}
The depth $d=-\min\delta_m$ is the largest fractional density deficit, rather than necessarily the deficit at the observer. The location parameter $r_0$ places the depression, and $r_w$ sets its radial width. The density contrast is zero at $r=r_0-r_w$ and $r=r_0+r_w$, provided these radii are positive and lie outside the small smoothed region at the centre. Thus $r_w$ is the distance to the outer zero crossing, not the full width or a full width at half maximum. For $r_0=0$ the profile is a central depression; allowing positive $r_0$ adds shell-like profiles to the analysis.

Compensation requires the total mass deficit and excess to cancel, $\int_0^\infty r^2U(r)dr=0$, and fixes the dimensionless $q_s$ for each value of $r_0/r_w$. This quantity is distinct from the mass coordinate $q(r)$ introduced below. We smooth the profile inside 100 Mpc with an even, mass-preserving polynomial to make it regular at the origin, and normalize the smoothed $U$ to unit maximum through $\mathcal N$. Appendix~\ref{app:profiles} gives this prescription. Shifting the original void-centred HSW form and imposing this smoothing are explicit modeling choices.

The exterior parameter $\Omega_m$ specifies the matter density far from the feature. The mean contrast inside radius $r$ is $\bar\delta_m(r)=3r^{-3}\int_0^r u^2\delta_m(t_0,u)\,du$; it differs from the local contrast at the shell. Set $R(t_0,r)=r$, $A_m=\Omega_m(1+\bar\delta_m)$, and $\Omega_\Lambda=1-\Omega_m$ in the exterior. A common bang time fixes the curvature shell by shell through the age integral in Eq.~\eqref{eq:age}. We solve the nonlinear dust-plus-$\Lambda$ shell dynamics and null cone. These fits are not perturbative: we do not use a series expansion in a small depth $d$. We define the mass coordinate
\begin{equation}
 q(r)=r[A_m(r)/\Omega_m]^{1/3}.
\end{equation}
At early times $R(t_i,r)\to\bar a(t_i)\,q(r)$, so $q$ replaces $R(t_i,r)/\bar a(t_i)$ in Eq.~\eqref{eq:transport}, and the rulers no longer depend on the choice of $t_i$:
\begin{equation}
\begin{aligned}
 \frac{\Lp}{\rd}&=(1+z)\frac{R}{q},&
 \frac{\Ll}{\rd}&=(1+z)\frac{R'}{q'},\\
 X_\perp&=\frac{q}{\rd},&
 X_\parallel&=\frac{cq'}{(1+z)\dot R'\rd}.
\end{aligned}\label{eq:desi_hsw_transport}
\end{equation}
The transverse cancellation of $R$ does not remove the geometry: both $q(r)$ and the light-cone map $r(z)$ depend on the profile. In the radial expression, using $c/(H_\parallel\rd)$ would omit longitudinal ruler evolution.

Our direct search domain is
\begin{equation}
\begin{aligned}
 0.1&\leq\Omega_m\leq0.5,&0&\leq d\leq0.999,\\
 0.2&\leq r_w/{\rm Gpc}\leq8,&0&\leq r_0/{\rm Gpc}\leq6.
\end{aligned}\label{eq:main_domain}
\end{equation}

These are search limits rather than posterior priors. The nearby best fits lie inside them; the competing broad D+S+A solution lies at the boundary $r_0=0$. The narrow shells, with $r_w\simeq0.24$--$0.27$ Gpc, are close to the lower limit $r_w=0.2$ Gpc. Positive density and mass, positive metric signature, and a regular, monotonic observed null cone are required; reported solutions also pass sampled $R'>0$ and $q'>0$ checks. Details of the origin prescription and numerical validation are in Appendices~\ref{app:profiles} and \ref{app:robust_methods}.

With an uncalibrated SN magnitude and a free $\rd$, $H_0=68.8$ km s$^{-1}$ Mpc$^{-1}$ is a scale convention. Lengths rescale as $H_0^{-1}$; $H_0\rd$, $H_0r_w$, and $H_0r_0$ are the invariant fitted combinations. Flat $\Lambda$CDM has three free parameters: $\Omega_m$, $\rd$, and the SN magnitude offset ($k=3$). The HSW model adds the depth, width, and position of the void, $d$, $r_w$, and $r_0$ ($k=6$). Both models are fitted to the same data with the same covariance, and both leave $\rd$ and the SN offset free.

\subsection{Reading the effect of depth, width, and location}\label{sec:teaching_profiles}
\begin{figure*}[tbp]
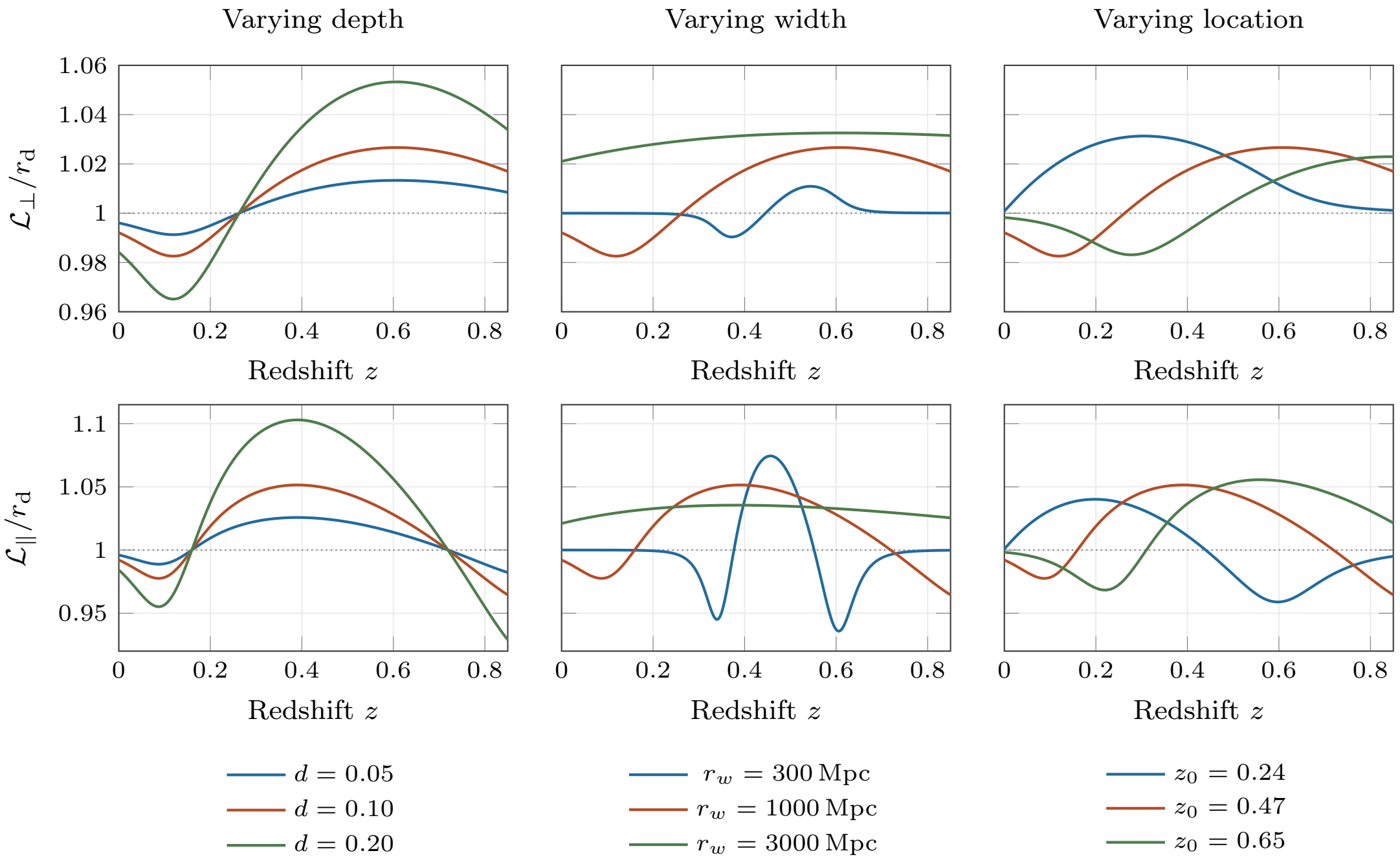

\centering


\caption{How a density depression changes the acoustic ruler. The upper and lower rows show the transverse and radial observed-comoving ruler lengths, respectively; the dotted lines are the FRW limit. Starting from the reference profile specified in Sec.~\ref{sec:teaching_profiles}, each column varies one parameter: depth, width, or trough location. The common emission-time correction and the unequal directional stretching are both included at first order. The curves are recomputed from the equations in Appendix~\ref{app:profiles}; their numerical checks are described in Appendix~\ref{app:illustrations}.}\label{fig:teaching_rulers}
\end{figure*}

\begin{figure*}[tbp]
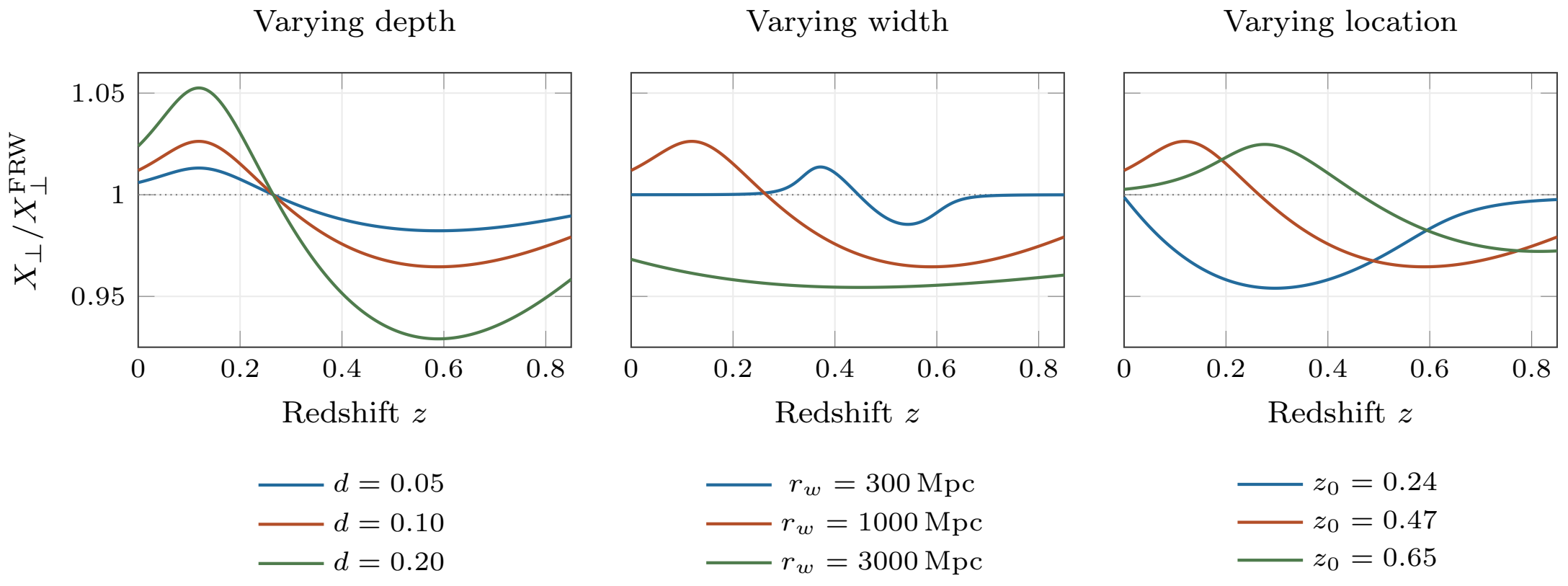

\centering


\caption{From a transported ruler to an angular BAO measurement. The ratio $X_\perp/X_\perp^{\rm FRW}$ includes both the changed distance and the changed transverse ruler for the same illustrative profiles as Fig.~\ref{fig:teaching_rulers}. Depth, width, and location are varied in turn. The horizontal dotted line is FRW. These first-order curves illustrate the common local transverse observable, not a fitted difference between angular and three-dimensional surveys.}\label{fig:teaching_angular}
\end{figure*}

Figures~\ref{fig:teaching_rulers} and \ref{fig:teaching_angular} provide a visual guide to the parameter dependence. They use the compensated first-order HSW-shaped template of Appendix~\ref{app:profiles}, with reference depth $d=0.1$, width $r_w=1000$ Mpc, and trough label $z_0=0.47$ in a flat background with $(\Omega_m,H_0)=(0.299,68.8\ {\rm km\,s^{-1}\,Mpc^{-1}})$. Here $r_0=\bar\chi(z_0)$, so $z_0$ labels the location in the reference background; it is not the exact observed redshift of a nonlinear fitted shell. These illustrations use the first-order equations, while the fits use the exact solution. No parameters in these figures have been fitted to the observations.

At linear order, the transverse ruler responds to the mean density contrast enclosed within a shell, while the radial ruler also responds to the density at the shell itself. Including the emission-time correction discussed in Sec.~\ref{sec:geometry}, Appendix~\ref{app:profiles} gives
\begin{align}
 \frac{\Lp}{\rd}-1&\simeq\delta\ln a-\frac{\bar\delta}{3},\label{eq:teaching_transverse_ruler}\\
 \frac{\Ll}{\rd}-1&\simeq\delta\ln a+\frac{2\bar\delta}{3}-\delta_m.
 \label{eq:teaching_linear_rulers}
\end{align}
Here $\delta_m$ and its enclosed mean $\bar\delta$ are evaluated at emission, and $\delta\ln a=Q-1$ to first order. The time shift $\delta\ln a$ stretches both rulers equally. Their ratio, $\Ll/\Lp-1\simeq\bar\delta-\delta_m$, is set by the difference between the mean density inside the shell and the density at the shell. Since $\bar\delta$ is an average over the whole sphere, it varies smoothly with distance, while $\delta_m$ varies as fast as the profile itself; the radial ruler therefore follows sharper features than the transverse one.

In Fig.~\ref{fig:teaching_rulers}, increasing the depth increases the departures from FRW in proportion to $d$ within the linear approximation. A narrow profile produces a more localized feature, while a broad profile changes the ruler over a larger redshift interval. Moving the trough moves the affected interval. These are useful physical expectations, not a reason to extrapolate the linear curves to nearly empty voids; the results in Sec.~\ref{sec:desi_hsw} use the nonlinear solution.

Figure~\ref{fig:teaching_angular} follows the next step in the measurement: dividing the geometric distance by the transverse ruler. It shows $X_\perp(z)/X_\perp^{\rm FRW}(z)$, so a curve below unity means that angular BAO would report a smaller distance-to-ruler ratio than in the reference cosmology. The figure combines the distance and ruler changes; it is not obtained by holding the distance fixed. An ideal local three-dimensional transverse measurement would follow the same curve. The distinction between the actual surveys enters through their different selections and estimators, as explained in Sec.~\ref{sec:geometry}.

The figures also show why the ruler-shape test and the direct fit are complementary. The ruler-shape test removes any constant factor, so a broad void, which changes the ruler by nearly the same amount at all redshifts, is invisible to it, whereas a void that makes the ruler vary across the observed redshifts remains detectable. The direct fit also uses the supernova distances themselves and the radial BAO, so it can constrain voids that the ratio test cannot. How strongly the data constrain voids is determined by the fits in Secs.~\ref{sec:desi_hsw} and \ref{sec:robustness}, not by these examples.

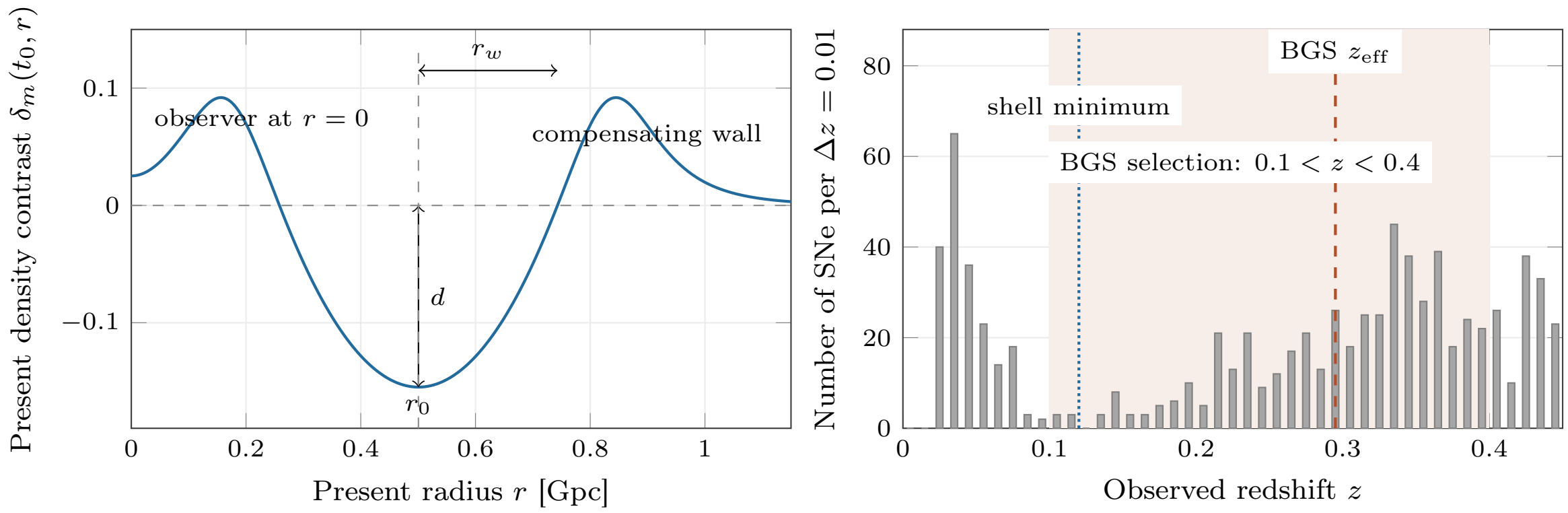
\begin{figure*}[tbp]
\centering
\begin{tikzpicture}
\begin{groupplot}[group style={group size=2 by 1,horizontal sep=1.1cm},reviewaxis,width=.45\textwidth,height=5.5cm]
\nextgroupplot[xlabel={Present radius $r$ [Gpc]},ylabel={Present density contrast $\delta_m(t_0,r)$},xmin=0,xmax=1.15,ymin=-.19,ymax=.15]
\addplot[black!45,dashed] coordinates {(0,0) (1.15,0)};
\addplot[plotblue,thick] coordinates {(0,0.02515443) (0.003833333,0.02522791) (0.007666667,0.02544815) (0.0115,0.02581451) (0.01533333,0.02632594) (0.01916667,0.02698096) (0.023,0.02777765) (0.02683333,0.02871368) (0.03066667,0.02978629) (0.0345,0.03099227) (0.03833333,0.03232801) (0.04216667,0.03378941) (0.046,0.03537198) (0.04983333,0.03707075) (0.05366667,0.03888031) (0.0575,0.04079478) (0.06133333,0.04280784) (0.06516667,0.04491269) (0.069,0.04710207) (0.07283333,0.0493682) (0.07666667,0.05170286) (0.0805,0.05409731) (0.08433333,0.05654232) (0.08816667,0.05902812) (0.092,0.06154447) (0.09583333,0.06408055) (0.09966667,0.06662503) (0.1035,0.06916543) (0.1073333,0.07168513) (0.1111667,0.07416466) (0.115,0.07658268) (0.1188333,0.07891608) (0.1226667,0.08114024) (0.1265,0.08322931) (0.1303333,0.08515658) (0.1341667,0.08689498) (0.138,0.08841755) (0.1418333,0.08969804) (0.1456667,0.09071154) (0.1495,0.09143508) (0.1533333,0.09184823) (0.1571667,0.09193368) (0.161,0.0916777) (0.1648333,0.09107055) (0.1686667,0.09010671) (0.1725,0.08878503) (0.1763333,0.0871087) (0.1801667,0.0850851) (0.184,0.08272556) (0.1878333,0.08004489) (0.1916667,0.07706098) (0.1955,0.07379423) (0.1993333,0.07026697) (0.2031667,0.06650293) (0.207,0.06252662) (0.2108333,0.05836288) (0.2146667,0.05403637) (0.2185,0.04957117) (0.2223333,0.04499048) (0.2261667,0.04031629) (0.23,0.03556926) (0.2338333,0.03076853) (0.2376667,0.02593169) (0.2415,0.02107468) (0.2453333,0.01621189) (0.2491667,0.01135614) (0.253,0.006518783) (0.2568333,0.001709779) (0.2606667,-0.003062203) (0.2645,-0.007789673) (0.2683333,-0.01246622) (0.2721667,-0.01708638) (0.276,-0.02164557) (0.2798333,-0.02613995) (0.2836667,-0.03056633) (0.2875,-0.03492213) (0.2913333,-0.03920522) (0.2951667,-0.04341394) (0.299,-0.04754694) (0.3028333,-0.0516032) (0.3066667,-0.05558195) (0.3105,-0.05948261) (0.3143333,-0.06330478) (0.3181667,-0.0670482) (0.322,-0.0707127) (0.3258333,-0.0742982) (0.3296667,-0.07780469) (0.3335,-0.0812322) (0.3373333,-0.08458081) (0.3411667,-0.0878506) (0.345,-0.09104168) (0.3488333,-0.09415417) (0.3526667,-0.09718817) (0.3565,-0.1001438) (0.3603333,-0.1030212) (0.3641667,-0.1058204) (0.368,-0.1085416) (0.3718333,-0.1111847) (0.3756667,-0.11375) (0.3795,-0.1162375) (0.3833333,-0.1186471) (0.3871667,-0.1209791) (0.391,-0.1232333) (0.3948333,-0.12541) (0.3986667,-0.127509) (0.4025,-0.1295304) (0.4063333,-0.1314742) (0.4101667,-0.1333405) (0.414,-0.1351293) (0.4178333,-0.1368405) (0.4216667,-0.1384742) (0.4255,-0.1400304) (0.4293333,-0.1415091) (0.4331667,-0.1429103) (0.437,-0.144234) (0.4408333,-0.1454802) (0.4446667,-0.1466488) (0.4485,-0.14774) (0.4523333,-0.1487537) (0.4561667,-0.1496898) (0.46,-0.1505485) (0.4638333,-0.1513297) (0.4676667,-0.1520333) (0.4715,-0.1526595) (0.4753333,-0.1532082) (0.4791667,-0.1536794) (0.483,-0.154073) (0.4868333,-0.1543892) (0.4906667,-0.1546279) (0.4945,-0.154789) (0.4983333,-0.1548727) (0.5021667,-0.1548789) (0.506,-0.1548075) (0.5098333,-0.1546587) (0.5136667,-0.1544324) (0.5175,-0.1541286) (0.5213333,-0.1537472) (0.5251667,-0.1532884) (0.529,-0.1527521) (0.5328333,-0.1521382) (0.5366667,-0.1514469) (0.5405,-0.1506781) (0.5443333,-0.1498318) (0.5481667,-0.1489079) (0.552,-0.1479066) (0.5558333,-0.1468278) (0.5596667,-0.1456714) (0.5635,-0.1444376) (0.5673333,-0.1431263) (0.5711667,-0.1417374) (0.575,-0.1402711) (0.5788333,-0.1387272) (0.5826667,-0.1371059) (0.5865,-0.135407) (0.5903333,-0.1336306) (0.5941667,-0.1317766) (0.598,-0.1298451) (0.6018333,-0.127836) (0.6056667,-0.1257494) (0.6095,-0.1235851) (0.6133333,-0.1213432) (0.6171667,-0.1190237) (0.621,-0.1166264) (0.6248333,-0.1141513) (0.6286667,-0.1115984) (0.6325,-0.1089677) (0.6363333,-0.1062589) (0.6401667,-0.1034722) (0.644,-0.1006072) (0.6478333,-0.09766406) (0.6516667,-0.09464255) (0.6555,-0.09154257) (0.6593333,-0.08836401) (0.6631667,-0.08510676) (0.667,-0.08177071) (0.6708333,-0.07835576) (0.6746667,-0.07486185) (0.6785,-0.07128893) (0.6823333,-0.06763701) (0.6861667,-0.06390616) (0.69,-0.06009651) (0.6938333,-0.05620832) (0.6976667,-0.05224198) (0.7015,-0.04819801) (0.7053333,-0.04407717) (0.7091667,-0.03988043) (0.713,-0.03560908) (0.7168333,-0.03126473) (0.7206667,-0.02684942) (0.7245,-0.02236568) (0.7283333,-0.01781657) (0.7321667,-0.01320583) (0.736,-0.008537921) (0.7398333,-0.003818157) (0.7436667,0.0009472093) (0.7475,0.005750867) (0.7513333,0.01058434) (0.7551667,0.01543788) (0.759,0.02030037) (0.7628333,0.02515922) (0.7666667,0.03000031) (0.7705,0.03480792) (0.7743333,0.03956474) (0.7781667,0.04425187) (0.782,0.0488489) (0.7858333,0.05333401) (0.7896667,0.0576842) (0.7935,0.0618755) (0.7973333,0.06588331) (0.8011667,0.06968281) (0.805,0.07324939) (0.8088333,0.07655917) (0.8126667,0.07958954) (0.8165,0.08231976) (0.8203333,0.08473152) (0.8241667,0.08680946) (0.828,0.08854169) (0.8318333,0.08992017) (0.8356667,0.09094103) (0.8395,0.09160472) (0.8433333,0.09191607) (0.8471667,0.09188417) (0.851,0.09152213) (0.8548333,0.09084673) (0.8586667,0.08987792) (0.8625,0.08863832) (0.8663333,0.08715257) (0.8701667,0.08544675) (0.874,0.08354773) (0.8778333,0.08148263) (0.8816667,0.07927822) (0.8855,0.07696053) (0.8893333,0.0745544) (0.8931667,0.07208318) (0.897,0.06956853) (0.9008333,0.06703021) (0.9046667,0.06448601) (0.9085,0.06195177) (0.9123333,0.0594413) (0.9161667,0.05696655) (0.92,0.05453764) (0.9238333,0.052163) (0.9276667,0.04984948) (0.9315,0.04760254) (0.9353333,0.04542633) (0.9391667,0.0433239) (0.943,0.04129727) (0.9468333,0.03934762) (0.9506667,0.03747537) (0.9545,0.03568032) (0.9583333,0.03396172) (0.9621667,0.03231841) (0.966,0.03074885) (0.9698333,0.02925122) (0.9736667,0.02782348) (0.9775,0.02646344) (0.9813333,0.02516875) (0.9851667,0.02393702) (0.989,0.02276577) (0.9928333,0.02165254) (0.9966667,0.02059486) (1.0005,0.01959028) (1.004333,0.01863638) (1.008167,0.01773081) (1.012,0.01687127) (1.015833,0.01605554) (1.019667,0.01528146) (1.0235,0.01454695) (1.027333,0.01385002) (1.031167,0.01318875) (1.035,0.01256132) (1.038833,0.01196596) (1.042667,0.01140101) (1.0465,0.01086488) (1.050333,0.01035605) (1.054167,0.009873085) (1.058,0.009414612) (1.061833,0.008979334) (1.065667,0.008566021) (1.0695,0.008173506) (1.073333,0.007800683) (1.077167,0.007446505) (1.081,0.007109981) (1.084833,0.006790175) (1.088667,0.006486199) (1.0925,0.006197216) (1.096333,0.005922431) (1.100167,0.005661097) (1.104,0.005412503) (1.107833,0.005175982) (1.111667,0.004950899) (1.1155,0.004736656) (1.119333,0.004532689) (1.123167,0.004338464) (1.127,0.004153474) (1.130833,0.003977244) (1.134667,0.003809322) (1.1385,0.003649281) (1.142333,0.003496719) (1.146167,0.003351254) (1.15,0.003212527)};
\draw[<->,black] (axis cs:0.5005552,0) -- (axis cs:0.5005552,-0.1548857) node[midway,right,font=\scriptsize] {$d$};
\draw[dashed,black!45] (axis cs:0.5005552,-.19) -- (axis cs:0.5005552,.13);
\draw[<->,black] (axis cs:0.5005552,.115) -- (axis cs:0.7429074,.115) node[midway,above,font=\scriptsize] {$r_w$};
\node[font=\scriptsize,anchor=south] at (axis cs:0.5005552,-.188) {$r_0$};
\node[font=\scriptsize,anchor=west] at (axis cs:.02,.075) {observer at $r=0$};
\node[font=\scriptsize,anchor=east] at (axis cs:1.12,.06) {compensating wall};
\nextgroupplot[xlabel={Observed redshift $z$},ylabel={Number of SNe per $\Delta z=0.01$},xmin=0,xmax=.45,ymin=0,ymax=88]
\path[fill=plotred!10] (axis cs:.1,0) rectangle (axis cs:.4,88);
\addplot[ybar,bar width=1.9pt,fill=black!35,draw=black!50] coordinates {(0.005,0) (0.015,0) (0.025,40) (0.035,65) (0.045,36) (0.055,23) (0.065,14) (0.075,18) (0.085,3) (0.095,2) (0.105,3) (0.115,3) (0.125,0) (0.135,3) (0.145,8) (0.155,3) (0.165,3) (0.175,5) (0.185,6) (0.195,10) (0.205,5) (0.215,21) (0.225,13) (0.235,21) (0.245,9) (0.255,12) (0.265,17) (0.275,21) (0.285,13) (0.295,26) (0.305,18) (0.315,25) (0.325,25) (0.335,45) (0.345,38) (0.355,28) (0.365,39) (0.375,18) (0.385,24) (0.395,22) (0.405,26) (0.415,10) (0.425,38) (0.435,33) (0.445,23)};
\draw[plotred,thick,dashed] (axis cs:.295,0) -- (axis cs:.295,88);
\draw[plotblue,thick,densely dotted] (axis cs:0.1200342,0) -- (axis cs:0.1200342,88);
\node[font=\scriptsize,anchor=north,fill=white] at (axis cs:.295,88) {BGS $z_{\rm eff}$};
\node[font=\scriptsize,anchor=north,fill=white] at (axis cs:0.1200342,75.67999999999999) {shell minimum};
\node[font=\scriptsize,anchor=north,fill=white] at (axis cs:.23,63.36) {BGS selection: $0.1<z<0.4$};
\end{groupplot}
\end{tikzpicture}
\caption{Reading the primary DESI+SN (D+S) profile. Left: the present density contrast of the nearby narrow shell. The observer is central; $r_0$ locates an underdense spherical shell, $d$ is its fractional depth, and $r_w$ reaches its outer zero crossing. The central contrast is $+0.0252$. Right: the DES-Dovekie redshift distribution near the feature and the DESI Bright Galaxy Survey (BGS) selection interval. The shell minimum appears at $z=0.1200$, within that interval but below its effective redshift $0.295$. Evaluating one local prediction at $z_{\rm eff}$ can therefore miss variation sampled by the survey window. The BGS shading shows the published selection range, not a measured pair-response weight.}\label{fig:profile_guide}
\end{figure*}

Figure~\ref{fig:profile_guide} illustrates the parameters for the primary fitted profile. Its central density contrast is positive even though the surrounding shell is underdense. A preferred central overdensity in a different profile parametrization therefore need not conflict with the underdense-shell description used here.

\subsection{BAO and supernova inputs}
The DESI DR2 data vector $\bm y_{\rm D}$ contains 13 measurements with a Gaussian likelihood: $D_{\mathrm V}/\rd$ at $z=0.295$, and $\DM/\rd$ and $\DH/\rd$ at six effective redshifts~\cite{DESI,DESIdata}. We keep the published ordering, in which the final pair lists the radial measurement first, and use the published covariance matrix $C_{\rm D}$ rather than one reconstructed from rounded error bars. The DESI contribution is
\begin{equation}
 \chi^2_{\rm D}=(\bm y_{\rm D}-\bm X)^T C_{\rm D}^{-1}
 (\bm y_{\rm D}-\bm X),\label{eq:desichi}
\end{equation}
where $\bm X$ contains the model predictions $X_V$, $X_\perp$, and $X_\parallel$ for the corresponding measurements. Its interpretation follows the local-summary assumptions in Sec.~\ref{sec:geometry}.

The legacy angular compilation contains 15 points over $0.110\leq z\leq2.225$~\cite{Nunes,Pantos}; Table~\ref{tab:angular} lists the measured angles. At $0.57\leq z\leq0.65$ we use the five original SDSS DR11 measurements in Table~3 of Carvalho et al., arXiv version~4~\cite{CarvalhoDR11}, rather than the different values reproduced in the Nunes compilation. The direct density-profile fits use the first fourteen points ($z\leq0.65$), including when DESI is added; the fixed-profile benchmark fits in the technical supplement use all fifteen. We fit the measured angles $\bm\theta$ (in degrees) directly, since their errors are Gaussian in $\theta$ but not in $1/\theta$. With covariance $C_\theta$, the angular contribution is
\begin{equation}
 \chi^2_\theta=(\bm\theta-\bm\theta_{\rm th})^T
 C_\theta^{-1}(\bm\theta-\bm\theta_{\rm th}),\qquad
 \theta_{\rm th}^{\rm(deg)}=\frac{180/\pi}{X_\perp},
 \label{eq:anglechi}
\end{equation}
where $\theta_{\rm th}$ is the predicted angle.

We treat the angular measurements as independent, with a diagonal covariance, although they may be correlated through their common sky area, selection and large-scale structure. Their correlation with DESI is likewise unknown and set to zero. The choice of angular catalogue also matters: thin-shell measurements obtained with different methods for extracting the BAO feature can differ in normalization and scatter~\cite{Menote}. Catalogues built from overlapping galaxy samples must therefore not be combined as independent data.

The primary SN likelihood uses all 1820 released DES-Dovekie magnitudes and their full statistical-plus-systematic precision matrix, which is the inverse covariance~\cite{Popovic,DovekieData}. The separate Pantheon+ fits test the effect of changing the SN compilation. They use corrected magnitudes and the full covariance~\cite{Pantheon,PantheonData}, with $z_{\rm HD}>0.01$ and the distance-ladder calibrators excluded: 1580 light-curve rows for 1466 distinct identifiers. Some supernovae appear more than once, observed by different surveys; we keep the published covariance between these repeated entries, so they are not counted as independent. We leave the SN absolute magnitude free, as a constant offset fitted to the data, so no external calibration of the SN luminosity is needed. Appendix~\ref{app:robust_methods} records the data-field and precision-matrix conventions.

The ruler-shape diagnostic reconstructs the Dovekie distance--redshift relation and compares it with a separate Pantheon+ reconstruction. The catalogues overlap, and a reconstruction reuses its parent SN observations. We therefore compare these analyses separately rather than multiplying their likelihoods as independent evidence.

\subsection{Removing the unknown SN normalization}
For each trial cosmology we choose the constant magnitude offset that minimizes the SN $\chi^2$. This procedure, called profiling, allows the data to determine the shape of the distance--redshift relation without assuming an absolute SN luminosity.
The predicted apparent magnitude of supernova $i$ is
\begin{equation}
 m_i^{\rm th}=5\log_{10}\frac{\DL(z_i)}{\rm Mpc}+25+\mathcal M,
\end{equation}
where $\mathcal M$ is the SN absolute magnitude, treated as a free constant. Since $H_0$ is fixed only as a convention, $\mathcal M$ also absorbs any change in the overall distance scale.

With $v_i=m_i-5\log_{10}[\DL(z_i)/{\rm Mpc}]-25$, minimizing over $\mathcal M$ gives
\begin{align}
 \widehat{\mathcal M}&=\frac{\bm1^TC_{\rm SN}^{-1}\bm v}
 {\bm1^TC_{\rm SN}^{-1}\bm1},\\
 \chi^2_{\rm SN}&=\bm v^TC_{\rm SN}^{-1}\bm v-
 \frac{(\bm1^TC_{\rm SN}^{-1}\bm v)^2}
 {\bm1^TC_{\rm SN}^{-1}\bm1}.\label{eq:snchi}
\end{align}
where $\bm1$ is the vector with all entries equal to one.

An overall distance rescaling is therefore unobservable to this SN likelihood. The release distinguishes the heliocentric redshift $z_{\rm HEL}$ from the corrected redshift $z_{\rm HD}$ used for the Hubble diagram. To retain that convention, the prediction is $\DL=(1+z_{\rm HEL})\DM(z_{\rm HD})$. In flat FRW, $\DM=(c/H_0)\int_0^{z_{\rm HD}}du/E(u)$; in the conditional LTB application it is the computed central transverse distance. A fully self-consistent inhomogeneous reanalysis would revisit the release's peculiar-velocity corrections, selection effects, and covariance construction.

\subsection{Combined likelihood and its interpretation}
For a physical LTB parameter vector
\begin{equation}
 \bm p=(d,r_w,r_0,\bm\lambda_{\rm shape},
 \bm p_{\rm exterior},\rd),\label{eq:parameters}
\end{equation}
$\bm\lambda_{\rm shape}$ denotes any additional profile-shape parameters and $\bm p_{\rm exterior}$ the background parameters. In the main HSW-shaped fit there are no additional freely varied shape parameters and only $\Omega_m$ varies in the exterior; the SN offset is profiled separately. No absolute-magnitude or sound-horizon prior is imposed. The primary direct likelihood is $\chi^2_{\rm D}+\chi^2_{\rm SN}$. The secondary D+S+A fit adds $\chi^2_\theta$ under the independence assumptions above. Geometry and both rulers must be recomputed from the same $\bm p$. Varying a ruler phenomenologically while leaving all distances and radial BAO predictions unchanged is a different model.

The ratio likelihood in Sec.~\ref{sec:shape} is an alternative compression of SN and transverse BAO information. It is particularly useful for interpreting ruler evolution and removing a constant calibration. Because it reuses the parent measurements, it must not be added to their direct likelihoods as independent evidence. The same warning applies to a shape statistic using DESI transverse data and a full DESI likelihood containing those components.

A best-fit point, a parameter scan, and a confidence region have different meanings. A fixed-parameter scan changes one quantity while holding the others fixed. A profile likelihood instead reoptimizes the other parameters at each trial value. Only with an appropriate statistical calibration can likelihood differences be converted into confidence statements.

For a single parameter, we report a nominal $\Delta\chi^2=1$ interval as the range over which $\chi^2$ rises by at most 1 above its minimum, with the other parameters refitted at each value. It is a $1\sigma$ interval only under the appropriate local Gaussian approximation. For the joint depth and width $(d,r_w)$, the Supplemental Material~\cite{SupplementalMaterial} shows the levels $\Delta\chi^2=2.30$ and $6.18$, which correspond to $68\%$ and $95\%$ only if the likelihood is Gaussian, so they are approximate here. These rules fail near $d=0$: without a void, its width and location have no effect on the observables, so the data cannot constrain them. For this reason, Section~\ref{sec:robustness} uses simulations, rather than these rules, to assess the significance of the main fit.

\section{Constraining ruler evolution and radial-profile shape}\label{sec:shape}
\subsection{A calibration-independent ruler ratio}
Let an uncalibrated SN reconstruction give
\begin{equation}
 \widehat\DM^{\rm SN}(z)=A\DM(z),\label{eq:SNrecon}
\end{equation}
where $A$ is a redshift-independent normalization. The conventional transverse BAO distance proxy is $\widehat\DM^{\rm BAO}(z)=\rd X_\perp(z)=\rd\DM(z)/\Lp(z)$. Hence
\begin{equation}
 \frac{\widehat\DM^{\rm SN}(z)}{\widehat\DM^{\rm BAO}(z)}
 =A\frac{\Lp(z)}{\rd}.\label{eq:ratio}
\end{equation}
Taking this ratio at two redshifts removes $A$ and the common $\rd$ normalization:
\begin{equation}
 \frac{\widehat\DM^{\rm SN}(z_b)/\widehat\DM^{\rm SN}(z_a)}
 {X_\perp(z_b)/X_\perp(z_a)}
 =\frac{\Lp(z_b)}{\Lp(z_a)}.\label{eq:double_ratio}
\end{equation}
For example, if the double ratio on the left is $1.03$, the effective transverse ruler at $z_b$ is 3\% longer than at $z_a$, regardless of the unknown absolute SN luminosity or initial ruler length. This is an illustrative interpretation, not an additional measurement. In FRW with a constant comoving ruler the ratio is unity. Thus BAO and SNe together measure a change in effective ruler length, rather than its absolute calibration. The relation assumes reciprocity, a constant SN absolute magnitude after correction, unbiased transverse BAO recovery, and compatible effective redshift weighting. Evolution of SN luminosity or BAO calibration can mimic ruler evolution and must be treated separately when relaxing these assumptions.

Equation~\eqref{eq:double_ratio} applies equally to an ideal angular or three-dimensional transverse BAO estimate. The same construction does not work for the radial ruler: radial BAO measures $\Hl\Ll$, and supernovae provide no independent measurement of $\Hl$ to divide it out. The radial ruler is therefore constrained only through the LTB model, in the direct fit of Sec.~\ref{sec:desi_hsw}. 

\subsection{A ruler-shape likelihood and the parameters it measures}
Here \emph{ruler shape} means the redshift dependence of the effective ruler, whereas \emph{density-profile shape} describes how matter density varies with radius. The former constrains the latter through the chosen LTB dynamics. Taking the logarithm of Eq.~\eqref{eq:ratio} at each BAO redshift $z_i$ gives
\begin{equation}
\begin{gathered}
 y_i\equiv\ln\frac{\widehat\DM^{\rm SN}(z_i)}{\rd X_\perp(z_i)}
 =c+\eta_i(\bm p)+e_i,\\
 \eta_i(\bm p)=\ln\frac{\Lp(z_i;\bm p)}{\rd},
\end{gathered}
\end{equation}
where $c=\ln A$ is the same unknown constant at every redshift, $\eta_i$ is the model prediction for parameters $\bm p$, and $e_i$ is the measurement noise, with covariance $C_y$. For an LTB void, $\eta_i$ is determined by the void parameters $\bm p$; the rest of this section describes how the ruler shape, and therefore the statistic below, depends on them. We remove $c$ below, so that the statistic responds only to how the ruler changes with redshift. If $C_S$ and $C_X$ are the covariances of the logarithmic SN distances and logarithmic BAO scales, then
\begin{equation}
 C_y=C_S+C_X-C_{SX}-C_{XS}.\label{eq:cy}
\end{equation}
The cross terms vanish only under an independence assumption. The reconstruction induces correlations among SN distances at different redshifts, even if the requested evaluation points are distinct.

Minimizing $\chi^2$ over the constant $c$, as for $\mathcal M$ in Sec.~\ref{sec:data}, gives
\begin{align}
 P_y&=C_y^{-1}-\frac{C_y^{-1}\bm1\bm1^TC_y^{-1}}
 {\bm1^TC_y^{-1}\bm1},\\
 \chi^2_{\rm shape}(\bm p)&=(\bm y-\bm\eta)^TP_y(\bm y-\bm\eta).
 \label{eq:shapechi}
\end{align}
Since $P_y\bm1=0$, a ruler rescaling that is the same at all redshifts does not change $\chi^2_{\rm shape}$. The Fisher matrix of this statistic is $F^{\rm shape}_{ab}=\bm J_a^TP_y\bm J_b$, where $\bm J_a$ has components $J_{ia}=\partial\eta_i/\partial p_a$; it measures how well the data constrain each parameter combination through the redshift dependence of the ruler alone. A parameter combination is identifiable here only if it changes the predicted redshift dependence. A combination that changes only the normalization cannot be measured by this test.

For a fixed shape, width, and location in linear theory, $\bm\eta=d\bm u$. The unconstrained best-fit depth and conditional curvature are
\begin{equation}
 \widehat d=\frac{\bm u^TP_y\bm y}{\bm u^TP_y\bm u},\qquad
 \sigma_d^2=\frac{1}{\bm u^TP_y\bm u}.\label{eq:depth_projection}
\end{equation}
Here $\bm u^TP_y\bm u$ is the Fisher matrix $F^{\rm shape}$ for the single parameter $d$, for which $\bm J=\bm u$.
Physical bounds and nonlinear evolution modify this simple Gaussian interpretation. If a broad profile makes $\bm u$ nearly constant over the observed interval, $\bm u^TP_y\bm u$ becomes small and depth is weakly constrained. At $d=0$ there is no density depression, so assigning it a width or a location has no observable meaning. These degeneracies explain why a null ruler-evolution test can still allow substantial, slowly varying inhomogeneity.

The likelihood in Eq.~\eqref{eq:shapechi} uses a Gaussian approximation in logarithmic observables and a smooth SN reconstruction. It is useful for interpreting the ruler shape, but is not identical to the native-angle likelihood in Eq.~\eqref{eq:anglechi}. When precise probabilities are needed, or when the errors are large enough that taking logarithms and ratios makes them noticeably non-Gaussian, it is better to fit the measured magnitudes and angles directly.

\subsection{What the current ruler-shape data say}\label{sec:shape_data}
We now measure the ruler shape. Its simplest summary is the slope $b$ of a linear trend, $\ln\Lp=c+b(z-z_*)$. Using the fourteen angular BAO measurements over $0.11\leq z\leq0.65$ and the DES-Dovekie SN distances, we find
\begin{equation}
 b\equiv\frac{d\ln\Lp}{dz}=0.0589\pm0.0941,\label{eq:slope}
\end{equation}
consistent with a constant ruler.
The SN distances are reconstructed by fitting $\ln[\widehat\DM^{\rm SN}/z]$ as a degree-four polynomial in $\ln(1+z)$ to the DES-Dovekie distance moduli with their full covariance; the coefficient covariance is propagated to the BAO redshifts, and the SN and BAO errors are treated as independent. Polynomial degrees three to six give slopes between $0.055$ and $0.062$~\cite{SupplementalMaterial}.

This constrains only a linear trend, not arbitrary functions of redshift. Equivalently, $\ln[\Lp(0.65)/\Lp(0.11)]=0.0318\pm0.0508$. The statistic $\chi^2_{\rm shape}$ is $12.516$ for a constant ruler and $12.124$ with the slope free. Fitting the measured angles directly, with the reconstruction coefficients refitted at each $b$, gives $b=0.0475$, with a $\Delta\chi^2=1$ interval $-0.0453<b<0.1405$. Using Pantheon+ instead of DES-Dovekie gives $b=0.0028\pm0.0949$. Both SN samples are therefore consistent with a constant transverse ruler.

The measured slope therefore limits voids that change the ruler steadily across the observed redshifts. It constrains other voids only weakly: a broad profile changes the ruler almost uniformly, which the unknown constant absorbs, and a narrow shell can lie between the sampled redshifts. These limits refer to the transverse ruler at the time of emission, not to the primordial sound horizon, and direct SN distances and radial BAO are needed to distinguish such models.

Fitting specific void profiles to the same ratio data, with the trough fixed at $z\simeq0.5$ and the exact nonlinear ruler, gives nominal $\Delta\chi^2=6.18$ upper limits on the depth between $0.21$ and $0.52$ for six profile shapes, tightest at intermediate widths, $650\lesssim r_w\lesssim1075\,$Mpc; for the HSW-shaped profile, $d<0.23$ at $r_w=825\,$Mpc~\cite{SupplementalMaterial}. No profile is significantly preferred over a constant ruler: the largest improvement, $\Delta\chi^2=3.5$ for the HSW-shaped profile, occurs by chance with probability $0.19$ in simulations that search all six families. Broad and very narrow voids remain weakly constrained, for the reasons given above.

The ratio can also cancel distance changes that remain visible in the supernova observations themselves. For example, the deep HSW-shaped void preferred by the ratio data ($d\simeq0.76$, $r_w=200\,$Mpc) fits the SN magnitudes and angles directly worse than no void~\cite{SupplementalMaterial}. This motivates the direct physical fit in the next section. The two analyses use the same measurements and are never counted as independent evidence.
\section{Exact nonlinear fits to DESI DR2 and supernovae}\label{sec:desi_hsw}
\subsection{Matched likelihood and main results}
The primary D+S fit combines the thirteen DESI DR2 measurements with the 1820 DES-Dovekie SNe. The secondary D+S+A fit adds fourteen angular BAO measurements. For each trial density profile we solve the light cone to $z=2.33$, the effective redshift of the highest-redshift DESI measurement (Lyman-$\alpha$), and evaluate the transverse, radial, and isotropic observables using Eq.~\eqref{eq:desi_hsw_transport} and $X_V=(zX_\perp^2X_\parallel)^{1/3}$. The likelihood retains the full released covariance. One common initial ruler length is fitted to all BAO measurements; the independent SN magnitude offset is fitted to the supernovae. Appendix~\ref{app:robust_methods} gives the analytic minimization over these scales.

We search from several starting profiles and repeat the optimization with independent seeds. The search finds several local minima: different parameter choices from which small changes worsen the fit. Table~\ref{tab:desi_hsw} lists the lowest located fit and a competing broad profile for each dataset. These are not different models: both are local minima of the same HSW-shaped model, and the labels only describe the profile at each minimum. We cannot guarantee that no better fit exists elsewhere in parameter space; the reported fits are the best that the search found. The flat-$\Lambda$CDM comparison uses exactly the same measurements and freedom in the two absolute scales.

\begin{table*}[tbp]
\caption{Matched fits and competing density profiles. D+S denotes DESI plus the 1820 DES-Dovekie SNe; D+S+A adds fourteen angular BAO measurements. $\chi^2_{\rm SN}$, $\chi^2_{\rm D}$, and $\chi^2_\theta$ are the SN, DESI, and angular-BAO contributions, and the total is their sum; D+S does not include $\chi^2_\theta$. The narrow-shell and broad-profile rows are exact HSW-shaped LTB fits with the same six fitted parameters. The primary broad solution is a shell at positive radius; the secondary broad solution is a central depression. Both LTB and flat $\Lambda$CDM refit $\Omega_m$, the common initial ruler, and the SN offset; LTB additionally fits $d,r_w,r_0$. Lengths use the common $H_0=68.8\ {\rm km\,s^{-1}\,Mpc^{-1}}$ scale convention without an absolute calibration prior. Entries are fitted points, not posterior means or confidence intervals. Differences quoted in the text are calculated before rounding, so subtracting displayed entries can differ in the last digit.}\label{tab:desi_hsw}
\small
\setlength{\tabcolsep}{3pt}
\begin{ruledtabular}
\begin{tabular}{llrrrrrrrrr}
Data & Model/profile & $\Omega_m$ & $d$ & $r_w$ [Gpc] & $r_0$ [Gpc] & $\widehat\rd$ [Mpc] & $\chi^2_{\rm SN}$ & $\chi^2_{\rm D}$ & $\chi^2_\theta$ & Total\\
D+S & Flat $\Lambda$CDM & 0.30605 & 0 & --- & --- & 146.616 & 1634.045 & 11.232 & --- & 1645.278\\
D+S & Narrow shell & 0.30233 & 0.15489 & 0.242 & 0.501 & 147.020 & 1626.196 & 10.559 & --- & 1636.755\\
D+S & Broad shell & 0.32329 & 0.04166 & 1.960 & 2.186 & 143.402 & 1630.512 & 6.890 & --- & 1637.402\\
D+S+A & Flat $\Lambda$CDM & 0.29770 & 0 & --- & --- & 147.919 & 1636.223 & 11.044 & 29.849 & 1677.116\\
D+S+A & Narrow shell & 0.29478 & 0.15557 & 0.271 & 0.538 & 148.227 & 1628.014 & 10.775 & 27.554 & 1666.342\\
D+S+A & Broad central & 0.35433 & 0.10239 & 4.406 & 0.000 & 139.229 & 1630.426 & 11.044 & 27.721 & 1669.191\\
\end{tabular}
\end{ruledtabular}
\end{table*}

For D+S, HSW improves the total by
\begin{equation}
 T_{\rm obs}\equiv\chi^2_{\Lambda{\rm CDM},\min}-\chi^2_{{\rm HSW},\min}
 =8.523.
\end{equation}
The SN contribution supplies $7.850$ of this gain and DESI supplies $0.673$. The Akaike information criterion, $\mathrm{AIC}=\chi^2+2k$, penalizes each fitted parameter in addition to the fit statistic. With three extra parameters, the descriptive Akaike difference is $\Delta{\rm AIC}=6-T_{\rm obs}=-2.523$ ($-2.490$ with the usual finite-sample correction). Adding the angular BAO measurements gives $T_{\rm obs}=10.774$ and $\Delta{\rm AIC}=-4.774$; its contributions are $8.210$ from SNe, $0.269$ from DESI, and $2.295$ from the angles. For the primary D+S comparison this is only a modest descriptive AIC preference. D+S+A is additionally affected by the angular-catalogue normalization tension: the separate Pantheon+ benchmark in Supplemental Material Sec.~S5~C gains $19.286$ in $\chi^2$ when an independent angular normalization is allowed. That benchmark uses a different SN sample and cannot be subtracted from the present LTB gain, but it shows why the secondary combination is not independent evidence for inhomogeneity. AIC only ranks the two models, however; it does not give the probability that the improvement is real, which we estimate with simulations in Sec.~\ref{sec:robustness}.

\subsection{Which previously favored geometry does DESI reject?}
A fit to SNe and angular BAO without DESI preferred a particular deep density depression. Keeping that geometry fixed while refitting the common ruler and SN offset to D+S+A gives $\chi^2_{\rm D}=4493.19$ and a total of $6169.72$. Within the local-summary prescription, DESI therefore strongly rejects that specific geometry, even after refitting the sound horizon and the SN magnitude offset. This illustrates the additional information supplied by the DESI transverse and radial measurements. It is not a bound on every possible deep density profile: changing the location and width can produce quite different predictions. The technical supplement records that DESI-free fit and its data selection: $(\Omega_m,d,r_w,r_0)=(0.3608,0.6966,0.6334\,{\rm Gpc},3.3790\,{\rm Gpc})$.

\subsection{Different profiles with comparable fit quality}
In the primary D+S fit, the \emph{nearby narrow shell} has $d\simeq0.155$, $r_w\simeq0.242$ Gpc, and $r_0\simeq0.501$ Gpc. Its density minimum is seen at $z=0.1200$, inside the DESI Bright Galaxy Survey (BGS) selection interval but below its effective redshift. The competing \emph{broad shell} is shallower and farther out: $(\Omega_m,d,r_w,r_0)=(0.3233,0.0417,1.960\,{\rm Gpc},2.186\,{\rm Gpc})$. Its total $\chi^2$ is only $0.647$ higher. Table~\ref{tab:desi_hsw} shows the trade-off: the broad shell fits DESI better and the supernovae less well. Near-degeneracy here means comparable fit quality for different physical profiles.

After adding angular BAO, the secondary D+S+A fit has a nearby narrow shell and a competing \emph{broad central depression}, with $(\Omega_m,d,r_w,r_0)=(0.3543,0.1024,4.406\,{\rm Gpc},0)$ and $\Delta\chi^2=2.848$. This broad solution is centred on the observer; it is not the distant broad shell of the primary fit. Throughout, profile labels refer to the stated dataset. An error estimate based only on the curvature near one fitted point would miss these other successful geometries.

To examine depth beyond one fitted point, we fix $d$ in turn and minimize over the other parameters. This depth profile gives D+S likelihood losses of $1.974$ at $d=0.25$, $4.520$ at $d=0.30$, and $6.983$ at $d=0.999$. At the last point the shell has moved to $r_0=4.993$ Gpc with width $0.2205$ Gpc. The corresponding D+S+A losses are $1.893$, $4.093$, and $10.011$. These curves show the fit cost of a specified depth within the search domain; they are not calibrated 95\% depth limits. This remote shell, at $z\approx1.8$, lies between the DESI effective redshifts $1.48$ and $2.33$, where the predictions are evaluated, so it barely affects them; the actual measurements, which average over $0.8<z<2.1$ and $1.8<z<4.2$, would include it. Averaging over the survey intervals is therefore particularly relevant here.

The nearby D+S minimum predicts
\begin{equation}
 \frac{\Lp(0.65)}{\Lp(0.11)}=0.9731,\qquad
 \frac{\Ll(0.65)}{\Ll(0.11)}=0.9005.
\end{equation}
For the primary broad shell the same transverse and radial ratios are $1.0112$ and $1.0097$. Thus the nearby shell predicts a roughly 10\% radial decrease between these redshifts, whereas the broad shell predicts a roughly 1\% increase, despite their similar fit quality. For the secondary D+S+A nearby shell the ratios are $0.9765$ and $0.9019$. These are predictions of fitted models with one fixed initial $\rd$, not direct measurements of ruler evolution or uncertainty bands. The $10\%$ decrease is an endpoint ratio anchored inside the nearby shell: $\Ll(0.11)/\rd=1.1104$, whereas $\Ll(0.65)/\rd\simeq1.0000$. It does not mean that the ruler at $z=0.65$ is $10\%$ shorter than the initial ruler. There is no radial DESI BAO measurement below $z=0.510$. Supplemental Material Fig.~S1 shows $\mathcal L_i/\rd$ directly. Figure~\ref{fig:robust_rulers} compares them with the results obtained after changing the SN sample.

\section{Robustness and statistical interpretation}\label{sec:robustness}
The direct fits constrain particular geometries, but several profiles remain comparably successful. We now ask which observations select those profiles, how the BAO approximations affect that selection, and how often statistical fluctuations in a homogeneous universe can produce the reported improvement.
\subsection{Supernova selection and catalogue dependence}\label{sec:sn_cuts}
Does the nearby narrow shell describe a feature selected mainly by nearby supernovae? We test this by removing DES-Dovekie objects below $z=0.05$ and $0.1$, refitting both LTB and flat $\Lambda$CDM on each retained sample. We also replace Dovekie with Pantheon+ to test dependence on the SN compilation. Related BAO--SN tests have shown sensitivity to the low-redshift Hubble diagram~\cite{Artola}. Motivated also by the DES-versus-legacy comparison of N\'ajera, Banik, and Desmond~\cite{NajeraDESOnly}, we perform an exact survey selection, retaining only \texttt{IDSURVEY=10}, both with and without \texttt{MUERR<1}. Here \texttt{MUERR} is the released distance-modulus uncertainty in magnitudes. Earlier $\Lambda$LTB work also demonstrated the importance of SN redshift coverage: Camarena et al.~\cite{CamarenaBubble} found support for a Hubble-bubble solution with nearby SNe that disappeared when the full SN sample was included. Their calibrated data combination and profile differ from ours, so both analyses agree on the dependence on the SN selection, not on the numerical results.

Each cut retains the covariance of the selected observations, with the discarded data marginalized over rather than held fixed. Appendix~\ref{app:robust_methods} describes the matrix operation and its independent check. The Pantheon+ fit uses the 1580 noncalibrator light-curve rows with $z_{\rm HD}>0.01$ and a free magnitude offset. Because the compilations overlap, they provide alternative analyses rather than independent confirmations. A nominal Dovekie cut at $z>0.023$ removes no objects---the minimum released redshift is $0.02509$---so only the higher cuts test the role of nearby SNe.

\begin{table*}[tbp]
\caption{Matched DESI+SN sensitivity fits. Each row refits both HSW and flat $\Lambda$CDM to the same SN selection and full covariance. $T$ is their improvement, so $\Delta{\rm AIC}=6-T$. The last two columns are predicted ruler ratios between $z=0.65$ and $0.11$, not measurements with confidence intervals. Dimensional radii use the common $H_0$ scale convention. Both DES-only rows use only the DES SNe (\texttt{IDSURVEY=10}). The parameter $r_0$ is the present-day areal radius of the density minimum, not an observer offset.}\label{tab:sn_robustness}
\small
\begin{ruledtabular}
\begin{tabular}{lrrrrrrrr}
SN sample or width restriction & $N_{\rm SN}$ & $\Omega_m$ & $d$ & $r_w$ [Gpc] & $r_0$ [Gpc] & $T$ & $\Lp$ ratio & $\Ll$ ratio\\
Dovekie, all & 1820 & 0.3023 & 0.1549 & 0.242 & 0.501 & 8.523 & 0.9731 & 0.9005 \\
Dovekie, $z>0.05$ & 1679 & 0.3050 & 0.0346 & 0.663 & 2.569 & 7.407 & 1.0013 & 1.0225 \\
Dovekie, $z>0.1$ & 1619 & 0.3014 & 0.0348 & 0.669 & 2.578 & 7.435 & 1.0013 & 1.0226 \\
DES only & 1623 & 0.3013 & 0.0348 & 0.669 & 2.578 & 7.438 & 1.0013 & 1.0226 \\
DES only, \texttt{MUERR<1} & 1541 & 0.3013 & 0.0348 & 0.669 & 2.578 & 7.437 & 1.0013 & 1.0226 \\
Pantheon+ & 1580 & 0.3113 & 0.0307 & 0.983 & 2.311 & 7.120 & 1.0079 & 1.0195 \\
Dovekie, $r_w\geq0.6$ Gpc & 1820 & 0.3233 & 0.0417 & 1.960 & 2.186 & 7.876 & 1.0112 & 1.0097 \\
\end{tabular}
\end{ruledtabular}
\end{table*}

Table~\ref{tab:sn_robustness} shows that an improvement of about seven in $\chi^2$ remains after either informative low-redshift cut and when Pantheon+ is used. The preferred geometry changes: the cuts select a shallower shell near $r_0=2.57$ Gpc, and Pantheon+ also selects a shallower, more distant profile. The radial ruler prediction changes from a decrease of about 10\% to a small increase. Thus the modest improvement over flat $\Lambda$CDM is shared by different profiles; it does not uniquely select the nearby $d\simeq0.155$ shell or its ruler history. With angular BAO also retained, the $z>0.1$ cut gives $T=3.633$ and Pantheon+ gives $T=6.277$, again with different preferred profiles.

The direct DES selection contains 1623 SNe; the additional \texttt{MUERR<1} cut removes 82 and leaves 1541. All 197 legacy SNe in the full release lie below $z=0.088$, whereas only four DES SNe lie below $z=0.1$. Thus the $z>0.1$ selection and the exact DES-only selection differ by four objects. Restoring them changes $T$ by only $0.0028$; the uncertainty cut changes it by $0.0013$. Both DES-only fits select $d\simeq0.0348$, $r_w\simeq0.669$ Gpc, and $r_0\simeq2.578$ Gpc. The predicted endpoint ratios are $\Lp(0.65)/\Lp(0.11)\simeq1.00130$ and $\Ll(0.65)/\Ll(0.11)\simeq1.02261$. These ratios describe best-fit ruler histories; they are not confidence intervals and do not imply a monotonic increase or decrease between the two redshifts.

The source of the joint improvement changes with the selection. Without the uncertainty cut, the BAO contribution improves by $7.5015$ in $\chi^2$, while the SN contribution worsens by $0.0636$; the cut gives essentially the same decomposition. Consequently, these best fits are not evidence that DES-only SNe themselves require an underdensity. With three additional shape parameters, $T\simeq7.44$ corresponds to only $\Delta{\rm AIC}\simeq-1.44$. The fitted shell has its minimum at $z=0.711$ and zero crossings at $0.496$ and $0.956$, close to the LRG redshifts $0.510$, $0.706$, and $0.934$. For LRG2 $X_\parallel$, the marginal residual $(y-X)/\sigma$ moves from $-1.62$ to $+0.41$. These one-entry pulls are descriptive: the total gain uses the full correlated likelihood. Supplemental Material Fig.~S2 compares the data with both models. No new simulations have been performed to calibrate the fit improvement for these selections, so the full-sample simulation fraction in Sec.~\ref{simulated_data} must not be applied to the DES-only fits. Supplemental Material Sec.~S10 gives the likelihood totals, numerical checks, and density/ruler comparison figure.

This is a likelihood using a subset of the published Dovekie release, retaining its calibration, bias corrections and covariance; it is not a rerun of the light-curve and selection pipeline using DES alone. The uncertainty cut tests sensitivity to high-uncertainty objects and does not itself establish that the remaining objects are all Type Ia supernovae. The comparison also changes the low-redshift coverage and statistical weight of the SN sample, so it cannot by itself identify a calibration error in the legacy sample. The nearby full-sample shell spans $z\simeq0.060$--$0.181$, across the transition between the legacy and DES samples. Efstathiou~\cite{Efstathiou2025} discussed a roughly $0.04$-mag low/high-redshift offset by comparing the earlier DES5Y and Pantheon+ compilations. That result is motivation to test selection sensitivity, not a measured offset in DES-Dovekie or a demonstrated explanation of our fitted shell. In particular, the sound-horizon-calibrated $H_0$ inference of Ref.~\cite{NajeraDESOnly} is a different test from our uncalibrated profile comparison.

A complementary fit imposes $r_0=0$, so the density minimum lies at the observer. Refitting the other parameters gives only $T\simeq0.885$ for either DES-only selection. Within this restricted family the additional shape freedom does not improve AIC over flat $\Lambda$CDM. The best fits found, which reach the limit of the depth search range, are reported in Supplemental Material Sec.~S10; this is not an exclusion of all central-void profiles.

We also require $r_w\geq0.6$ Gpc while keeping the full Dovekie sample. This excludes the narrowest profiles and selects the primary broad shell, retaining $T=7.876$. The gain therefore does not depend entirely on admitting a narrow nearby shell. The width cut is a sensitivity test, not an externally established prior on void size.

\subsection{Effects of the finite BAO ruler length}\label{sec:finite}
The local stretching laws in Eq.~\eqref{eq:transport} treat the ruler as infinitesimal compared with the scale on which the background varies. The nearby fitted widths are only $1.6$--$1.8$ times the initial ruler length, so we test the effect of its finite extent. The matter elements defining the ruler are still assumed to follow the background expansion; this is the passive-transport assumption.
This tests only one part of the effect described in Sec.~\ref{sec:survey}, the variation of the ruler along its own length, for a single ruler at each effective redshift; it does not include the averaging over the pairs of a bin or the template fit.

The endpoints have an initial physical separation $a_i\rd$, corresponding to a separation $\rd$ in the early spatial metric after dividing out $a_i^2$. We follow radial endpoints with a fixed mean observed redshift, or transverse endpoints at the same observed redshift, and calculate their finite redshift or angular separation. We divide out the corresponding finite-length FRW factor to recover the convention used by the published summaries. Appendix~\ref{app:robust_methods} defines this finite-separation diagnostic.

At the primary nearby minimum, the largest fractional change in a DESI prediction at the thirteen published entries is $4.12\times10^{-4}$, or about 0.041\%. For D+S+A it is $1.34\times10^{-3}$, or about 0.134\%. Refitting the geometry, common ruler, and SN offset leaves the nearby fits essentially unchanged (Table~\ref{tab:bao_robustness}). However, a dense redshift grid through the primary shell gives radial corrections as large as 2.02\%. A small correction at the representative DESI redshifts therefore does not guarantee a small correction everywhere within the survey intervals.

This check isolates stretching by a varying background over one finite ruler. It does not calculate changes in the full acoustic correlation caused by structure growth, tracer bias, redshift-space distortions, or reconstruction. Its success tests this geometric approximation, while the next subsection examines averaging over the redshift range of a measurement.

\subsection{Averaging BAO measurements over a redshift interval}\label{sec:windows}
A DESI BAO measurement combines tracer pairs from a finite redshift interval. If the LTB prediction varies substantially across that interval, its value at one effective redshift may not represent the measurement. The public thirteen-number likelihood does not provide the full pair-response weights needed for an LTB refit. We therefore compare two simple illustrative weights: equal weight across each published interval, and a triangular weight that favors redshifts near the quoted effective value.

For each weight we average the LTB-to-FRW prediction ratio and multiply by the FRW prediction at the effective redshift. This construction exactly preserves the flat-FRW limit. The weights are normalized, the full covariance is retained, and both radial and transverse entries are treated consistently. They are sensitivity tests, not replacements for the actual survey response; their intervals and mathematical definitions appear in Appendix~\ref{app:robust_methods}.
This is a rough version of the averaging between pairs described in Sec.~\ref{sec:survey}: the local predictions are averaged over each bin with simple weights, without following the shift of the BAO peak through the fit.

At the original D+S geometry the uniform weight changes an entry by up to 1.02\%, whereas the triangular weight gives at most 0.146\%. After refitting, the uniform weight selects the broad shell and the triangular weight retains the nearby narrow shell. In D+S+A the uniform weight reduces the gain from $10.774$ to $8.135$. For DES-only data, keeping the local best-fit geometry fixed while re-profiling the common ruler and SN offset gives $T=6.979$ and $7.250$ for the uniform and triangular weights; imposing \texttt{MUERR<1} gives $6.977$ and $7.249$. The flat reference is fitted to the same selection, and the window prescription preserves its predictions exactly. These are lower bounds on the improvement achievable by re-optimizing the geometry within each proxy, not newly located window-fit optima. These changes identify a part of the inference that requires survey-specific information: a preference between competing density profiles can depend on how the measurement samples redshift. The tests do not determine which illustrative weight best represents DESI.

\begin{table*}[tbp]
\caption{Sensitivity to the BAO prediction prescription, using the same data and full covariance. Each prescription exactly preserves the flat-FRW limit, and both models retain the same calibration freedom. Window rows are explicit proxy tests, not fits using measured DESI pair-response kernels. HSW parameters shown are for the full-sample D+S refits. The last column instead holds the DES-only local-fit geometry fixed and re-profiles only the ruler and SN offset; these values are not geometry-refitted window optima. A dash denotes an unevaluated check.}\label{tab:bao_robustness}
\begin{ruledtabular}
\begin{tabular}{lrrrrrr}
BAO prescription & $T$ (D+S) & $d$ & $r_w$ [Gpc] & $r_0$ [Gpc] & $T$ (D+S+A) & $T_{\rm DES\ only}^{\rm fixed}$\\
Local prediction at $z_{\rm eff}$ & 8.523 & 0.1549 & 0.242 & 0.501 & 10.774 & 7.438 \\
Finite ruler length & 8.529 & 0.1549 & 0.243 & 0.501 & 10.834 & --- \\
Uniform redshift weight & 7.922 & 0.0435 & 1.984 & 2.188 & 8.135 & 6.979 \\
Triangular redshift weight & 8.569 & 0.1574 & 0.239 & 0.496 & 10.756 & 7.250 \\
\end{tabular}
\end{ruledtabular}
\end{table*}

\paragraph*{Why the BGS interval needs special attention.}
For the primary nearby solution, the local dilations relative to its exterior with the same initial ruler range over $0.9410\lesssim\alpha_\perp\lesssim0.9995$ and $0.8525\lesssim\alpha_\parallel\lesssim1.1994$ within $0.1<z<0.4$. These are unweighted local extrema, not inferred biases of the DESI measurement. They show why the small finite-ruler correction at $z_{\rm eff}=0.295$ cannot certify either the use of a single effective redshift in the baseline fits or the linearized response of Sec.~\ref{sec:survey} throughout the interval. The pair construction and its Taylor expansion are checked separately in Appendix~\ref{app:pair_checks}.

As a separate control, we omit the single BGS summary and refit both cosmologies to the remaining twelve DESI entries and the same SNe. The nearby minimum persists, with $(\Omega_m,d,r_w,r_0)=(0.30149,0.15757,0.23882\,{\rm Gpc},0.49626\,{\rm Gpc})$ and an improvement $T=8.571$. With the angular BAO retained, the matched improvement is $10.683$. Thus the local nearby minimum is not driven by the one compressed BGS point. This control does not determine what the missing BGS survey response would do; the uniform-window proxy already shows that a different averaging prescription can favor the competing broad shell.

\subsection{Statistical significance from simulated homogeneous data}
\label{simulated_data}
Is the improvement $T_{\rm obs}=8.523$ of the LTB fit over flat $\Lambda$CDM significant? For three additional parameters, the usual rule (Wilks' theorem) would compare it with a $\chi^2$ distribution with three degrees of freedom. This rule does not apply here: at $d=0$ there is no void, so its width and location have no effect on the data.

We therefore ask directly how often a homogeneous universe produces such an improvement. We generate 5000 simulated datasets from the best-fitting flat $\Lambda$CDM model, with Gaussian errors and the full SN and DESI covariance, and fit each with flat $\Lambda$CDM and with LTB profiles, with the same calibration freedom (Appendix~\ref{app:robust_methods}). In 369 of them, a fraction $0.0738$ (95\% binomial simulation interval $[0.0667,0.0814]$), the LTB fit improves $\chi^2$ by at least $T_{\rm obs}$. Because the LTB search in each simulation covers a restricted set of profiles, a complete search could only find larger improvements, so $0.0738$ is a lower limit on the true probability. The observed improvement, in contrast, was found by a continuous search from several starting points. If that search missed a better fit, the true $T_{\rm obs}$ would be larger and the fraction smaller, so this lower limit assumes that the reported fit is the best one. Since even the lower end of the interval exceeds $5\%$, the improvement does not establish a detection of inhomogeneity.

This result assumes the fitted $\Lambda$CDM parameters and Gaussian errors; we have not calibrated the test at every possible homogeneous cosmology. It applies to the primary D+S fit and does not provide confidence regions for the void parameters; neither these simulations nor the simulations without DESI are used as the significance of D+S+A.

\subsection{Uncertainty in profile shape and ruler evolution}

\begin{figure*}[tbp]
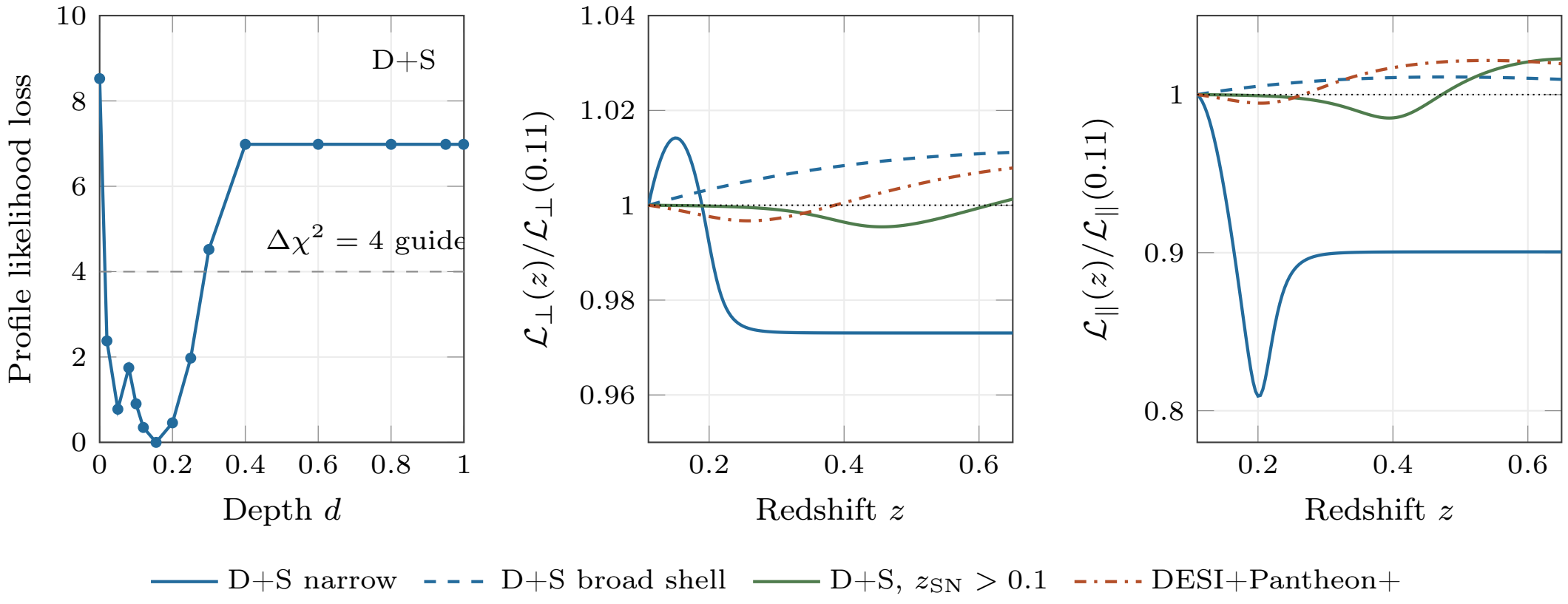

\centering
\par\smallskip\pgfplotslegendfromname{robustrulerlegend}
\caption{Depth constraint and predicted ruler evolution. Left: increase of $\chi^2$ above the best D+S fit when the depth $d$ is fixed and all other parameters (exterior density, width, location, and calibrations) are refitted; lines join the lowest located minima. The dashed line at $\Delta\chi^2=4$ is a visual guide, not a calibrated confidence limit. At large depth the loss stays near seven because the void moves to $r_0\approx5$ Gpc, between the DESI effective redshifts. Middle and right: transverse and radial rulers predicted by the fits, normalized at $z=0.11$. Blue solid and dashed: the nearby narrow shell and the distant broad shell, which fit the D+S data almost equally well ($\Delta\chi^2=0.647$). Green: fit without the SNe at $z<0.1$, to test whether the nearby SNe select the narrow shell; red dash-dotted: fit with Pantheon+ instead of DES-Dovekie (Sec.~\ref{sec:sn_cuts}). The dotted line is FRW. These are predictions of fitted models, not confidence bands: the narrow shell has a radial endpoint ratio below unity, whereas the other comparably good fits have endpoint ratios slightly above unity. These ratios do not imply monotonic histories.}\label{fig:robust_rulers}
\end{figure*}

Figure~\ref{fig:robust_rulers} (left) shows the likelihood cost of fixing the depth while refitting the other parameters. It is about $2$ at $d=0.25$ and $4.5$ at $d=0.3$, and remains near $7$ for a nearly empty shell that moves to $r_0\approx5$ Gpc between the sampled DESI redshifts. These costs constrain specified configurations, but do not provide a calibrated upper confidence limit on depth over all allowed locations and widths; the line at $\Delta\chi^2=4$ is only a guide. The middle and right panels show a related ambiguity: nearby and broad minima separated by less than one in $\chi^2$ predict a radial decrease of about $10\%$ and an increase of about $1\%$ between $z=0.11$ and $0.65$. The sign of that model-predicted evolution is therefore not selected uniquely. This does not remove the independent constraint on the smooth transverse trend in Eq.~\eqref{eq:slope}. The curves are predictions of particular fits, not confidence bands.

The analyses provide complementary constraints. The transverse BAO--SN ratio bounds a specified smooth trend in effective ruler length through Eq.~\eqref{eq:slope}. The direct fits use the distance information as well and exclude particular geometries, including the previously favored deep profile once DESI is added. At the same time, allowing the shell location and width to vary leaves other successful profiles, and the SN sample and redshift averaging affect which profile is preferred. These results support constraints within the stated model and measurement assumptions, rather than a universal upper limit on depth or a measured nonzero ruler evolution.

\section{A secondary consistency test: angular and three-dimensional BAO}\label{sec:discrepancy}
\subsection{Why the two transverse measurements share one local ruler}
An angular BAO analysis measures the apparent angle of the acoustic feature. A three-dimensional analysis fits its transverse and radial scales, allowing separate rescalings of a reference template. In the local approximation both transverse measurements therefore probe the same distance divided by the same transverse ruler. This equality assumes that the template can represent the local acoustic ellipse, that the fit recovers its peak without bias, and that the measurements have matching effective redshift weights:
\begin{equation}
 \widehat X_\perp^{\,3D}
 =\alpha_\perp\frac{\DM^{\fid}}{\rd^{\fid}}
 =\frac{\DM}{\Lp}
 =\frac{1}{\theta_{\rm BAO}}
 =\widehat X_\perp^{\,\theta}.\label{eq:common}
\end{equation}
The reference (fiducial) cosmology used to convert coordinates cancels in this expression. This does not assert that a finite-survey pipeline is insensitive to all template errors~\cite{Perez}; it states what the ideal local fit measures. A common change in the local ruler can substantially shift the common observable, as in Fig.~\ref{fig:teaching_angular}, but cannot shift one transverse estimator relative to the other. 

The comparison here is between \emph{two transverse estimators}. It differs from the transverse--radial distance comparison of Camarena et al.~\cite{CamarenaDESI2025}. In LTB, transverse and radial observables depend on different expansion rates and ruler lengths; no equality between $\DM/\Lp$ and $c/(\Hl\Ll)$ is asserted. Equation~\eqref{eq:common} therefore does not rule out an inhomogeneous explanation of a transverse--radial consistency problem. It applies only to the proposed explanation of a difference between two transverse estimators, under the stated assumptions of local fits and matching redshift weights.

The angular value at $z=0.510$ is $4.81\pm0.17$ degrees. Propagating the error to first order for $x_{\rm angular}\equiv(180/\pi)/\theta_{\rm deg}$ gives $x_{\rm angular}=11.912\pm0.421$, compared with the DESI transverse summary $x_{\rm DESI}=13.588\pm0.168$, a difference of $3.70\sigma$. In the calculation below we use $y_{\rm D}=13.58758434$ and $\sigma_{\rm D}=0.16836678$ from the pinned vector and covariance, rather than rounded summary-table values. The label $\DM/\rd$ refers to the conventional FRW interpretation of these quantities. Since the angular errors are Gaussian in $\theta$ rather than in $1/\theta$, we also compare the two measurements in the measured angle. In our local LTB interpretation the common theoretical prediction is $x=\DM/\Lp=q/\rd$, so we fit the two measurements with one common $x$:
\begin{equation}
 \chi^2(x)=\left[\frac{4.81-(180/\pi)/x}{0.17}\right]^2
 +\left[\frac{x-y_{\rm D}}{\sigma_{\rm D}}\right]^2.
\end{equation}
The minimum, at $x=13.42$, is $\chi^2=11.10$ (fitting the two values separately gives $\chi^2=0$), corresponding to $3.33\sigma$. The two estimates of the tension, $3.70\sigma$ and $3.33\sigma$, differ only in how the angular error is treated.

Our conclusion is that \emph{common local LTB ruler transport does not explain the discrepancy}. It applies when the transverse estimators recover the same local observable with matching weights. Beyond that limit, an angular estimate samples a thin selection, whereas DESI LRG1 spans $0.4<z<0.6$ and fits pairs distributed through that interval. Their responses can differ. Establishing an explanation of the observed discrepancy would require its sign and magnitude from both actual survey selections and template fits, including effects beyond the illustrative geometric response in Eq.~\eqref{eq:xiobs}. The diagnostics of Sec.~\ref{sec:robustness} do not perform that calculation, so they neither demonstrate such an explanation nor exclude every inhomogeneous estimator effect.

\section{Discussion and conclusions}\label{sec:discussion}
Combining BAO and SNe in an inhomogeneous cosmology requires calculating how the acoustic ruler stretches as well as how far away it is. In a central $\Lambda$LTB model the transverse and radial BAO observables are $\DM/\Lp$ and $c/(\Hl\Ll)$. The distance follows the observer's light cone, and the two ruler lengths follow the different directional expansion rates. Even with a single initial sound horizon $\rd$, the observed ruler can vary with redshift and have different lengths along and across the line of sight.

The BAO--SN ratio provides a direct interpretation of the ruler information. Comparing this ratio at two redshifts removes the unknown absolute scale and isolates the change in transverse ruler length. For the stated smooth reconstruction and angular-BAO likelihood, the fitted trend $d\ln\Lp/dz=0.0589\pm0.0941$ is consistent with a constant ruler. This is a constraint on a specified redshift trend. It does not by itself reconstruct the density profile: a broad inhomogeneity may change the ruler almost uniformly, while a localized shell may lie between the sampled redshifts. 

The direct fit retains the SN distance information and adds radial BAO to test physical profiles. DESI supplies an important restriction on those profiles. The particular deep geometry favored by SNe and angular BAO alone gives $\chi^2_{\rm D}=4493.19$ when tested with DESI, even after the absolute scales are refitted. Other geometries remain viable. For the primary DESI+DES-Dovekie comparison, the best located exact HSW-shaped fit improves $\chi^2$ by $8.523$ over flat $\Lambda$CDM with the same calibration freedom and three additional shape parameters. Most of that gain comes from SNe. The nearby narrow shell and the broader, shallower shell farther away differ by only $0.647$ in total $\chi^2$, because they balance the BAO and SN contributions differently. The combined data therefore rule out some specific voids, but cannot yet single out one: the narrow nearby shell and the broad distant shell fit almost equally well.

The robustness tests distinguish reproducible fit improvement from unique profile reconstruction. Low-redshift SN cuts, the exact DES-only selections, the Pantheon+ comparison, and the removal of BGS retain a modest improvement, while the preferred density profile can change. The two DES-only fits agree closely with the $z>0.1$ selection and with each other; their gain is predominantly from BAO and their preferred underdensity lies on a shell around the central observer. Finite-ruler corrections are small at the DESI effective redshifts; the local dilations of the nearby solution nevertheless vary strongly inside the BGS interval, and illustrative window weights can exchange the preference between nearby and broad shells. In homogeneous simulations, 369 of 5000 restricted searches reach the reported improvement, giving a tail-fraction lower-bound estimate of 0.0738. Thus the data constrain specified geometries and a smooth transverse ruler trend within the stated assumptions, without establishing a $5\%$-level detection of inhomogeneity or a unique sign of the model-predicted radial evolution.

Compared with Zhang and Zhang~\cite{ZhangZhang}, who use first-order LTB distances, a profile centred on the observer with a fixed relative width, and a CMB calibration of the sound horizon, we solve the LTB background exactly, vary the depth, width and location of the void independently, leave the sound horizon free, and test the results against changes in the SN sample and in the averaging of the BAO predictions. The two profile parametrizations differ, so their central contrast and our shell depth cannot be compared directly. The different stretching of the ruler along and across the line of sight is already established~\cite{Zumalacarregui,GarciaBellido,ValkenburgMarraClarkson,CamarenaCopernican}; here we use it to show both which voids BAO and SNe exclude and which remain allowed when the location and width of the void are free.

Camarena et al.~\cite{CamarenaDESI2025} report departures from FLRW at $2.8\sigma$ with Pantheon+ and $5.2\sigma$ with DESY5. Their CMB-calibrated profile-likelihood analyses and our uncalibrated homogeneous-mock test use different SN releases, density families, and statistics; the quoted significances therefore do not test the same hypothesis. Supplemental Material, Sec.~S10 explains why the reported fit gains are not directly interchangeable.

\subsection*{Scope and limitations}
The exact background solution does not make the compressed-data likelihood exact. The repeatable result is a modest fit gain, about $7$--$8.5$, across several SN choices; it does not uniquely select the nearby shell. The DES-only shallow shell retains a gain close to seven under both illustrative window weights, but this is not validation with measured survey kernels. The full-sample mock bound does not calibrate DES-only significance, and the reported profile losses are not universal confidence limits. A matched comparison with $w_0w_a$ or a Gaussian density prior would require an additional specified analysis; neither is inferred from the present flat-$\Lambda$CDM comparison.

A next step toward survey-level constraints is to fit the acoustic correlation using the actual pair selections, redshift-space effects, and reconstruction, and to revisit the SN peculiar-velocity and selection corrections within LTB. Statistically calibrated confidence regions or a specified posterior would then quantify profile uncertainty beyond the fits shown here. External CMB information and the kinematic Sunyaev--Zel\'dovich effect can supply additional constraints. The latter probes the CMB dipole seen by scattering electrons and has strongly constrained large LTB void scenarios~\cite{Bull}; its prediction must be recomputed for the present compensated $\Lambda$LTB profiles before applying a numerical bound. Neither likelihood is included here.

Finally, the ideal local angular and three-dimensional transverse BAO measurements both probe $\DM/\Lp$. Common LTB ruler evolution changes them together, so it cannot explain their discrepancy by shifting only one. This secondary result identifies a limitation of a common-ruler explanation. A comparison of actual surveys must also account for their different redshift weights and measurement procedures.

\begin{acknowledgments}
We thank Valerio Marra, David Camarena, and Indranil Banik for helpful comments and suggestions concerning related work, supernova selection, and the interpretation of the density profile. OpenAI Codex assisted with manuscript organization, scientific exposition, and revisions directed by the authors. The tool and model information, numerical role, and verification procedures are described in Appendix~\ref{app:robust_methods}. The authors retain responsibility for the scientific content.
\end{acknowledgments}

\section*{Data availability}
Public inputs are pinned to the revisions in Refs.~\cite{DESIdata,PantheonData,DovekieData}; the native angles are listed in Appendix~\ref{app:inputs}. The executable numerical package is publicly available on Zenodo~\cite{NumericalArchive}, version 1.2.0, DOI \href{https://doi.org/10.5281/zenodo.24374395}{10.5281/zenodo.24374395}. The baseline archive contains the exact-background and full-covariance likelihood code, recorded fits, and pair-response and likelihood checks. The numerical addendum includes the DES-only and $r_0=0$ fits, the fixed-geometry redshift-window tests, and the code and results for the ruler histories and data/model residual figures. The accompanying READMEs give the reproduction commands. A checksum-verified downloader retrieves the pinned public inputs from their original providers; third-party data are not redistributed. The package reproduces the conditional local-summary analysis defined in Sec.~\ref{sec:geometry}. Supporting derivations and auxiliary controls are supplied in the accompanying Supplemental Material~\cite{SupplementalMaterial}.

\appendix
\section{Profile definitions and nonlinear construction}\label{app:profiles}
The growing-mode continuity equation gives a regular radial peculiar velocity $v=-f\bar a\bar H r\bar\delta/3$. At linear order,
\begin{align}
 \frac{\delta\Hp}{\bar H}&=-\frac{f}{3}\bar\delta,&
 \frac{\delta\Hl}{\bar H}&=f\left(\frac23\bar\delta-\delta_m\right),\\
 g\equiv\frac{\Hl-\Hp}{\bar H}&=f(\bar\delta-\delta_m),&
 \bar\delta'&=-\frac{3g}{fr}.
\end{align}
Integrating $H_i-\bar H$ along a comoving shell, with a negligible early contrast, yields
\begin{align}
 \frac{L_\perp}{\rd}&=1-\frac13\bar\delta,&
 \frac{L_\parallel}{\rd}&=1+\frac23\bar\delta-\delta_m.
\end{align}
All contrasts here refer to emission time. Multiplying by $Q=1+\delta\ln a$ gives the observed-comoving rulers in Eqs.~\eqref{eq:teaching_transverse_ruler}--\eqref{eq:teaching_linear_rulers}.

The emission-time correction follows directly from the perturbed null cone. Write $t(z)=\bar t(z)+\delta t(z)$ and $\delta\ln a=\bar H\delta t$. The background time-shift terms cancel in its derivative, leaving $d(\delta\ln a)/dz=(\delta\Hl/\bar H)/(1+z)$. For completeness, linearizing Eq.~\eqref{eq:distance} around flat $\Lambda$CDM gives
\begin{equation}
\begin{split}
 \frac{\delta\DM}{\bar\DM}=\frac{1}{\bar\chi}\int_0^{\bar\chi}d\chi'\bigg[
 &-\frac{\delta\Hl}{\bar H}+E+\frac32\Omega_m(z')\delta\ln a\\
 &+\frac{\Hl-\Hp}{\bar H}\frac{\bar H\bar D_A}{c}\bigg].
\end{split}\label{eq:linear_distance}
\end{equation}
For a simultaneous bang time, the growing-mode curvature at first order is
\begin{equation}
 E(r)=-\frac12\left(\frac{H_0r}{c}\right)^2
 \bar\delta^{(0)}(r)\left(\Omega_{m0}+\frac{2f_0}{3}\right).
\end{equation}
The distance perturbation does not enter the ratio likelihood because it cancels there; it must be retained in a direct physical distance fit.

For the HSW-shaped kernel in Eq.~\eqref{eq:main_hsw}, compensation fixes $q_s$ through
\begin{equation}
 \int_{-r_0/r_w}^{\infty}\frac{(x+r_0/r_w)^2(1-x^2)}{1+(|x|/q_s)^{9.5}}\,dx=0.
\end{equation}
The integration is over the complete radial half-line. For the exact solution we replace $U$ inside $r_c=100$ Mpc by $U_c(r)=\sum_{j=0}^3c_{2j}(r/r_c)^{2j}$. Matching its value and first two derivatives at $r_c$, and preserving $\int_0^{r_c}r^2U\,dr$, determines the four coefficients. This even central polynomial makes the origin regular. If its maximum changes, we renormalize the complete kernel so that $d$ remains the physical trough depth. Mass compensation survives both operations. The teaching curves use the unpatched first-order template; the exact fitted profiles use this regular construction.

In the present areal gauge $R(t_0,r)=r$, write $a_s=R/r$, $\tau=H_0t$, $A_m=\Omega_m(1+\bar\delta_m)$, and $k=2Ec^2/(H_0^2r^2)$. The shell dynamics and common-age condition are
\begin{align}
 \left(\frac{da_s}{d\tau}\right)^2&=\frac{A_m}{a_s}+k+(1-\Omega_m)a_s^2,\\
 \tau_0&=\int_0^1\frac{da_s}{\sqrt{A_m/a_s+k+(1-\Omega_m)a_s^2}}.
 \label{eq:age}
\end{align}
Here $\tau_0$ is the age of the flat exterior. The age equation fixes each shell's curvature; shell evolution and null-cone integration then give the geometric observables. The early dust limit satisfies $R/\bar a\to q=r(A_m/\Omega_m)^{1/3}$, yielding the ruler laws in Eq.~\eqref{eq:desi_hsw_transport}. Radiation and generation of the acoustic feature are not evolved by this dust+$\Lambda$ solver.

For the current variable-$\Omega_m$ fits, the baseline universal shell table has 501 logarithmic mass nodes plus the homogeneous node and 401 time nodes. Four-node exterior interpolation accelerates optimization. Final checks remove that interpolation, use 1001/801 nodes, a 2.5-Mpc radial spacing, and null-cone relative tolerance $2\times10^{-10}$. An independent implementation evolves 2401 physical radial shells with 801 time samples. Sampled regularity requires positive density, $q'$, $R'$, and $1+2E$ over the domain. Search bounds are $0.1\leq\Omega_m\leq0.5$, $0\leq d\leq0.999$, $0.2\leq r_w/{\rm Gpc}\leq8$, and $0\leq r_0/{\rm Gpc}\leq6$.

\section{Robustness methods and validation}\label{app:robust_methods}
\subsection{Common ruler and exact-background checks}
For a fixed geometry write the DESI prediction as $\bm b/\rd$ and the angular prediction as $\bm t\rd$. The BAO statistic is
\begin{equation}
\chi^2_{\rm BAO}(\rd)=\frac{B_2}{\rd^2}-\frac{2B_1}{\rd}
 +A_2\rd^2-2A_1\rd+\mathrm{constant},\label{eq:desi_shared_ruler}
\end{equation}
where $B_2=\bm b^TC_{\rm D}^{-1}\bm b$, $B_1=\bm b^TC_{\rm D}^{-1}\bm y_{\rm D}$, $A_2=\bm t^TC_\theta^{-1}\bm t$, and $A_1=\bm t^TC_\theta^{-1}\bm\theta$. We check every positive root of $A_2\rd^4-A_1\rd^3+B_1\rd-B_2=0$ and choose the minimum statistic. For D+S, $\widehat\rd=B_2/B_1$. Direct scalar minimization verifies this profiling. Finite-pair corrections depend on $\rd$, so those refits optimize it explicitly with the geometric parameters, keeping the SN offset analytic.

Removing exterior-density interpolation and refining the shell and radial calculation changes the baseline D+S and D+S+A minima by less than $5\times10^{-5}$. Independent physical-radial-shell integrations and direct covariance-likelihood evaluations reproduce the baseline totals within $10^{-4}$ and the tested radial predictions within $10^{-6}$ fractionally through $z=2.33$. The full DESI covariance is positive definite. At the same fitted predictions, diagonal-only DESI statistics would be $11.201$ and $11.812$, compared with the full-covariance $10.559$ and $10.775$. Varying the central regularization radius from 100 Mpc to 50 or 200 Mpc changes the baseline fixed-geometry totals by less than $0.04$; this is a model-prescription test distinct from integration convergence. The new SN and window minima and depth-profile points also pass exact-exterior refinement and sampled regularity checks.

\subsection{Finite-pair construction}
In the early dust limit the spatial line element, divided by the squared exterior scale factor, is
\begin{equation}
 d\sigma_i^2=\frac{dq^2}{g(q)}+q^2d\Omega^2,\qquad g(q)=1+2E[r(q)].
\end{equation}
The radial distance coordinate $s$ in this rescaled metric (physical distance divided by the early scale factor) obeys $ds/dr=q'/\sqrt{g}$. Combining this with the exact null equation gives
\begin{equation}
 \frac{ds}{dz}=\frac{cq'}{(1+z)\dot R'}\equiv h(z).
\end{equation}
We solve $s(z+\Delta z/2)-s(z-\Delta z/2)=\rd$ for the finite radial separation at fixed mean observed redshift. The local limit is $\Delta z=\rd/h(z)$.

For transverse endpoints on the same shell $q_c=q[r(z)]$, the initial spatial geodesic has a turning radius $b$. Setting $u=\sqrt{q^2-b^2}$ and $u_c=\sqrt{q_c^2-b^2}$ removes the turning-point square-root singularity. Its length and angle are
\begin{align}
 \rd&=2\int_0^{u_c}\frac{du}{\sqrt{g(\sqrt{b^2+u^2})}},\\
 \theta&=2\int_0^{u_c}\frac{b\,du}{(b^2+u^2)\sqrt{g(\sqrt{b^2+u^2})}}.
\end{align}
Solve the first equation for $u_c$ and evaluate the second. In flat space $\theta_F=2\arcsin[\rd/(2q_F)]$. The diagnostic predictions are
\begin{equation}
 X_\perp^{\rm pair}=\frac{q_F(z)}{\rd}\frac{\theta_F}{\theta},\qquad
 X_\parallel^{\rm pair}=\frac{h_F(z)}{\rd}\frac{\Delta z_F}{\Delta z},
\end{equation}
with the flat comparison evaluated at the same exterior $\Omega_m$ and $\rd$. This convention removes finite-pair effects already implicit in the FRW-calibrated summary. It returns the original LTB local observables as the pair length tends to zero. It is an explicitly prescribed passive-geometric diagnostic, not a calculation of a clustering peak's recovered dilation.

We use 48-point quadrature and a 6001-point redshift antiderivative. Doubling both resolutions changes the sampled predictions by less than $3\times10^{-12}$ fractionally. The homogeneous limit agrees within $8\times10^{-15}$. Reducing the ruler by a factor 100 recovers the local predictions within $1.3\times10^{-7}$ at the tested data redshifts.

\subsection{Window proxies}
For each DESI entry $i$, let $b_i(z)=\rd X_i(z)$ be its ruler-independent local numerator. In LTB, the transverse numerator is $b_\perp=q$, the radial one is $b_\parallel=cq^{\prime}/[(1+z)\dot R^{\prime}]$, and $b_V=(z b_\perp^2b_\parallel)^{1/3}$. The transverse numerator reduces to $\DM$ only in the homogeneous limit. We average the ratio of the LTB to the flat FRW prediction over the redshift interval of the entry with a normalized weight $W_i$, and rescale it by the FRW value at the effective redshift:
\begin{equation}
 b_i^W=b_i^F(z_{{\rm eff},i})
 \int dz\,W_i(z)\frac{b_i^{\rm LTB}(z)}{b_i^F(z)}.
\end{equation}
The flat FRW prediction uses the same exterior density, so that without a void $b_i^W=b_i^F(z_{{\rm eff},i})$. The intervals are $[0.1,0.4]$, $[0.4,0.6]$, $[0.6,0.8]$, $[0.8,1.1]$, $[1.1,1.6]$, $[0.8,2.1]$, and $[1.8,4.2]$ for BGS, LRG1, LRG2, LRG3+ELG1, ELG2, QSO, and Ly$\alpha$, respectively~\cite{DESI}. One weight is uniform; the other rises linearly from zero at the lower edge to its peak at $z_{\rm eff}$ and falls linearly to zero at the upper edge. We integrate each side of the peak separately with 48-point quadrature, and extend the light cone to $z=4.2$. Neither weight is claimed to be the survey's actual pair-response kernel. For samples that provide both $\DM/\rd$ and $\DH/\rd$, the same weight is used for both, and their published covariance is retained. The angular BAO predictions are kept local.
\subsection{DESI null simulations and SN-cut validation}
In each simulation, the LTB search starts from a fixed set of void models, the union of two Cartesian grids. The broad grid uses $\Omega_m=0.24,0.26,\ldots,0.40$, $d\in\{.025,.05,.10,.20,.40,.70,.95\}$, $r_w/{\rm Gpc}\in\{.20,.35,.60,1.2,2.5,5,8\}$, and $r_0/{\rm Gpc}\in\{0,.3,.6,1,2,3,4,5,6\}$. The denser grid, around the nearby narrow shell, uses $\Omega_m=0.28,0.29,\ldots,0.34$, $d\in\{.05,.10,.15,.20,.25,.30\}$, $r_w/{\rm Gpc}\in\{.20,.25,.30,.40\}$, and $r_0/{\rm Gpc}\in\{.25,.40,.50,.60,.75\}$. After removing duplicates the set contains 4797 models, all of which satisfy the regularity conditions of Sec.~\ref{sec:model} at the sampled times. It is a chosen subset of models, not a set of posterior draws.

Each simulated SN dataset is the best-fitting flat $\Lambda$CDM magnitudes plus random noise with the full SN covariance. A common shift of all magnitudes is omitted from the noise, since the free offset $\mathcal M$ would absorb it anyway. Independent DESI noise, drawn from the DESI covariance for all thirteen measurements, is added to the corresponding predictions. In each simulation we refit flat $\Lambda$CDM ($\Omega_m$ and both absolute normalizations) and evaluate the set of void models above; for D+S the common ruler $\rd$ is fitted analytically for every trial model, and flat $\Omega_m$ is first minimized on a fine grid with local interpolation. For each simulation we first record the best improvement over flat $\Lambda$CDM among the models of the set. If it lies between 7 and $T_{\rm obs}$, the simulation is close to the threshold, so we refine it with a continuous search over the void parameters; 251 of the 5000 simulations are refined in this way. Every simulation that reaches $T_{\rm obs}$, either on the set or after refinement, is then recomputed more accurately: the flat model is minimized directly, the LTB shells are evaluated with the exact exterior at finer radial resolution, and the regularity conditions are checked. To guard against numerical errors, $0.002$ is subtracted from its improvement before the comparison with $T_{\rm obs}$. Because every simulation is treated by the same rule, each is an independent trial, and the number reaching $T_{\rm obs}$ follows a binomial distribution. The random seed, 271027, and all checks of candidate simulations are recorded. Since a restricted search can only find a smaller improvement than a complete one, $T_{\rm restricted}\leq T_{\rm full}$ for each simulation, and the estimated fraction is a lower limit on the true probability. Its binomial interval describes only the finite number of simulations; it does not calibrate the test for other null-model parameters or provide confidence regions for the void parameters.

For the low-redshift SN cuts, an independent Schur-complement calculation from the original precision matrix agrees with the retained covariance described in the next subsection to a relative accuracy better than $10^{-10}$, which confirms that the excluded supernovae are marginalized over rather than held fixed. As in the baseline, the released SN redshift corrections and covariance model are retained; changing the catalogue or removing nearby supernovae is a sensitivity test, not a complete rederivation of those corrections in LTB.

\subsection{Input conventions and statistical implementation}
The DES-Dovekie release stores the upper triangle of a precision matrix, i.e.\ the inverse covariance. We symmetrize it and use it directly; we do not invert it as if it were a covariance. For an SN cut, we first invert the full precision matrix, select the covariance submatrix of the retained supernovae, and invert that submatrix. Selecting the precision submatrix directly would instead treat the omitted supernovae as if their values were known (conditioning on them). The Schur-complement check above verifies the marginalization convention. Pantheon+ uses \texttt{m\_b\_corr}, $z_{\rm HD}>0.01$, and \texttt{IS\_CALIBRATOR}=0, with an independent free magnitude offset. Repeated light curves retain their covariance. The DESI vector retains the released radial-before-transverse ordering of its final pair.

For a change $\Delta\bm X$ in the thirteen DESI predictions, we also report the error-weighted norm $(\Delta\bm X^TC_{\rm D}^{-1}\Delta\bm X)^{1/2}$. This is the size of the change after accounting for correlated measurement errors, not a detection significance. At the best D+S fit (the nearby narrow shell) it is 0.0142 for the finite-ruler correction, 1.06 for the uniform redshift weight, and 0.152 for the triangular weight. The D+S+A finite-ruler value is 0.0461.

\subsection{Checks of the pair description and the BGS control}\label{app:pair_checks}
We verify the local-dilation identities and the profiled linear response derived in the Supplemental Material~\cite{SupplementalMaterial}. At depths $d=0.004,0.002,0.001$, the maximum errors of the first-order radial dilation against the exact background are $1.19\times10^{-5}$, $2.97\times10^{-6}$ and $7.43\times10^{-7}$, consistent with quadratic truncation error. A direct joint linear least-squares fit and nuisance profiling agree to $6\times10^{-16}$; the pure-radial response has no transverse leakage.

For the nearby solution, an unexpanded radial pair integral at $z=0.20$, $s=147.02$ Mpc and $\mu=1$ differs from its leading short-pair expansion by about $0.00812$ in $\alpha_\parallel$. Both endpoints lie inside the BGS interval. Halving and quartering the pair separation reduce the largest scanned remainder to $0.000620$ and $0.0000411$. The result is stable under radial-grid and derivative-step refinement. Numerical pair predictions should therefore use the unexpanded expressions. Independently integrating geodesics of the initial acoustic metric at ten redshifts from $0.11$ to $2.33$, for $\mu=0,0.5,1$, gives maximum fractional differences from the plane-parallel pair construction of $1.4\times10^{-5}$ for nearby and $1.9\times10^{-6}$ for broad. The flat-limit check agrees within $1.6\times10^{-10}$. These are sampled geometric checks, not bounds on a recovered survey peak or a validation of its linearized template response.

For the BGS omission test, we select the $12\times12$ covariance submatrix of the remaining DESI entries before inversion, thereby marginalizing the omitted datum. Both cosmologies refit their common ruler and SN offset, with the same SN sample and profile search bounds as the baseline. Three starts sample the nearby, broad-shell and broad-central basins. The lowest located totals are $1636.697$ versus $1645.268$ for D+S without BGS, and $1665.716$ versus $1676.399$ when angular BAO is added. Exact-exterior refinement changes these LTB totals by less than $7\times10^{-5}$. This test does not recalculate a survey response or a detection significance. Version 1.2.0 of the numerical release includes the inputs, scripts, minima, and convergence checks for this BGS-omission test.

\subsection{Independent checks of the illustrative curves}\label{app:illustrations}
Figures~\ref{fig:teaching_rulers} and \ref{fig:teaching_angular} are computed directly from the first-order equations of Appendix~\ref{app:profiles}. To verify this computation, we check its ingredients. The growth factor, which sets the time dependence of the density contrast, agrees with a numerical solution of the growth equation to $6\times10^{-12}$. The enclosed density is computed as $\bar\delta(r)=3\int_0^1u^2\delta_m(ur)\,du$, with $u=r'/r$, which avoids dividing a small enclosed mass by a small $r^3$ near the observer; without this, the curves would be in error by up to $0.4\%$ at the lowest plotted redshift. The recomputation does not alter any fitted numerical result. It checks the first-order illustration, not its accuracy at large density contrast.

An independent implementation, which evolves the radial shells separately from the fitting code (Appendix~\ref{app:profiles}), also reproduces the primary narrow and broad solutions and the secondary narrow solution. The independent recomputations reproduce the two narrow solutions within $10^{-4}$ in the total statistic. The full-covariance calibration and the identity $X_\parallel=c/(H_\parallel\Ll)$ pass separate checks. These tests show that the likelihood is computed correctly for the model as defined; they do not test the local approximation for the BAO measurements described in Sec.~\ref{sec:survey}.

\paragraph{AI-assisted implementation and verification.}
OpenAI Codex assisted with derivations, development and debugging of the Python likelihood and numerical-checking scripts, execution of fits and robustness tests, and preparation of plotting code under the authors' direction. GPT-6 was used for the final computational review and manuscript revision; exact model-version identifiers for the earlier sessions were not retained. The authors directed the scientific questions and modeling choices and are responsible for reviewing and verifying all scientific claims and results. Verification includes the independent-integration comparisons, limiting-case tests, covariance checks, and resolution tests documented above. These computational checks do not constitute external human peer review. The plotted curves are calculated from the stated equations and recorded results. An additional Claude review supplied by the authors (identified by them as Fable 5.2) was used to flag possible issues; accepted scientific changes were checked against primary references and recorded or recomputed numerical results. Version 1.2.0 of the numerical release includes the pair-response, BGS-omission, geometric-response, and likelihood cross-checks described in this appendix.

\section{Inputs and numerical conventions}\label{app:inputs}
\begin{table}[tbp]
\caption{Angular BAO measurements used in the angular likelihood. The five rows at $z=0.57$--$0.65$ follow Carvalho et al., Table~3, arXiv version~4~\cite{CarvalhoDR11}; the other rows follow the compilation in Ref.~\cite{Nunes}. The ruler-shape test and the direct HSW fits use the first fourteen rows; the fixed-profile benchmark fits in the supplement use all fifteen.}\label{tab:angular}
\begin{ruledtabular}
\begin{tabular}{rrr}
$z$ & $\theta$ [deg] & $\sigma_\theta$ [deg]\\
0.110 & 19.80 & 3.26\\
0.235 & 9.06 & 0.23\\
0.365 & 6.33 & 0.22\\
0.450 & 4.77 & 0.17\\
0.470 & 5.02 & 0.25\\
0.490 & 4.99 & 0.21\\
0.510 & 4.81 & 0.17\\
0.530 & 4.29 & 0.30\\
0.550 & 4.25 & 0.25\\
0.570 & 4.62 & 0.40\\
0.590 & 4.37 & 0.35\\
0.610 & 3.86 & 0.33\\
0.630 & 3.88 & 0.42\\
0.650 & 3.54 & 0.17\\
2.225 & 1.77 & 0.31\\
\end{tabular}
\end{ruledtabular}
\end{table}

An angle $\theta$ in degrees corresponds to $x=\DM/\Lp=(180/\pi)/\theta$, conventionally reported as $\DM/\rd$ under the FRW ruler assumption, with first-order error $\sigma_x=(180/\pi)\sigma_\theta/\theta^2$. A Gaussian error in $\theta$, however, is not a Gaussian error in $x$. The angular likelihood of Sec.~\ref{sec:data} therefore compares the measured angles with the predicted ones directly, which requires no transformation, whereas the ruler-shape test of Sec.~\ref{sec:shape} works with logarithms of distance ratios and treats their errors as Gaussian, which is an approximation. We keep the two explicitly distinct.

The input manifest gives the exact download URLs, byte counts, and SHA-256 hashes of all data files. The solver and validation settings are given in Appendices~\ref{app:profiles} and \ref{app:robust_methods}. Earlier profile scans, fixed-background benchmarks, Gaussian-prior calculations, and the derivation of the pair dilations are kept in the technical supplement and machine-readable result files; they are not additional independent evidence in the primary fit.



\begin{thebibliography}{99}
\bibitem{Zumalacarregui} M. Zumalac\'arregui, J. Garc\'ia-Bellido, and P. Ruiz-Lapuente, Tension in the void: Cosmic rulers strain inhomogeneous cosmologies, J. Cosmol. Astropart. Phys. \textbf{10} (2012) 009, \href{https://arxiv.org/abs/1201.2790}{arXiv:1201.2790}.\bibitem{GarciaBellido} J. Garc\'ia-Bellido and T. Haugb\o lle, Confronting Lema\^itre--Tolman--Bondi models with observational cosmology, J. Cosmol. Astropart. Phys. \textbf{04} (2008) 003, \href{https://arxiv.org/abs/0802.1523}{arXiv:0802.1523}.\bibitem{Dinda} B. R. Dinda, R. Maartens, and C. Clarkson, Calibration-independent consistency test of DESI DR2 BAO and SNIa, J. Cosmol. Astropart. Phys. \textbf{12} (2025) 025, \href{https://arxiv.org/abs/2509.19899}{arXiv:2509.19899}.\bibitem{Nunes} R. C. Nunes, S. K. Yadav, J. F. Jesus, and A. Bernui, Cosmological parameter analyses using transversal BAO data, Mon. Not. R. Astron. Soc. \textbf{497}, 2133 (2020), \href{https://arxiv.org/abs/2002.09293}{arXiv:2002.09293}.\bibitem{Pantos} I. Pantos and L. Perivolaropoulos, On the origin of the BAOtr--DESI tension (2026), \href{https://arxiv.org/abs/2604.11106}{arXiv:2604.11106}.\bibitem{DESI} M. Abdul Karim et al. (DESI Collaboration), DESI DR2 results II: Measurements of baryon acoustic oscillations and cosmological constraints, Phys. Rev. D \textbf{112}, 083515 (2025), \href{https://arxiv.org/abs/2503.14738}{arXiv:2503.14738}.\bibitem{DESIdata} Cobaya BAO data repository, \href{https://github.com/CobayaSampler/bao_data/tree/bb0c1c9009dc76d1391300e169e8df38fd1096db/desi_bao_dr2}{DESI DR2 mean vector and covariance}, revision \texttt{bb0c1c9}, used for the reported reference fits.\bibitem{Menote} R. Menote and V. Marra, Baryon acoustic oscillations in thin redshift shells from BOSS DR12 and eBOSS DR16 galaxies, Mon. Not. R. Astron. Soc. \textbf{513}, 1600 (2022), \href{https://arxiv.org/abs/2112.10000}{arXiv:2112.10000}.\bibitem{Pantheon} D. Brout et al., The Pantheon+ analysis: Cosmological constraints, Astrophys. J. \textbf{938}, 110 (2022), \href{https://arxiv.org/abs/2202.04077}{arXiv:2202.04077}.\bibitem{PantheonData} Pantheon+ and SH0ES collaboration, \href{https://github.com/PantheonPlusSH0ES/DataRelease/tree/c447f0fea703fcd0fff57de5000947b5ca81286b}{Public data release}, revision \texttt{c447f0f}, used for the reported direct fits.\bibitem{Popovic} B. Popovic et al. (DES Collaboration), The Dark Energy Survey Supernova Program: A reanalysis of cosmology results and evidence for evolving dark energy with an updated Type Ia supernova calibration, Mon. Not. R. Astron. Soc. (2026), \href{https://doi.org/10.1093/mnras/stag632}{doi:10.1093/mnras/stag632}, \href{https://arxiv.org/abs/2511.07517}{arXiv:2511.07517}.\bibitem{DovekieData} DES Collaboration, \href{https://github.com/des-science/DES-SN5YR/tree/c9a4fcafc4cbd19bd750dee47fc76194a45c181f/4_DISTANCES_COVMAT}{DES-Dovekie Hubble diagram and precision matrix}, revision \texttt{c9a4fca}.\bibitem{Haslbauer} M. Haslbauer, I. Banik, and P. Kroupa, The KBC void and Hubble tension contradict $\Lambda$CDM on a Gpc scale---Milgromian dynamics as a possible solution, Mon. Not. R. Astron. Soc. \textbf{499}, 2845 (2020), \href{https://arxiv.org/abs/2009.11292}{arXiv:2009.11292}.\bibitem{Hamaus} N. Hamaus, P. M. Sutter, and B. D. Wandelt, Universal density profile for cosmic voids, Phys. Rev. Lett. \textbf{112}, 251302 (2014), \href{https://arxiv.org/abs/1403.5499}{arXiv:1403.5499}.\bibitem{Novosyadlyj} B. Novosyadlyj, M. Tsizh, and Yu. Kulinich, Evolution of density and velocity profiles of dark matter and dark energy in spherical voids, Mon. Not. R. Astron. Soc. \textbf{465}, 482 (2017), \href{https://arxiv.org/abs/1610.07920}{arXiv:1610.07920}.\bibitem{Perez} A. P\'erez-Fern\'andez et al., Fiducial-cosmology-dependent systematics for the DESI 2024 BAO analysis, J. Cosmol. Astropart. Phys. \textbf{01} (2025) 144, \href{https://arxiv.org/abs/2406.06085}{arXiv:2406.06085}.\bibitem{Bull} P. Bull, T. Clifton, and P. G. Ferreira, The kSZ effect as a test of general radial inhomogeneity in LTB cosmology, Phys. Rev. D \textbf{85}, 024002 (2012), \href{https://arxiv.org/abs/1108.2222}{arXiv:1108.2222}. \bibitem{ZhangZhang} Y. Zhang and P. Zhang, Testing $\Lambda$LTB as an alternative to dynamical dark energy (2026), \href{https://arxiv.org/abs/2610.05729v1}{arXiv:2610.05729v1}.\bibitem{Artola} M. Artola, S. Fay, and L. Perivolaropoulos, Sensitivity of calibration-free DESI DR2 BAO--supernova constraints to a distance-duality tilt (2026), \href{https://arxiv.org/abs/2610.16496v2}{arXiv:2610.16496v2}.\bibitem{NumericalArchive} M. Magklara and L. Perivolaropoulos, \emph{BAO and Type Ia Supernovae in LTB Cosmology: Executable Numerical Package}, version 1.2.0, Zenodo (2026), \href{https://doi.org/10.5281/zenodo.24374395}{doi:10.5281/zenodo.24374395}.\bibitem{SupplementalMaterial} See the Supplemental Material appended to this manuscript for fixed-location profile scans, DESI-free controls, auxiliary benchmarks, a Gaussian density prior, the corrected derivation of local, pair and fitted dilations, and the DES-only robustness analysis. It includes Refs.~\cite{Zumalacarregui,GarciaBellido,Dinda,Nunes,Pantos,DESI,DESIdata,Menote,Pantheon,PantheonData,Popovic,DovekieData,Haslbauer,Hamaus,Novosyadlyj,Perez,Bull,ZhangZhang,Artola,DESIValidation,MarraPaakkonen,ValkenburgMarraClarkson,CamarenaCopernican,CamarenaBubble,PedrottiMiscalibration,NajeraDESOnly,BanikPotential,CamarenaDESI2025}.\bibitem{DESIValidation} U. Andrade et al. (DESI Collaboration), Validation of the DESI DR2 Measurements of Baryon Acoustic Oscillations from Galaxies and Quasars (2025), \href{https://arxiv.org/abs/2503.14742}{arXiv:2503.14742}.

\bibitem{MarraPaakkonen} V. Marra and M. P\"a\"akk\"onen, Observational constraints on the $\Lambda$LTB model, J. Cosmol. Astropart. Phys. \textbf{12} (2010) 021, \href{https://arxiv.org/abs/1009.4193}{arXiv:1009.4193}.
\bibitem{ValkenburgMarraClarkson} W. Valkenburg, V. Marra, and C. Clarkson, Testing the Copernican principle by constraining spatial homogeneity, Mon. Not. R. Astron. Soc. \textbf{438}, L6 (2014), \href{https://arxiv.org/abs/1209.4078}{arXiv:1209.4078}.
\bibitem{CamarenaCopernican} D. Camarena, V. Marra, Z. Sakr, and C. Clarkson, The Copernican principle in light of the latest cosmological data, Mon. Not. R. Astron. Soc. \textbf{509}, 1291 (2022), \href{https://arxiv.org/abs/2107.02296}{arXiv:2107.02296}.
\bibitem{CamarenaBubble} D. Camarena, V. Marra, Z. Sakr, and C. Clarkson, A void in the Hubble tension? The end of the line for the Hubble bubble, Class. Quantum Grav. \textbf{39}, 184001 (2022), \href{https://arxiv.org/abs/2205.05422}{arXiv:2205.05422}.
\bibitem{PedrottiMiscalibration} D. Pedrotti, L. A. Escamilla, V. Marra, L. Perivolaropoulos, and S. Vagnozzi, BAO miscalibration cannot rescue late-time solutions to the Hubble tension, Phys. Rev. D \textbf{113}, 043507 (2026), \href{https://arxiv.org/abs/2510.01974}{arXiv:2510.01974}.
\bibitem{NajeraDESOnly} J. A. N\'ajera, I. Banik, and H. Desmond, Does the inverse distance ladder rule out late-time solutions to the Hubble tension? A model-independent test using DES-Dovekie supernovae and DESI DR2 BAO (2026), \href{https://arxiv.org/abs/2610.23657}{arXiv:2610.23657}.
\bibitem{BanikPotential} I. Banik, J. A. N\'ajera, and H. Desmond, Constraints on the gravitational potential from DESI DR2 BAO and its implications for the local void scenario, Universe \textbf{12}, 235 (2026), \href{https://arxiv.org/abs/2606.02712}{arXiv:2606.02712}.
\bibitem{CamarenaDESI2025} D. Camarena, K. Greene, J. Houghteling, and F.-Y. Cyr-Racine, DESIgning concordant distances in the age of precision cosmology: the impact of density fluctuations, Phys. Rev. D \textbf{112}, 083526 (2025), \href{https://doi.org/10.1103/1gj5-3x4m}{doi:10.1103/1gj5-3x4m}, \href{https://arxiv.org/abs/2507.17969}{arXiv:2507.17969}.

\bibitem{CarvalhoDR11} G. C. Carvalho, A. Bernui, M. Benetti, J. C. Carvalho, E. de Carvalho, and J. S. Alcaniz, Measuring the transverse baryonic acoustic scale from the SDSS DR11 galaxies, Astropart. Phys. \textbf{119}, 102432 (2020), \href{https://doi.org/10.1016/j.astropartphys.2020.102432}{doi:10.1016/j.astropartphys.2020.102432}. The angular values used here are those in Table~3 of \href{https://arxiv.org/abs/1709.00271v4}{arXiv:1709.00271v4}.
\bibitem{Efstathiou2025} G. Efstathiou, Evolving Dark Energy or Supernovae Systematics?, \href{https://arxiv.org/abs/2408.07175v3}{arXiv:2408.07175v3} (2025).

\end{thebibliography}
\end{document}


\title{Supplemental Material for ``BAO and Type Ia Supernovae in LTB Cosmology: Constraints on Radial Inhomogeneity and Acoustic-Ruler Evolution''}
\author{Myrto Magklara}
\author{Leandros Perivolaropoulos}
\email[Contact author: ]{leandros@uoi.gr}
\affiliation{Department of Physics, University of Ioannina, GR-45110 Ioannina, Greece}
\date{19 November 2026}
\maketitle

This supplement contains fixed-location profile scans, DESI-free controls, auxiliary benchmarks, and the derivation of the local, pair and fitted dilations used in Sec.~II~F of the main paper. Section~\ref{sec:des_only} gives the DES-only likelihood analysis and its diagnostic checks. The primary inference is the DESI+DES-Dovekie analysis in that paper; nominal contour thresholds in the fixed-location controls do not provide calibrated confidence regions for it. The survey-response derivation is in Sec.~\ref{app:pair_average}, and the numerical robustness methods are in Appendix~B of the main paper. Equations, sections, and tables prefixed by S belong to this supplement. Unprefixed references refer to the main paper. All bibliographic entries cited here also appear in its reference list.

\renewcommand{\thesection}{S\arabic{section}}
\renewcommand{\theequation}{S\arabic{equation}}
\renewcommand{\thetable}{S\arabic{table}}
\renewcommand{\thefigure}{S\arabic{figure}}
\section{Profile definitions and nonlinear construction}\label{app:profiles}
This section defines the six density profiles used in the ruler-shape scans of Secs.~\ref{sec:conditional_shape} and \ref{sec:nonlinear_shape}, and their exact nonlinear construction. The first-order relations between the density profile, the expansion rates, and the rulers are given in Appendix~A of the main text. We write the present radial contrast as
\begin{equation}
 \delta_m^{(0)}(r)=-d\frac{\mathcal W(r;r_w,r_0)}{\mathcal N},\qquad
 \mathcal N=\max_{r\geq0}\mathcal W>0.\label{eq:depth}
\end{equation}
Thus $d=-\min\delta_m^{(0)}$ is the physical depth and nonnegative density requires $0\leq d\leq1$. This condition does not guarantee the validity of linear perturbation theory. The unnormalized coefficient is $\delta_0=-d/\mathcal N$.

The six displaced kernels are defined with $x=(r-r_0)/r_w$:
\begin{align}
 \mathcal W_{\rm DoG}&=e^{-x^2/2}-A e^{-x^2/4},\\
 \mathcal W_{\rm exp}&=e^{-|x|}-A e^{-|x|/\sqrt2},\\
 \mathcal W_{\rm MB}&=x^2e^{-x^2/2}-A(x/\sqrt2)^2e^{-x^2/4},\\
 \mathcal W_{\rm HSW}&=\frac{1-x^2}{1+(|x|/q_s)^{9.5}},\\
 \mathcal W_{\rm sig}&=\frac{1-\tanh[(|x|-1)/(2h)]}{1+\tanh[1/(2h)]},
 \quad h=\frac16,\\
 \mathcal W_{\rm GP}&=(1-x^2)e^{-x^2}.
\end{align}
These radial shapes are motivated by void-profile studies~\cite{Haslbauer,Hamaus,GarciaBellido,Novosyadlyj}; displacing and normalizing them about a chosen observer-centred shell is an additional modeling choice. In all cases use Eq.~\eqref{eq:depth}. For DoG and exponential, $\mathcal N=1-A$; for HSW, sigmoid, and Gaussian--polynomial, $\mathcal N=1$. For Maxwell--Boltzmann the maximum must be found from the complete compensated kernel, which vanishes at $x=0$.

Compensation uses $\int_0^\infty r^2\mathcal W(r)dr=0$. For DoG, $A=I(r_w)/I(\sqrt2r_w)$, where
\begin{align}
 I(\sigma)&=\sqrt{2\pi}\sigma(r_0^2+\sigma^2)\Phi_+(\sigma)
 +r_0\sigma^2e^{-r_0^2/(2\sigma^2)},\\
 \Phi_+(\sigma)&=\frac12\left[1+\operatorname{erf}\left(\frac{r_0}{\sqrt2\sigma}\right)\right].
\end{align}
For the exponential profile, $A=J(r_w)/J(\sqrt2r_w)$ with $J(\sigma)=2\sigma r_0^2+4\sigma^3-2\sigma^3e^{-r_0/\sigma}$. Maxwell--Boltzmann uses the ratio of the two positive component integrals. HSW compensation instead fixes $q_s$ through
\begin{equation}
 \int_{-r_0/r_w}^{\infty}
 \frac{(x+r_0/r_w)^2(1-x^2)}{1+(|x|/q_s)^{9.5}}\,dx=0.
\end{equation}
At $r_0=1922.233$ Mpc, widths $320,500,1294$ Mpc give $q_s=1.58174,1.55277,1.41078$, respectively. A fixed $q_s=0.92$ is not compensated in this convention. Compensation is not imposed on the sigmoid or Gaussian--polynomial templates: the sigmoid has no overdense wall, and the wall of the Gaussian--polynomial profile carries only part of the missing mass. 

Compensation is an integral mass condition, not a guarantee of origin regularity or a globally matched exact spacetime. Several displaced kernels have a nonzero radial derivative at the origin, while the exponential and sigmoid have shell-centred cusps. The exact continuations of Sec.~\ref{sec:nonlinear_shape} use the following mass-preserving central prescription. For the normalized kernel $U$, set $U_c(r)=\sum_{j=0}^3c_{2j}(r/r_c)^{2j}$ inside $r_c$. The four linear constraints are
\begin{equation}
\begin{aligned}
 \sum_{j=0}^3c_{2j}&=U(r_c),\\
 \sum_{j=0}^3 2jc_{2j}&=r_cU'(r_c),\\
 \sum_{j=0}^3 2j(2j-1)c_{2j}&=r_c^2U''(r_c),\\
 \sum_{j=0}^3\frac{c_{2j}}{2j+3}&=\int_0^1 ds\,s^2U(r_cs).
\end{aligned}
\end{equation}
If needed, divide the whole patched kernel by its new maximum so that $d$ remains the physical depth. This rescaling preserves compensation. The even compensated benchmark of Eq.~\eqref{eq:regular_profile} uses a different origin prescription and remains a separate family.

For the nonlinear scans, curvature is solved from the common-age condition with 128-point quadrature. A universal shell table in $(\bar a,\ln A_m)$ accelerates the calculation without expanding in density: the baseline has 1001 logarithmic mass nodes plus the exact exterior mass and 641 time nodes over $0.02\leq\bar a\leq1$. Density integration uses 5-Mpc spacing in the observed radial domain; the null cone uses an adaptive eighth-order integrator. The initial depth grid has spacing 0.05, with additional points near unity, followed by scalar minimization and root refinement. The scan enforces $0\leq d<1$, stopping at $0.999999$. Widths have 25-Mpc spacing. Refinement uses 2001/1281 shell nodes and 2.5-Mpc radial spacing. A separate integration of individual radial shells checks the HSW results. Machine-readable grids, convergence differences, and sampled regularity minima accompany the calculation.
\section{Conditional six-family ruler constraints}\label{sec:conditional_shape}
\subsection{Linear profiles and their predicted ruler evolution}
Section~\ref{sec:conditional_shape} applies the ruler-shape likelihood of Sec.~IV of the main text to six void profiles: difference of Gaussians (DoG), exponential, Maxwell--Boltzmann, HSW-shaped, sigmoid, and Gaussian--polynomial (Mexican-hat), defined in Sec.~\ref{app:profiles}. The location of the trough is fixed at $r_0=\bar\chi(0.5)$ and the exterior at the flat reference $(\Omega_m,H_0)=(0.299,68.8\,\mathrm{km\,s^{-1}\,Mpc^{-1}})$, while the depth $d$ and width $r_w$ vary, over $0\leq d\leq1$ and $200\leq r_w\leq4000$ Mpc. In this subsection we derive the first-order prediction of each profile for the transverse ruler; the fits are in Sec.~\ref{sec:shape_numbers}. Section~\ref{sec:nonlinear_shape} repeats the calculation with the exact nonlinear solution, and Sec.~\ref{sec:benchmark} gives a separate even compensated benchmark.

At first order the density contrast grows as $\delta_m(r,t)=D_+(t)\delta_m^{(0)}(r)$, with $D_+(t_0)=1$, and its mean inside radius $r$ is $\bar\delta^{(0)}(r)=3r^{-3}\int_0^r du\,u^2\delta_m^{(0)}(u)$. The radial expansion rate is then
\begin{equation}
 \frac{\delta\Hl}{\bar H}
 =fD_+\left(\frac23\bar\delta^{(0)}-\delta_m^{(0)}\right),\qquad
 f=\frac{d\ln D_+}{d\ln\bar a},
\end{equation}
and the shift of the emission time at fixed observed redshift is
\begin{equation}
 \delta\ln a(z)=\int_0^z\frac{\delta\Hl(z')}{\bar H(z')}
 \frac{dz'}{1+z'}.
\end{equation}
Taking the logarithm of the first-order transverse ruler of the main text, $\Lp/\rd-1\simeq\delta\ln a-\bar\delta/3$, with $\bar\delta=D_+(z)\bar\delta^{(0)}$ evaluated at $r=\bar\chi(z)$, gives the model prediction for the shape likelihood, $\eta(z)=\ln(\Lp/\rd)$:

\begin{equation}
 \eta(z)=-\frac{D_+(z)}{3}\bar\delta^{(0)}(\bar\chi(z))
 +\delta\ln a(z).\label{eq:linear_ruler}
\end{equation}
The first term is the stretching of the transverse ruler by the enclosed underdensity, and the second is the shift of the emission time; omitting the second would give the background-comoving ruler instead of the observed one. The distance cancels in the SN/BAO ratio of Sec.~IV~A of the main text, but the relation between the profile and the ruler still depends on the assumed background and growing mode.

\subsection{Conditional numerical shape constraints}\label{sec:shape_numbers}
This subsection fits the six profiles of Sec.~\ref{sec:conditional_shape} to the ratio of SN and angular BAO distances. We first describe the SN reconstruction, then the model-independent linear trend, and finally the profile scans.

The analysis uses the fourteen angular BAO measurements at $z\leq0.65$ and all 1820 supernovae of the DES-Dovekie Hubble diagram~\cite{Popovic,DovekieData}, with the full statistical-plus-systematic covariance, from the public data at a fixed repository version~\cite{DovekieData}. The SN distances are reconstructed with the fit
\begin{equation}
 \ln\frac{\widehat\DM^{\rm SN}(z)}{zD_*}
 =\sum_{k=0}^4 a_k[\ln(1+z)]^k,\label{eq:poly}
\end{equation}
where $D_*$ is an arbitrary fixed reference length absorbed by $a_0$. The fitted quantity for supernova $i$ is $(\ln10/5)(\mu_i-25)-\ln(1+z_{{\rm HEL},i})-\ln z_{{\rm HD},i}$, and the magnitude covariance is multiplied by $(\ln10/5)^2$. A rescaled orthogonal polynomial basis and a whitened QR solve improve the numerical conditioning without changing the fitted polynomial. The covariance of the coefficients is propagated as a full matrix to the BAO redshifts. The SN-only fit gives $\chi^2=1623.928$ for 1815 nominal degrees of freedom. The reconstructed distances have fractional errors of $0.31$--$0.53\%$ at the fourteen BAO redshifts in the release's arbitrary normalization; relative to $z=0.45$, the error at $z=0.11$ is $0.69\%$. These errors are strongly correlated and smaller than the angular errors, and are retained in the likelihood.

For the degree-four reconstruction, a constant ruler gives $\chi^2_{\rm shape}=12.516$ for thirteen nominal degrees of freedom, and a linear trend $\eta(z)=b(z-z_*)$, with the constant removed, gives
\begin{equation}
 \frac{d\ln\Lp}{dz}=b=0.0589\pm0.0941,\label{eq:slope_appendix}
\end{equation}
with $\chi^2=12.124$ (Sec.~IV~C of the main text). This corresponds to $\ln[\Lp(0.65)/\Lp(0.11)]=0.0318\pm0.0508$ and constrains only a linear trend. Polynomial degrees three to six give constant-ruler statistics between $12.385$ and $12.633$ and slopes between $0.0547$ and $0.0620$. Fitting the measured angles directly instead, with the five polynomial coefficients refitted with their full Gaussian likelihood and a free angular normalization, gives a constant-ruler contribution of $12.445$ relative to the SN-only minimum and $b=0.0475$, with $\Delta\chi^2=1$ interval $-0.0453<b<0.1405$; this check includes the uncertainty of the reconstruction rather than treating its mean curve as exact. A degree-four Pantheon+ reconstruction, using $0.01<z_{\rm HD}<1.2$ and excluding calibrators (1561 light-curve rows), gives a constant-ruler $\chi^2=12.343$ and $b=0.0028\pm0.0949$; since the angular data are shared, this is a sensitivity comparison rather than an independent confirmation.

We then scan each of the six profiles, using Eq.~\eqref{eq:linear_ruler} including the emission-time term, over $200\leq r_w\leq4000$ Mpc in steps of 25 Mpc and $0\leq d\leq1$, with the depth optimized continuously. For each profile we measure $\Delta\chi^2$ from its own best fit in the $(d,r_w)$ plane. Table~\ref{tab:shape} gives the strongest depth limit, $d_{6.18}$: the smallest, over all widths, of the largest depth inside the $\Delta\chi^2=6.18$ contour, which would enclose $95\%$ for a Gaussian likelihood in two parameters. It also gives the improvement of the best fit over no void, $\Delta\chi^2_0=\chi^2(d=0)-\chi^2_{\rm min}$. These limits depend on the width range searched and are not calibrated one-parameter $95\%$ limits, nor constraints from the full DESI+SN fit of the main text.

\begin{table*}[t]
\caption{Depth constraints from the ratio of DES-Dovekie and angular BAO distances, with fixed location and exterior. $d_{6.18}$ is the strongest depth limit from the $\Delta\chi^2=6.18$ contour, reached at the listed width; $\Delta\chi^2_0$ is the improvement of the best fit over no void within the scanned widths. ``Log'' uses the likelihood of logarithmic distance ratios; ``angle'' fits the measured angles, refitting the SN polynomial coefficients. Depths are from the linear profiles.}\label{tab:shape}
\begin{ruledtabular}
\begin{tabular}{lrrrrr}
Profile & $d_{6.18}$ (log) & $d_{6.18}$ (angle) & $r_w$ [Mpc] & $\Delta\chi^2_0$ (log) & $\Delta\chi^2_0$ (angle)\\
DoG & 0.341 & 0.326 & 600 & 1.898 & 1.843\\
Exponential & 0.633 & 0.612 & 725 & 1.540 & 1.434\\
Maxwell--Boltzmann & 0.241 & 0.253 & 1075 & 0.420 & 0.282\\
HSW-shaped & 0.232 & 0.219 & 825 & 3.465 & 3.390\\
Sigmoid & 0.387 & 0.373 & 675 & 1.602 & 1.424\\
Gaussian--polynomial & 0.379 & 0.364 & 975 & 2.232 & 2.088\\
\end{tabular}
\end{ruledtabular}
\end{table*}

No profile is clearly preferred over no void. For five profiles, no void lies within $\Delta\chi^2=2.30$ of the best fit (the nominal $68\%$ contour); for HSW it does not, but it lies within $6.18$ for all six. The HSW best fit is deep, $d\simeq0.708$, and lies at the smallest width scanned, $r_w=200$ Mpc, outside the regime where the linear profiles are reliable. This improvement is not evidence for a void: the width has no meaning at $d=0$, the depth is bounded, six profiles are examined, and the contour levels are not calibrated. The best fits of the DoG, exponential, sigmoid, and Gaussian--polynomial profiles also lie at the smallest width, so the reference minimum, and with it the contours, depend on the width range.

Intermediate widths are the most constrained because they make the ruler vary across the observed redshifts, whereas broad profiles change it almost uniformly, which the unknown calibration absorbs. Section~\ref{sec:nonlinear_shape} repeats these scans with the exact nonlinear solution, and Sec.~\ref{sec:direct_hsw} tests the preferred HSW profile against the SN magnitudes and angles directly, since the ratio cancels distance changes that the SN data themselves constrain. No robust preferred width or location is established; the direct fits of Sec.~V of the main text let both vary. Increasing the light-cone quadrature from 4001 to 16001 points changes the depth limits by less than $2\times10^{-7}$, so the differences between the two likelihoods are not numerical.

\section{Exact nonlinear ruler-shape test and direct check}\label{sec:nonlinear_shape}
We now replace the linear ruler predictions of Sec.~\ref{sec:conditional_shape} by exact dust-plus-$\Lambda$ LTB solutions for all six profiles, with the same fixed exterior, location, and width range. The present density is specified in the areal gauge $R(t_0,r)=r$, and its enclosed mass defines $A_m(r)=\Omega_m[1+\bar\delta_m(t_0,r)]$. A simultaneous bang time fixes the curvature through the age equation~\eqref{eq:age}; we then integrate the expansion of each shell and the central null geodesics. The ruler is therefore computed from the full nonlinear evolution of the density profile, not by extrapolating the linear prediction of Eq.~\eqref{eq:linear_ruler}.

Near the observer the profiles use the central regularization of Sec.~\ref{app:profiles}; the exponential and sigmoid profiles remain piecewise smooth at their cusps in the trough. We check that the density and mass are positive, that the metric has the correct signature, and that the observed null cone is regular and monotonic. At the reported minima and contour intersections we also check $R'>0$ on a radial--time grid covering the profile and $0.02\leq\bar a\leq1$, and $q'>0$ in the early dust limit. These numerical checks do not prove the absence of shell crossing at all times.

With the common initial ruler of Eq.~\eqref{eq:massruler}, the exact prediction for the shape likelihoods is
\begin{equation}
 \begin{aligned}
 \eta(z)&=\ln\left[\frac{(1+z)R(t(z),r(z))}{q(r(z))}\right],\\
 q(r)&=r\left[\frac{A_m(r)}{\Omega_m}\right]^{1/3}.
 \end{aligned}
 \label{eq:nonlinear_m}
\end{equation}
Because this uses the observed redshift directly, it already contains the shift of the emission time that the linear prediction adds separately as $\delta\ln a$. The shape likelihoods use the same DES-Dovekie reconstruction and angular data as Table~\ref{tab:shape}. Since the distance cancels in the SN/BAO ratio, these fits do not check whether the distances of the void model itself agree with the supernovae; Sec.~\ref{sec:direct_hsw} makes that check.

\begin{table*}[t]
\caption{Exact nonlinear counterparts of Table~\ref{tab:shape}, with the same fixed exterior, location, and $200\leq r_w\leq4000$ Mpc range. $d_{6.18}$ is the strongest depth limit from the $\Delta\chi^2=6.18$ contour around each profile's own best fit, reached at the listed width (two values give the widths for the log and angle likelihoods; a single value means they coincide); it is not a one-parameter $95\%$ limit. $\Delta\chi^2_0$ is the improvement of the best fit over no void. ``Log'' uses the likelihood of logarithmic distance ratios; ``angle'' fits the measured angles, refitting the SN polynomial coefficients. Both use the exact ruler.}\label{tab:nonlinear_shape}
\begin{ruledtabular}
\begin{tabular}{lrrrrr}
Profile & $d_{6.18}$ (log) & $d_{6.18}$ (angle) & $r_w$ [Mpc] & $\Delta\chi^2_0$ (log) & $\Delta\chi^2_0$ (angle)\\
DoG & 0.326 & 0.312 & 650 & 1.661 & 1.607\\
Exponential & 0.524 & 0.507 & 925 & 1.362 & 1.260\\
Maxwell--Boltzmann & 0.209 & 0.218 & 1075 & 0.407 & 0.273\\
HSW-shaped & 0.229 & 0.217 & 825 & 3.454 & 3.362\\
Sigmoid & 0.339 & 0.328 & 725 & 1.459 & 1.291\\
Gaussian--polynomial & 0.348 & 0.335 & 1025 & 1.920 & 1.788\\
\end{tabular}
\end{ruledtabular}
\end{table*}

Comparing Tables~\ref{tab:shape} and \ref{tab:nonlinear_shape}, the linear depth limits are not uniformly accurate: the nonlinear changes depend on the profile shape and can move the width at which the strongest limit occurs. The HSW-shaped limit, however, is stable, changing from $0.232$ to $0.229$ in the logarithmic likelihood and from $0.219$ to $0.217$ with the angles, still at $r_w=825$ Mpc. Its best fit moves from $d=0.708$ to about $0.764$ in the logarithmic likelihood, at $r_w=200$ Mpc, with improvement $\Delta\chi^2_0=3.454$; with the angles the best fit is near $d=0.738$ and the improvement is $3.362$. The exact solution therefore neither removes the modest preference of the ratio test for this HSW void nor turns it into a detection. The best width lies at the lower end of the range; widths of $100$, $125$, $150$, and $175$ Mpc give smaller improvements, $0.650$, $1.813$, $2.597$, and $3.203$. At $r_w=200$ Mpc, however, the void is barely larger than the BAO ruler, so the local-ruler approximation is doubtful there, even with an exact background.

Since the width of the void has no meaning at $d=0$, the usual conversion of $\Delta\chi^2$ into a significance does not apply. We therefore generate 20,000 Gaussian realizations with the same covariance of the logarithmic ruler data and repeat the HSW grid search in each. On the discrete simulation grid the observed statistic is $3.45248$, slightly below the depth-optimized value $3.454$ in Table~\ref{tab:nonlinear_shape}. Comparing simulated and observed statistics on that same grid, an improvement at least as large as $3.45248$ occurs with probability $p=0.085$ (Monte Carlo error $0.002$); searching all six profiles in each realization raises this to $p=0.190\pm0.003$. These probabilities assume the stated profiles, exterior, location, width range, covariance approximation, and search grid. They give the significance of the improvement, not the coverage of the two-parameter contours.

The nonlinear shell and null-cone calculations pass resolution and independent-integration checks. For the HSW best fit, doubling the shell-table resolution and halving the radial integration step changes the log likelihood by less than $5\times10^{-5}$, and an independent integration of the radial shells agrees with the refined prediction for $\eta(z)$ within $5\times10^{-8}$. At the HSW depth limit the corresponding difference in the ruler is below $2\times10^{-9}$. Varying $r_c$ between 50 and 200 Mpc changes the log-ruler statistic by less than $2\times10^{-6}$ for the tested profiles, and the small-depth limit recovers Eq.~\eqref{eq:linear_ruler}. These checks concern the numerical solution of the background; the local passive-ruler approximation remains a separate assumption.

\subsection{A direct nonlinear HSW fit to SN magnitudes and BAO angles}\label{sec:direct_hsw}
The SN/BAO ratio discards the distance information. To test the HSW profile preferred by the ratio against the distances themselves, we fit the exact model directly to all 1820 DES-Dovekie magnitudes, with their full covariance, and to the same fourteen measured angles. Table~\ref{tab:direct_hsw} summarizes the results. The SN magnitude offset and the initial ruler $\rd$ are fitted separately. In degrees, the predicted angles are $\theta_i^{\rm th}=\rd t_i$ with $t_i=(180/\pi)/q(r(z_i))$, so that
\begin{equation}
 \widehat\rd=\frac{\sum_i t_i\theta_i/\sigma_i^2}
 {\sum_i t_i^2/\sigma_i^2}.
\end{equation}
The statistic is the sum of the SN and angular contributions; the SN reconstruction is not used. In this first calculation $H_0$, $\Omega_m$, and $r_0$ remain fixed and DESI is not included; Sec.~\ref{sec:free_hsw} frees $\Omega_m$ and $r_0$ and compares with flat $\Lambda$CDM fitted in the same way.

\begin{table*}[t]
\caption{Direct exact HSW fits to DES-Dovekie magnitudes and the fourteen angular BAO measurements. Each row fits the SN offset and $\rd$. The middle row evaluates the profile preferred by the ruler-shape test, without refitting its depth and width; the last row is the best direct fit on the 25-Mpc width grid. The quoted $\rd$ assumes the fixed $H_0=68.8$ and is not an independent absolute calibration.}\label{tab:direct_hsw}
\begin{ruledtabular}
\begin{tabular}{lrrrrrr}
Profile & $d$ & $r_w$ [Mpc] & $\widehat\rd$ [Mpc] & $\chi^2_{\rm SN}$ & $\chi^2_\theta$ & $\chi^2_{\rm total}$\\
No void & 0 & --- & 156.003 & 1635.839 & 12.415 & 1648.255\\
Ruler-shape minimum & 0.7646 & 200 & 155.504 & 1644.864 & 9.820 & 1654.684\\
Direct joint minimum & 0.0882 & 1575 & 151.365 & 1630.760 & 11.935 & 1642.695\\
\end{tabular}
\end{ruledtabular}
\end{table*}

The deep, narrow HSW profile preferred by the ratio fits the data directly worse than no void, by $6.429$: its better fit to the angles does not make up for its poorer fit to the SN distances. The best direct fit is instead a shallow, broad void, $d=0.0882$ and $r_w=1575$ Mpc, which improves the total by $5.560$. Refining the width continuously gives $d=0.0871$ and $r_w=1563$ Mpc and changes the minimum by only $0.0008$. An independent integration of the radial shells reproduces this minimum within $2\times10^{-5}$ in $\chi^2$, and refining the resolution changes it by less than $7\times10^{-5}$.

Most of this improvement, however, comes from the fixed exterior density rather than from the void. Fitting the same data with flat $\Lambda$CDM and $\Omega_m$ free gives $\Omega_m=0.3308$ and $\chi^2=1643.668$, already $4.587$ below the fixed-$\Omega_m$ reference and only $0.973$ above the best void. The two fits are not nested, so this is not a significance test, but it shows that the apparent gain of the void depends strongly on fixing the exterior cosmology.

Relative to the best direct fit, the strongest depth limit from the $\Delta\chi^2=6.18$ contour is $d=0.0304$ at $r_w=550$ Mpc. This holds at that width only: larger depths are allowed at other widths, including the best-fit width. No void lies inside this contour, since its improvement is only $5.560$. We assign no significance or confidence coverage to this direct fit, which would require dedicated simulations. Section~\ref{sec:free_hsw} lets the exterior density and location vary in the same likelihood, and Sec.~V of the main text adds the DESI measurements.
\section{Direct fits without DESI, with free exterior density and location}\label{sec:free_hsw}
In Sec.~\ref{sec:direct_hsw} the exterior density and the location of the void were fixed. Here both are free. Flat $\Lambda$CDM and the exact HSW-shaped model are fitted to the same 1820 DES-Dovekie SN magnitudes and fourteen angular BAO measurements, with the same covariances, and with the SN offset and $\rd$ fitted separately in both. The homogeneous model fits $\Omega_m$; the HSW model additionally fits the present depth $d$, width $r_w$, and location $r_0$ of the underdense shell, with the observer at the centre. The deep void found here is the DESI-free fit that DESI rejects in Sec.~V~B of the main text. We search
\begin{equation}
\begin{aligned}
0.1&\leq\Omega_m\leq0.5,&0&\leq d\leq0.999,\\
0.2&\leq r_w/{\rm Gpc}\leq4,&0&\leq r_0/{\rm Gpc}\leq6.
\end{aligned}\label{eq:free_domain}
\end{equation}
These are search limits, not priors; the width range is the one of Sec.~\ref{sec:conditional_shape}. We use $\Omega_\Lambda=1-\Omega_m$ in the exterior and, whenever $\Omega_m$ changes, recompute its age, the curvature for a simultaneous bang time, the evolution of the shells, and the null cone. The HSW compensation is recalculated for each $r_0/r_w$, with the same central regularization as before. The location $r_0$ is a present areal radius, varied independently, rather than the distance to a fixed redshift recomputed in each FRW background.

With uncalibrated SNe and free $\rd$, $H_0=68.8$ km s$^{-1}$ Mpc$^{-1}$ is a scale convention. The dimensional fitted quantities rescale as $H_0^{-1}$; the corresponding combinations $H_0\rd$, $H_0r_w$, and $H_0r_0$ carry the information. No external $H_0$, sound-horizon, or density prior is added. Thus both models receive the same calibration freedom. Counting the two profiled normalizations, flat $\Lambda$CDM has $k=3$ parameters and HSW has $k=6$.

Two global searches with independent random seeds (differential evolution), several local refinements, and scans in depth and location find several separate minima. The lowest one and two nearby alternatives are listed in Table~\ref{tab:free_hsw}; they are fitted points, not estimates with Gaussian errors. The nearby narrow void is only $0.327$ above the deep one, although its depth, location, and exterior density are very different, and a broader void lies $1.935$ above it. The uncertainty cannot therefore be described by the curvature of the likelihood around a single minimum.

\begin{table*}[t]
\caption{Direct fits to SN magnitudes and angular BAO measurements, without DESI, with the exterior density and location free. All rows use the same SN and angular likelihoods and fit the same two calibrations. The HSW rows are separate local minima; the first is the lowest found within Eq.~\eqref{eq:free_domain}. Differences quoted in the text are evaluated before rounding the entries. Radii and $\widehat\rd$ use the stated $H_0$ scale convention.}\label{tab:free_hsw}
\begin{ruledtabular}
\begin{tabular}{lrrrrrrrr}
Model & $\Omega_m$ & $d$ & $r_w$ [Gpc] & $r_0$ [Gpc] & $\widehat\rd$ [Mpc] & $\chi^2_{\rm SN}$ & $\chi^2_\theta$ & $\chi^2_{\rm total}$\\
Flat $\Lambda$CDM & 0.33076 & 0 & --- & --- & 154.554 & 1631.421 & 12.247 & 1643.668\\
HSW: deep mode & 0.36075 & 0.69661 & 0.63342 & 3.37905 & 154.918 & 1625.457 & 11.926 & 1637.383\\
HSW: nearby mode & 0.31849 & 0.13199 & 0.23903 & 0.49209 & 154.877 & 1625.261 & 12.450 & 1637.710\\
HSW: broad mode & 0.34647 & 0.08937 & 1.03739 & 0.94940 & 148.875 & 1626.663 & 12.656 & 1639.319\\
\end{tabular}
\end{ruledtabular}
\end{table*}

The improvement of the best void over flat $\Lambda$CDM is
\begin{equation}
T_{\rm obs}\equiv\chi^2_{\Lambda{\rm CDM},\min}-\chi^2_{{\rm HSW},\min}
=6.285,
\label{eq:free_improvement}
\end{equation}
of which $5.964$ comes from the SNe and only $0.321$ from the angular BAO. The Akaike score, ${\rm AIC}=\chi^2+2k$, which penalizes each additional parameter, gives $\Delta{\rm AIC}\equiv{\rm AIC}_{\rm HSW}-{\rm AIC}_{\Lambda{\rm CDM}}=-0.285$: the gain barely offsets the penalty of six for the three extra parameters. With the usual correction for a finite number of measurements (1834), $\Delta{\rm AIC}_c=-0.252$. Neither score indicates a meaningful preference for the void, and neither replaces a significance test. The usual rule for converting $\Delta\chi^2$ into a significance with three degrees of freedom does not apply, since at $d=0$ the width and location have no meaning and the depth lies at the edge of its range.

To test the significance, we generate 5000 simulated datasets from the best-fitting flat $\Lambda$CDM model, with Gaussian errors. In each, flat $\Lambda$CDM is refitted ($\Omega_m$ and both calibrations), and the HSW model is fitted by searching a fixed set of 4032 exact models within Eq.~\eqref{eq:free_domain}, again fitting both calibrations. In 505 of them the improvement reaches the observed $6.285$, a fraction of $0.101$ (95\% binomial interval $[0.093,0.110]$). Because the HSW search in each simulation is restricted to this set, a complete search could only find larger improvements, so $0.101$ is a lower limit on the true probability; it assumes the fitted $\Lambda$CDM model and Gaussian errors. Since even the lower end of the interval exceeds $0.05$, the improvement is not a significant detection of a void. The set of models and the treatment of the calibrations are given in Sec.~\ref{app:free_hsw}. These simulations test a different statistic from the ruler-shape probabilities of Sec.~\ref{sec:nonlinear_shape}, which are not reused here.

The likelihood as a function of depth and location shows the remaining degeneracies. Even for a nearly empty void, $d=0.999$, refitting the other parameters gives $\Omega_m\simeq0.370$, $r_w\simeq0.653$ Gpc, and $r_0\simeq3.465$ Gpc, only $0.492$ above the best fit, so this search gives no useful upper limit on the depth. The depth limits found at fixed location, $d=0.229$ and $d=0.0304$, therefore do not hold once the exterior density and location are free. We assign no confidence levels to these one-dimensional likelihood curves; that would require a calibration that accounts for the boundaries and the multiple minima.

The deep void has its density minimum at redshift $z(r_0)=1.063$, beyond the highest angular measurement, at $z=0.65$, and close to the highest SN redshift, $z=1.144$; only five SNe lie beyond it. Together with the fact that the gain comes almost entirely from the SNe, this means its depth should not be read as that of a well-measured void. Its transverse ruler gives $\Lp(0.65)/\Lp(0.11)=0.9582$, a $4.18\%$ decrease, whereas the almost equally good nearby void gives $0.9760$. These are predictions of particular fits, not detections of ruler evolution or of a change in the primordial $\rd$; the linear trend of Sec.~\ref{sec:shape_numbers} remains consistent with no evolution.

The solver with free parameters reproduces the homogeneous limit and the fixed-parameter HSW results. Removing the interpolation in $\Omega_m$, increasing the shell resolution, and halving the radial spacing change $\chi^2$ by less than $3\times10^{-5}$ at the three HSW minima. An independent integration of the radial shells reproduces the deep minimum within $2\times10^{-7}$ in $\chi^2$, and its ruler prediction within $5\times10^{-10}$ in $\ln\Lp$. All three voids have positive density and the correct metric signature, and satisfy $q'>0$ at early times and $R'>0$ at the sampled points. These checks concern the exact background solution only. They do not test the local-ruler approximation, which is least reliable for the nearby void, whose width, $r_w\simeq239$ Mpc, is only about $1.5$ times the fitted ruler. The comparison also assumes the released SN corrections and a diagonal angular covariance. The DESI measurements are added in Sec.~V of the main text; a fit to the full galaxy correlation function remains beyond the present calculation.
\section{An illustrative fixed profile and flat-$\Lambda$CDM reference fits}\label{sec:benchmark}
This section is an illustrative example with one fixed density profile, not an additional test of the results of the main text. With the width, location, exterior cosmology, and $\rd$ fixed, it shows how the geometric distance and the transverse ruler respond differently to the same void (Sec.~\ref{sec:benchmark_geometry}) and how the SN, angular BAO, and DESI data respond to its depth (Sec.~\ref{sec:benchmark_depth}); the corresponding constraints with all parameters free are given in Sec.~\ref{sec:free_hsw} and in Sec.~V of the main text. Section~\ref{sec:benchmark_frw} gives flat-$\Lambda$CDM reference fits, including a free normalization of the angular BAO measurements.

\subsection{A regular compensated density profile}
We use an exact dust-plus-$\Lambda$ solution with fixed exterior parameters
\begin{equation}
 (H_0,\Omega_m,\Omega_\Lambda,\rd)=(68.8,0.299,0.701,147.09),
\end{equation}
with $H_0$ in km\,s$^{-1}$\,Mpc$^{-1}$ and $\rd$ in Mpc. Let $r_0=\bar\chi(0.5)=1922.233$ Mpc, $w=1294$ Mpc, and $s=\sqrt2w$. Define
\begin{align}
 g_\sigma(r)&=e^{-(r-r_0)^2/(2\sigma^2)}+e^{-(r+r_0)^2/(2\sigma^2)},\\
 U(r)&=\frac{g_w(r)-A g_s(r)}{\max_{r\geq0}[g_w(r)-A g_s(r)]},\\
 A&=\frac{w(r_0^2+w^2)}{s(r_0^2+s^2)},\qquad
 \delta_m(t_0,r)=-dU(r).\label{eq:regular_profile}
\end{align}
The second Gaussian in $g_\sigma$, a mirror image at $-r_0$, makes the profile even in $r$, so that its radial derivative vanishes at the origin. The identity $\int_0^\infty dr\,r^2g_\sigma=\sqrt{2\pi}\sigma(r_0^2+\sigma^2)$ shows that the value of $A$ makes the profile compensated. Here $A=0.539016$, the maximum in the denominator of $U$ is $0.415611$, and the density minimum lies near 2013.45 Mpc. The mass deficit and excess cancel only asymptotically, not within a finite radius. This profile differs from the displaced DoG profile of Sec.~\ref{app:profiles}, although its width is also labelled $w$.

In the present areal gauge $R(t_0,r)=r$, integrate the density to obtain $M$. With $a_r=R/r$, $\tau=H_0t$, and $k=2Ec^2/(H_0^2r^2)$, define
\begin{equation}
 A_m(r)=\Omega_m[1+\bar\delta_m(t_0,r)].
\end{equation}
The shell equation and simultaneous-bang-time condition are
\begin{align}
 \left(\frac{da_r}{d\tau}\right)^2&=\frac{A_m}{a_r}+k+\Omega_\Lambda a_r^2,\\
 \tau_0&=\int_0^1\frac{da_r}{\sqrt{A_m/a_r+k+\Omega_\Lambda a_r^2}},
 \label{eq:age}
\end{align}
where $\tau_0$ is the exterior age. Solving Eq.~\eqref{eq:age} for each shell fixes the curvature, after which the shell evolution and null cone determine all geometric observables.

In the early dust growing-mode limit, the mass coordinate
\begin{equation}
 q(r)=r[A_m(r)/\Omega_m]^{1/3}
\end{equation}
satisfies $R/\bar a\to q$. The local rulers follow the established LTB stretching prescription~\cite{Zumalacarregui,ValkenburgMarraClarkson,CamarenaCopernican,CamarenaDESI2025} and become
\begin{equation}
 \ell_\perp=\rd\frac{R}{q},\qquad
 \ell_\parallel=\rd\frac{R'}{q'},\qquad
 X_\perp=\frac{q(r(z))}{\rd}.\label{eq:massruler}
\end{equation}
Although $R$ cancels in $X_\perp$, the result still depends on the model, since both the map $z\mapsto r(z)$ and the mass coordinate $q$ depend on the profile. The dust solver does not evolve radiation, baryons, or the formation of the acoustic feature: the background is exact within dust+$\Lambda$, but the BAO prediction still assumes a common early ruler and local passive transport.

\subsection{Geometry versus ruler evolution}\label{sec:benchmark_geometry}
Table~\ref{tab:exact} gives the benchmark at $z=0.51$. For $d=0.30$ the geometric distance decreases by $3.20\%$, whereas the observed transverse ruler increases by $9.45\%$, so their ratio, which transverse BAO measures, decreases by $11.55\%$. The shift inferred from BAO is therefore not the same as the change of the distance: SNe and BAO can respond differently to a void even when both trace the same spacetime correctly.

\begin{table}[t]
\caption{The fixed benchmark profile at $z=0.51$, with fixed width, location, and exterior parameters. The last column is the local transverse BAO observable, common to angular and three-dimensional measurements, not a difference between them.}\label{tab:exact}
\begin{ruledtabular}
\begin{tabular}{ccccc}
$d$ & $\DM/\bar\DM$ & $\Lp/\rd$ & $Q$ & $X_\perp/(\bar\DM/\rd)$\\
0.10 & 0.98951 & 1.02831 & 1.00622 & 0.96227\\
0.30 & 0.96805 & 1.09451 & 1.01878 & 0.88446\\
\end{tabular}
\end{ruledtabular}
\end{table}

An independent shell and light-cone integration with 801 radial and 641 time points reproduces these values. Refining to 1601/1281 points at $d=0,0.10,0.30$ changes $X_\perp$ by less than $3\times10^{-10}$ and $X_\parallel$ by less than $2\times10^{-8}$ over $0.002\leq z\leq2.4$. In the flat limit the distances agree with a separate quadrature to within $2\times10^{-11}$. At $d=0.30$ the minimum of $R'$ on the sampled shell grid is $0.04255$, and the radial metric factor is positive along the sampled light cone. At $d=10^{-4}$, $d^{-1}\ln(\Lp/\rd)$ agrees with the linear prediction for this profile within $1.3\times10^{-5}$. These checks test the background integration and the linear limit; they do not test the BAO measurement model or prove the absence of shell crossing at all times.

\subsection{A joint flat-FRW reference likelihood}\label{sec:benchmark_frw}
For reference, we fit flat $\Lambda$CDM to DESI, the angular BAO measurements (all fifteen), and Pantheon+ SNe, separately and in combination. Table~\ref{tab:fits} summarizes the fits. The parameters are $\Omega_m$ and $H_0\rd$, with
\begin{equation}
 \frac{\DM}{\rd}=\frac{c}{H_0\rd}\int_0^z\frac{du}{E(u)},\qquad
 E^2=\Omega_m(1+z)^3+1-\Omega_m,
\end{equation}
and radiation neglected. To test whether the angular catalogue agrees in normalization with the other data, we also allow a constant factor $\varepsilon$ in its predictions,
\begin{equation}
 \theta_{\rm th}^{\rm(deg)}=\frac{180}{\pi}\varepsilon\frac{\rd}{\DM}.
 \label{eq:calibration}
\end{equation}
The factor $\varepsilon$ applies only to the angular measurements; it is a test of their relative normalization, not a physical LTB ruler. A constant factor cancels in the double ratio of Sec.~IV~A of the main text, so an angular catalogue can pass the ruler-shape test and still disagree in normalization with another catalogue.

\begin{table*}[t]
\caption{Flat-$\Lambda$CDM reference fits to DESI, the fifteen angular BAO measurements (fitted in the measured angles), and Pantheon+ SNe, with the SN offset fitted. $H_0\rd$ is in km\,s$^{-1}$. A dash denotes a quantity not used or not constrained. These fitted models are distinct from the fixed exterior of the benchmark profile.}\label{tab:fits}
\begin{ruledtabular}
\begin{tabular}{lrrrrrr}
Data & $\Omega_m$ & $H_0\rd$ & $\varepsilon$ & $\chi^2_{\rm D}$ & $\chi^2_\theta$ & $\chi^2_{\rm SN}$\\
DESI & 0.29746 & 10153.98 & --- & 10.271 & --- & ---\\
Angular BAO & 0.39600 & 10442.08 & 1 & --- & 12.259 & ---\\
Pantheon+ & 0.33260 & --- & --- & --- & --- & 1386.405\\
DESI + angular & 0.28766 & 10255.86 & 1 & 12.284 & 27.748 & ---\\
DESI + Pantheon+ & 0.30441 & 10099.83 & --- & 10.904 & --- & 1388.892\\
All, fixed normalization & 0.29560 & 10193.45 & 1 & 11.084 & 29.918 & 1390.743\\
All, free normalization & 0.30471 & 10097.50 & 1.06159 & 10.959 & 12.661 & 1388.837\\
\end{tabular}
\end{ruledtabular}
\end{table*}

The combined data prefer an angular normalization about $6\%$ higher than unity: freeing $\varepsilon$ lowers the total $\chi^2$ from $1431.744$ to $1412.458$, with $1.04750<\varepsilon<1.07573$ at $\Delta\chi^2=1$. This depends on the assumed angular covariance. With a common correlation $\rho_c$ between the angular measurements, $C_{ij}=\sigma_i\sigma_j[(1-\rho_c)\delta_{ij}+\rho_c]$, the best fit becomes $\widehat\varepsilon=1.0533$, $1.0435$, and $1.0377$ for $\rho_c=0.1$, $0.3$, and $0.5$; these are illustrative choices, not measured correlations. The fit shows a normalization difference between the angular catalogue and the other data, but does not identify its cause and is not evidence for inhomogeneity.

\subsection{A fixed-profile LTB depth sequence}\label{sec:benchmark_depth}
For the benchmark profile we evaluate 15 depths, from $d=0$ to $0.35$ in steps of $0.025$, keeping the width, location, exterior cosmology, and $\rd$ fixed. The DESI predictions use the local observables $X_\perp$, $X_\parallel$, and $X_V$ of Sec.~II~E of the main text at the effective redshifts, the angular BAO are fitted in the measured angles, and the SN likelihood uses Pantheon+ with the offset fitted. Some of the contributions are
\begin{center}
\begin{tabular}{crrr}
\hline
$d$ & $\chi^2_{\rm D}$ & $\chi^2_\theta$ & $\chi^2_{\rm SN}$\\
0 & 10.923 & 34.018 & 1389.964\\
0.10 & 88.198 & 17.115 & 1386.811\\
0.175 & 297.385 & 13.029 & 1386.843\\
0.30 & 988.265 & 28.228 & 1391.774\\
\hline
\end{tabular}
\end{center}
The angular BAO alone are fitted best at $d=0.175$ on this grid, but DESI strongly disfavors that depth for this fixed profile. The combined $\chi^2$ improves slightly at a shallower depth: at $d=0.025$ it is about $7.00$ below its value at $d=0$. Neither the angular preference nor the DESI penalty is therefore a constraint on the depth by itself; that requires fitting the other parameters as well, as in Sec.~\ref{sec:free_hsw} and Sec.~V of the main text.

The SN data change little with depth, because they see only the shape of the distance--redshift relation, with a free normalization, whereas the BAO observables also contain the evolved rulers. Conversely, a profile that passes the ruler-shape test of Sec.~IV of the main text need not agree with the radial BAO or with the full distance data. The scan illustrates how the different data constrain a void together, but it does not fit the other parameters and does not exclude all profiles.
\section{Numerical details of the fits and simulations without DESI}\label{app:free_hsw}
This section gives the numerical details of the fits with free exterior density and location, and of their significance test, in Sec.~\ref{sec:free_hsw}.

\paragraph*{Exact solutions.}
The shell solution is tabulated as a function of the exterior scale factor $\bar a$ and $u=\ln[A_m/\Omega_m]$, for exterior densities spaced by $0.01$. Each exterior satisfies its own age equation with $\Omega_\Lambda=1-\Omega_m$, and four-node cubic interpolation in $\Omega_m$ speeds up the search. The baseline table has 501 logarithmic mass nodes plus the exact homogeneous node, and 401 time nodes; the reported minima are recomputed with 1001/801 nodes at their exact $\Omega_m$, without interpolation. The mass of the central polynomial is integrated analytically. With $\bar U(r)=3r^{-3}\int_0^rdu\,u^2U(u)$ the mean of the normalized profile inside radius $r$, the enclosed mass is $S\equiv A_m/\Omega_m=1-d\bar U$, and its derivative is evaluated as $S'=-3d(U-\bar U)/r$, which keeps the local density and the enclosed mass consistent. An independent integration of 2401 individual radial shells, which uses the age condition and the physical time directly, without the tabulated solutions, checks the results. Varying the radius of the central regularization from 100 Mpc to 50 or 200 Mpc changes the likelihood of the deep minimum by less than $1.2\times10^{-5}$.

\paragraph*{Simulated data.}
The simulations use the full SN covariance and the same diagonal angular errors as the fit to the real data. With the Cholesky factorization $P=U_P^TU_P$ of the SN precision matrix, the SN noise is drawn in the whitened space, with the component along $U_P\bm1$, a common shift of all magnitudes that the free offset would absorb, removed. The simulated SN data are centred on the distances of the best-fitting flat $\Lambda$CDM model, and the simulated angles, with independent noise, on its predictions $\widehat\rd\,\bm t_0$. In each simulation $\rd$ is refitted to the measured angles for every model, rather than held at the value used to generate the data, and the flat-$\Lambda$CDM density $\Omega_m$ is refitted. A spline on a fine grid in $\Omega_m$, used for this refit, agrees with direct quadrature within $5\times10^{-10}$ in the minimized $\chi^2$ for ten test simulations.

\paragraph*{Set of void models.}
In each simulation the HSW model is fitted by searching the fixed set
\begin{align}
\Omega_m&=0.28,0.30,\ldots,0.42,\nonumber\\
d&\in\{.02,.05,.10,.20,.40,.70,.95\},\nonumber\\
r_w/{\rm Gpc}&\in\{.20,.35,.60,1.0,1.8,3.2\},\nonumber\\
r_0/{\rm Gpc}&\in\{.2,.4,.6,1,1.5,2,2.5,3,3.5,4,5,6\}.
\end{align}
All 4032 models satisfy the regularity conditions at the sampled points. Recomputing their predictions at the exact $\Omega_m$ with refined shell and radial resolution does not change the number of simulations that reach the threshold; the largest change in the 5000 simulated improvements is $5.6\times10^{-4}$. We use these refined predictions for the final result.

\paragraph*{Why the result is a lower limit.}
Let $T_{\cal G}$ be the improvement over flat $\Lambda$CDM found within this set of models (or zero, if larger), and $T_{\rm full}$ the improvement that a complete continuous search would find in the same simulation. A restricted search can only find a smaller improvement, so $T_{\cal G}\leq T_{\rm full}$ in every simulation, and the probability under flat $\Lambda$CDM satisfies
\begin{equation}
\Pr_0(T_{\rm full}\geq T_{\rm obs})\geq
\Pr_0(T_{\cal G}\geq T_{\rm obs}).
\end{equation}
As a numerical safety margin, much larger than the refinement errors measured at the reported minima, $0.001$ is subtracted from each simulated improvement before the comparison. The threshold $T_{\rm obs}=6.285$ is the improvement of the continuous search on the real data; the best model within the fixed set fits the real data less well, with $\chi^2=1638.456$ instead of $1637.383$, so using the continuous value makes the threshold harder to reach and keeps the estimate a lower limit. With random seed 27102026, 505 of the 5000 simulations reach the threshold. The quoted interval describes only the finite number of simulations, not the uncertainty of a fully searched probability. The result assumes the fitted flat-$\Lambda$CDM parameters and Gaussian errors; it does not calibrate the test for other homogeneous models or account for unknown systematic errors in the data. The lower limit avoids relying on the usual $\Delta\chi^2$ rule, which does not apply here, without requiring a complete continuous search in every simulation.


\section{Local dilations, pair dilations, and fitted dilations}\label{app:pair_average}
This section derives the geometrical pair description and illustrative fit response used in Sec.~II~F of the main paper. The baseline likelihood uses local predictions at the published effective redshifts. The construction below is not a numerical evaluation of the actual DESI estimator or a complete LTB two-point function. It separates the assumptions needed for such an evaluation.

\paragraph*{Local dilations and the fiducial convention.}
Let $\chi_{\fid}(z)=\DM^{\fid}(z)$ be the comoving distance of a fixed flat fiducial cosmology, with $d\chi_{\fid}/dz=c/H^{\fid}$. The mass coordinate $q(r)$ is defined in Sec.~III~A of the main paper. Combining the local transported rulers with the fiducial conversions gives
\begin{equation}
\begin{aligned}
 \alpha_\perp&=\frac{\rd^{\fid}}{\rd}\frac{q[r(z)]}{\chi_{\fid}(z)},\\
 \alpha_\parallel&=\frac{\rd^{\fid}}{\rd}
 \frac{H^{\fid}(z)q'[r(z)]}{(1+z)\dot R'[t(z),r(z)]}.
\end{aligned}\label{eq:fixed_fid_alpha}
\end{equation}
Thus the survey fiducial model and $\rd^{\fid}$ need not equal the exterior FRW model or the initial ruler of the LTB model being fitted.  They remain fixed in an application to a specified catalogue and template. The ruler-only transverse factor $\rd/\Lp$ must also include the distance ratio $\DM/\chi_{\fid}$ to predict a dilation.

For the next derivation only, choose the fiducial cosmology to be the exterior and $\rd^{\fid}=\rd$, and write $\chi=\chi_{\fid}$. Across the line of sight a small separation contains $q\,\Delta\theta/\rd$ initial rulers, compared with $\chi\,\Delta\theta/\rd$ assigned by the fiducial model. Along the line of sight the corresponding counts are $X_\parallel dz$ and $d\chi/\rd$. Hence
\begin{equation}
 \alpha_\perp=\frac{q}{\chi},\qquad
 \alpha_\parallel=\frac{\bar H q'}{(1+z)\dot R'}.
 \label{eq:local_alpha}
\end{equation}
The null geodesic gives $dr/dz=c\sqrt{1+2E}/[(1+z)\dot R']$, so
\begin{equation}
\begin{aligned}
 \frac{dq}{d\chi}&=\sqrt{1+2E}\,\alpha_\parallel,\\
 \alpha_\perp(\chi)&=\frac{1}{\chi}\int_0^\chi
 \sqrt{1+2E(\chi')}\,\alpha_\parallel(\chi')\,d\chi'.
\end{aligned}\label{eq:aperp_integral}
\end{equation}
The central boundary condition is $q(0)=0$. For a different fixed fiducial convention, the first identity acquires the factor $\rd/\rd^{\fid}$ on its right; the second retains its form with $\chi_{\fid}$, because that factor cancels. Spatial curvature cannot in general be dropped from this relation.

To first order about the exterior, the rulers of main Appendix~A give
\begin{align}
 \alpha_\parallel-1&=-(1+f)\Big(\tfrac23\bar\delta-\delta_m\Big)
 -\Big(1-\tfrac32\Omega_m\Big)\delta\ln a,\\
 \alpha_\perp-1&=\frac{1}{\chi}\int_0^\chi
 [\alpha_\parallel(\chi')-1+E(\chi')]\,d\chi'.
\end{align}
Here $f=f(z)$ and $\Omega_m=\Omega_m(z)$ are the exterior growth rate and matter fraction at that redshift, and $\delta\ln a$ is the change of emission scale factor at fixed observed redshift. The radial expression combines the perturbation of the local expansion, $f(2\bar\delta/3-\delta_m)$, the background-rate change $-3\Omega_m\delta\ln a/2$, and the radial-ruler change $\delta\ln a+2\bar\delta/3-\delta_m$. The nonlinear numerical fits use the exact local expressions, not this perturbative expansion.

\paragraph*{Pair dilations.}
In small-angle, plane-parallel geometry, let $\chi_{1,2}=\chi_c\mp s_\parallel/2$, with $s_\parallel=s|\mu|$. In the matched fiducial convention, the number of initial rulers along a purely radial passive pair is $\int X_\parallel dz=\int\alpha_\parallel d\chi/\rd$, while the fiducial count is $s_\parallel/\rd$. The projected transverse component has count $(q_1+q_2)\Delta\theta/(2\rd)$, compared with $(\chi_1+\chi_2)\Delta\theta/(2\rd)$. Their ratios give
\begin{equation}
\begin{aligned}
 \alpha_\parallel^{\rm pair}&=\frac{1}{s_\parallel}
 \int_{\chi_1}^{\chi_2}\alpha_\parallel\,d\chi,\\
 \alpha_\perp^{\rm pair}&=
 \frac{\chi_1\alpha_\perp(\chi_1)+\chi_2\alpha_\perp(\chi_2)}{2\chi_c}.
\end{aligned}\label{eq:pair_alpha}
\end{equation}
The factors $\rd^{\fid}/\rd$ are already included when the general local dilations in Eq.~\eqref{eq:fixed_fid_alpha} are used. The radial integral is exact for a purely radial passive pair in the initial metric $d\sigma_i^2=dq^2/(1+2E)+q^2d\Omega^2$. Its use as a radial component of an oblique pair, together with the transverse endpoint component, is approximate. A finite oblique acoustic geodesic can also depend on the metric between the endpoints. In the limit $s_\parallel\to0$ both expressions reduce to their local values; a strictly transverse pair has no measurable radial component. Main Appendix~B separately follows a ruler of fixed initial length rather than a pair of fixed fiducial separation.

\paragraph*{Short-pair expansion.}
For a smooth profile, symmetry about $\chi_c$ removes odd powers of $s_\parallel$. Expanding Eq.~\eqref{eq:pair_alpha} gives
\begin{align}
 \alpha_\parallel^{\rm pair}-\alpha_\parallel(\chi_c)
 &=\frac{s_\parallel^2}{24}\alpha_\parallel''+O(s_\parallel^4),\\
 \alpha_\perp^{\rm pair}-\alpha_\perp(\chi_c)
 &=\frac{s_\parallel^2}{8}
 \left(\alpha_\perp''+\frac{2\alpha_\perp'}{\chi_c}\right)
 +O(s_\parallel^4).
\end{align}
The derivatives in these two equations are with respect to fiducial distance along the light cone, with $z=z(\chi)$; they do not hold observed redshift fixed. A line average of $\alpha_\perp$ would give $s_\parallel^2\alpha_\perp''/24$ and would miss the endpoint geometry. These series require the pair extent to be small compared with the relevant variation scales, not just with $\chi_c$. The main-paper numerical checks find a remainder of about $0.00812$ in the radial dilation for a 147-Mpc pair at $z=0.20$ in the nearby solution. Unexpanded integrals and endpoints should be used outside the controlled Taylor regime.

\paragraph*{Average over pairs.}
For a fixed separation and orientation, define the normalized midpoint weight $p(\chi_c|s,\mu)$ by the survey's weighted pair counts. Then
\begin{equation}
 \langle\delta\alpha_i^{\rm pair}\rangle_{s,\mu}
 =\int d\chi_c\,p(\chi_c|s,\mu)
 \delta\alpha_i^{\rm pair}(\chi_c,s|\mu|).
 \label{eq:pair_mean}
\end{equation}
For a smooth selection and small-angle geometry its leading pair-volume factor is $n(\chi_1)n(\chi_2)\chi_c^2$, where $n$ is number density per fiducial volume, not $dN/d\chi$. The actual weight also contains the angular mask, endpoint selections, and estimator weights, and excludes pairs with an endpoint outside the bin. Its normalization is imposed separately for each $(s,\mu)$. This average is first order in the dilation perturbations, but need not be first order in bin width: symmetric weighting can cancel a first moment. It therefore does not establish a universal hierarchy between variation within a pair and variation among pairs.

\paragraph*{From pair geometry to a correlation residual.}
For illustration take $\xi^{\rm temp}=b^2\xi_{\rm lin}(\tilde{s})+{\rm broadband}$, with $\tilde{s}=s[\alpha_\perp^2(1-\mu^2)+\alpha_\parallel^2\mu^2]^{1/2}$. At unit dilations its derivatives are
\begin{equation}
 h_\perp=b^2s\xi_{\rm lin}'(s)(1-\mu^2),\qquad
 h_\parallel=b^2s\xi_{\rm lin}'(s)\mu^2.
\end{equation}
When bias, growth or template response changes across the bin, retain the midpoint dependence inside the integral:
\begin{equation}
 v_{\rm geom}(s,\mu)=\int d\chi_c\,p(\chi_c|s,\mu)
 \sum_i h_i(s,\mu;\chi_c)\delta\alpha_i^{\rm pair}.
\end{equation}
{In the approximation that retains only geometrical remapping, the bin-averaged model correlation function obeys
\begin{equation}
\begin{aligned}
 \xi^{\rm model}_{\rm geom}(s,\mu)
 &=\xi^{\rm temp}_0(s,\mu)+v_{\rm geom}(s,\mu)\\
 &\quad+\mathcal O\!\left((\delta\alpha^{\rm pair})^2\right),
\end{aligned}
 \label{eq:xiobs_supp}
\end{equation}
valid to first order in the pair-dilation perturbations $\delta\alpha_i^{\rm pair}$. Here $\xi^{\rm temp}_0(s,\mu)$ is the unit-dilation template averaged over $\chi_c$ with the same normalized distribution $p(\chi_c|s,\mu)$ and the same midpoint-dependent template parameters used in $v_{\rm geom}$. These weights and parameters are held fixed while the pair dilations are varied. This relation excludes the additional physical residuals discussed below and measurement noise.}
Only a midpoint-independent response gives $v_{\rm geom}=\sum_i h_i\langle\delta\alpha_i^{\rm pair}\rangle$. Projecting this residual onto the fitted multipoles gives a data vector $\bm v_{\rm geom}$. A complete forward model can contain an additional residual $\bm v_{\rm other}$ from inhomogeneous clustering, redshift-space effects and reconstruction. These contributions are not evaluated here. The sizeable local-dilation range in the nearby solution's BGS interval also means that the accuracy of a first-order template expansion must be tested with the nonlinear remapping and actual pair weights.

\paragraph*{The fitted dilations.}
Write the model, linearized about a specified reference, as $\bm m=\bm m_0+D\delta\bm\alpha+N\delta\bm\beta$, with dilation derivatives $D$ and nuisance derivatives $N$, including amplitude and broadband terms. For a fixed data covariance $C_\xi$ and unconstrained linear nuisance parameters,
\begin{align}
 \mathcal W&=C_\xi^{-1}-C_\xi^{-1}N
 (N^TC_\xi^{-1}N)^{-1}N^TC_\xi^{-1},\\
 F&=D^T\mathcal WD,\qquad
 K=F^{-1}D^T\mathcal W,\\
 \delta\bm\alpha^{\rm fit}&=K(\bm v_{\rm geom}+\bm v_{\rm other}).
 \label{eq:fit_K}
\end{align}
The inverses require full rank on the fitted subspaces. For a Gaussian nuisance prior, add its precision to $N^TC_\xi^{-1}N$ before inversion; a displaced prior mean also contributes to the right-hand side. The matrix $C_\xi$ is the covariance of the fitted two-point-function data, not the $13\times13$ covariance of the released distance summaries. The latter cannot determine $D$, $N$ or $K$. Actual DESI fitting further uses tracer-dependent reconstruction, damping and nuisance treatment~\cite{DESIValidation}; the unrestricted two-dilation illustration is not the BGS isotropic fit or a universal description of all tracers.

In the stated linear fit $KD=I$. A residual $\bm v=D\delta\bm\alpha$ therefore returns exactly those dilations, and a pure radial shift produces no transverse bias merely because the fitted errors are correlated. Constant pair dilations, with no other residual, are a sufficient condition for this result, not a necessary one: a nonzero residual $\bm u$ with $K\bm u=0$ also leaves the fitted dilations unchanged. For the unconstrained nuisance case the response obeys
\begin{equation}
 |\delta\alpha_i^{\rm fit}|\leq
 \sqrt{(F^{-1})_{ii}}\sqrt{\bm v^T\mathcal W\bm v}.
\end{equation}
This inequality requires a specified residual; background shear or the covariance of fitted dilations alone supplies no survey-independent numerical bound. The derivation explains what must supplement the local likelihood, while the baseline and the finite-ruler/window diagnostics remain the distinct calculations described in the main paper.

\section{A profile-specific Gaussian density prior}\label{app:prior}
The BAO--SN constraints above are observational likelihood statements. A theoretical density prior, which describes how likely a given void is in a homogeneous universe with Gaussian initial fluctuations, is a distinct ingredient. We give its form here for completeness; it is not applied to the voids of this paper or used in any of its results. For a fixed linear template $u(r)$ and coefficient $\delta_0$, use a power spectrum and template defined at the same epoch. With
\begin{equation}
 \widetilde u(k)=4\pi\int_0^\infty dr\,r^2u(r)j_0(kr),
\end{equation}
the Fourier-band inverse-covariance norm is
\begin{equation}
 I=\frac{1}{2\pi^2}\int_0^{k_{\max}}dk\,
 \frac{k^2|\widetilde u(k)|^2}{P(k)},\qquad
 \sigma_{\delta_0}^2=I^{-1}.
\end{equation}
The matched coefficient has this variance at fixed location, width, and shape. A deterministic pattern $\delta=\delta_0u$ has relative Gaussian density $\exp(-\delta_0^2I/2)$. This is not the probability of an entire observed shell, an observational exclusion significance, or a correction for searching over profiles. A Gaussian-smoothed mean-density variance cannot replace $I^{-1}$ for an arbitrary template. A nonlinear present-day profile also requires a mapping to an initial linear field before such a prior can be applied consistently.

\section{Inputs and numerical conventions}\label{app:inputs}
The angular BAO measurements, and the distinction between fitting them in the measured angles and in logarithmic distance ratios, are given in Appendix~C of the main text. The five SDSS DR11 entries follow Ref.~\cite{CarvalhoDR11}, Table~3, arXiv version~4, rather than the alternative values in the Nunes compilation. The ruler-shape test and the direct HSW fits use the first fourteen angular measurements; the fits of Sec.~\ref{sec:benchmark} use all fifteen.

\paragraph*{Exact solver.}
The independent exact-background solver uses a radial grid of 801 points over 0--12000 Mpc and 641 time points over exterior scale factors 0.05--1; the refined comparison uses 1601/1281 points. The mass integral and the age integral, after a square-root change of variable, use 96-point Gauss--Legendre quadrature. The simultaneous-bang-time condition is solved for the curvature to a residual below $7\times10^{-15}$ in $H_0t$. The shells are integrated with the eighth-order Runge--Kutta method DOP853, with relative tolerance $3\times10^{-11}$ and absolute tolerance $3\times10^{-13}$ in $R/r$. Cubic interpolation gives $a_r(t,r)$ and its radial derivative, and differentiating the shell equation then gives $\dot R'$ consistently. The light cone uses relative tolerance $3\times10^{-10}$, absolute tolerances $3\times10^{-12}$ in $H_0t$ and $3\times10^{-7}$ Mpc in radius, and a maximum redshift step of 0.01. Outputs are sampled every 0.002 in redshift up to $z=2.4$.

\paragraph*{Linear profiles.}
The linear calculations of Sec.~\ref{sec:conditional_shape} use the exact growing mode of flat dust-plus-$\Lambda$, normalized to unity today. The enclosed density and the emission-time correction are computed with cumulative Simpson quadrature on 4001 points over $0\leq z\leq0.65$; 16001 points provide a convergence check. Compensation integrals extend over the full half-line, not only to the highest BAO redshift, and the HSW compensation reproduces the values of $q_s$ in Sec.~\ref{app:profiles}. At each width the logarithmic likelihood is quadratic in the depth, so its minimum within $0\leq d\leq1$ and its contour intersections are obtained analytically; the checks in the measured angles refit the SN polynomial coefficients numerically. The 25-Mpc spacing of the width grid is separate from the integration accuracy: quoted widths are grid points at which the limits occur, not measured widths.

\paragraph*{Data conventions.}
Flat-FRW distances use a separate 96-point Gaussian quadrature. The Pantheon+ covariance is restricted to the selected supernovae before inversion, as described in Appendix~B of the main text. The DES-Dovekie reconstruction uses the full released precision matrix, in the row order of the Hubble diagram. All fits keep the full covariance and fit the common normalization. Strong numerical convergence does not remove the remaining limitations: the likelihood conventions, the assumptions about the data, and the linear or local approximations for the BAO ruler. Points of fixed-parameter scans are not posterior samples.

\paragraph*{DESI fits of the main text.}
For the fits including DESI, the solver with free $\Omega_m$ is extended to $z=2.33$. The radial observable is computed from the same $q'$ and $\dot R'$ that determine the radial ruler, which checks $X_\parallel=c/(\Hl\Ll)$ directly. A Cholesky factor of the full $13\times13$ DESI covariance whitens the residuals, and all six pairs of transverse and radial measurements keep their released ordering and covariance. Fitting the common ruler $\rd$ through all positive stationary points of its $\chi^2$ (Appendix~B of the main text) agrees with an independent bounded minimization. At zero depth, the predictions recover the flat-FRW transverse, radial, and isotropic BAO observables to better than $10^{-12}$. The refined minima use exact shell tables at the fitted exterior density, a 2.5-Mpc radial spacing, and a null-cone relative tolerance of $2\times10^{-10}$. An independent solution with 2401 physical radial shells and 801 time samples provides the checks of the radial derivative and of the likelihood quoted in Sec.~V of the main text. The searches at fixed depth start from several nearby, broad, and distant profiles. Their likelihood curves show the loss at each depth, without confidence levels, and do not reuse the simulations without DESI.
\section{DES-only selection and a central-depression restriction}\label{sec:des_only}
\subsection{Data selection and the marginalized covariance}
The DES-only comparison addresses the survey-selection question raised in Ref.~\cite{NajeraDESOnly}. In the covariance-ordered DES-Dovekie Hubble diagram, the DES supernovae (\texttt{IDSURVEY=10}) comprise 1623 of the 1820 objects in the full release. Applying \texttt{MUERR<1} additionally removes 82 objects, leaving 1541. The 197 legacy objects all have $z_{\rm HD}\leq0.08704$. The four DES objects below $z_{\rm HD}=0.1$ have redshifts $0.07334$, $0.07374$, $0.09333$, and $0.09394$ and all pass the uncertainty cut. The $z>0.1$ selection therefore retained 1619 DES SNe and no legacy objects.

Let $P=C^{-1}$ be the full released statistical-plus-systematic precision matrix, reconstructed from its upper triangle, and partition the observations into retained ($K$) and discarded ($J$) rows. We first recover $C=P^{-1}$ and form
\begin{equation}
 P_K=(C_{KK})^{-1}
     =P_{KK}-P_{KJ}P_{JJ}^{-1}P_{JK}.
 \label{eq:des_covariance}
\end{equation}
The right-hand side, the Schur complement of the precision matrix, gives an independent check of the left-hand side. Keeping $P_{KK}$ alone would condition on the discarded measurements, treating them as known, instead of marginalizing them. The two sides of Eq.~\eqref{eq:des_covariance} agree to relative errors $5.33\times10^{-16}$ and $1.07\times10^{-15}$ for the two selections. Both covariance blocks are positive definite. Row indices and object identifiers are saved with the fit outputs; a separate metadata table with a different ordering is not used for covariance selection.

We retain the published Dovekie calibration and bias-correction pipeline. We perform new cosmological fits to this survey subset, but do not repeat the light-curve analysis, calibration, or bias-correction pipeline using DES data alone. The uncertainty cut is useful for testing high-uncertainty observations, but is not by itself a classification of which objects are SNe Ia. Neither selection diagnoses the origin of a DES--legacy difference without modelling calibration, selection and the low-redshift coverage and statistical weight of each sample.

\subsection{Matched fits with a free shell location}
Both samples are combined with the same 13 DESI DR2 entries and full $13\times13$ covariance. Legacy angular BAO are not included. The radial and transverse rulers are transported consistently from a single initial $\rd$. Predictions are evaluated at the DESI effective redshifts, retaining the main text's conditional assumption that the compressed BAO measurements can be interpreted in this LTB model. Both cosmologies refit $\Omega_m$ and minimize over the same positive initial ruler length and SN magnitude offset; no CMB ruler or absolute-distance prior is imposed. The scale $H_0=68.8\,$km\,s$^{-1}$\,Mpc$^{-1}$ only fixes dimensional units. The LTB fit additionally varies $d$, $r_w$, and $r_0$ within the main-paper bounds.

\begin{table*}[t]
\caption{Matched full-covariance DESI+SN fits. $T=\chi^2_{\rm flat}-\chi^2_{\rm LTB}$ compares models on the same observations. Raw $\chi^2$ differences between rows with different SN selections are not model-selection statistics. The first two rows give the full-sample and redshift-cut reference fits. The LTB and flat models have six and three fitted parameters, including the two profiled calibrations.}\label{tab:des_totals}
\begin{ruledtabular}
\begin{tabular}{lrrrrrr}
SN selection & $N_{\rm SN}$ & $\Omega_m^{\rm flat}$ & $\chi^2_{\rm flat}$ & $\chi^2_{\rm LTB}$ & $T$ & $\Delta{\rm AIC}$\\
Full DES+legacy & 1820 & 0.306046 & 1645.277834 & 1636.754802 & 8.523031 & $-2.523031$\\
$z>0.1$ & 1619 & 0.298852 & 1425.530327 & 1418.095197 & 7.435130 & $-1.435130$\\
DES only & 1623 & 0.298761 & 1427.101912 & 1419.664016 & 7.437896 & $-1.437896$\\
DES only, \texttt{MUERR<1} & 1541 & 0.298758 & 1426.767375 & 1419.330776 & 7.436599 & $-1.436599$\\
\end{tabular}
\end{ruledtabular}
\end{table*}

\begin{table*}[t]
\caption{Best-fit geometry and transported ruler ratios for the two DES-only selections. Ratios compare $z=0.65$ to $0.11$ and are not confidence intervals. Radii are present areal radii in the common scale convention; $r_0$ locates the underdense shell around an observer fixed at $r=0$.}\label{tab:des_parameters}
\begin{ruledtabular}
\begin{tabular}{lrrrrrr}
SN selection & $\Omega_m$ & $d$ & $r_w$ [Gpc] & $r_0$ [Gpc] & $\Lp$ ratio & $\Ll$ ratio\\
DES only & 0.301275 & 0.034784 & 0.669127 & 2.578201 & 1.001296 & 1.022613\\
DES only, \texttt{MUERR<1} & 0.301272 & 0.034782 & 0.669088 & 2.578231 & 1.001296 & 1.022612\\
\end{tabular}
\end{ruledtabular}
\end{table*}

Tables~\ref{tab:des_totals} and \ref{tab:des_parameters} show that the two selections give nearly identical results and closely reproduce the $z>0.1$ selection. Without the uncertainty cut, the joint LTB optimum has $\chi^2_{\rm SN}=1416.871883$ and $\chi^2_{\rm DESI}=2.792133$, compared with $1416.808261$ and $10.293650$ at the matched flat optimum. Thus the component gains are $T_{\rm SN}=-0.063622$ and $T_{\rm DESI}=7.501518$. With the cut they are $-0.064786$ and $7.501385$. These are contributions at the joint optima, not separate SN-only fits. They do not show that DES-only SNe require a void. With three additional shape parameters, the modest AIC advantage is about 1.44; no new simulations have been performed to calibrate the fit improvement, and no posterior intervals or confidence regions have been computed for these selections.

\begin{figure*}[t]
\centering
\includegraphics[width=\textwidth]{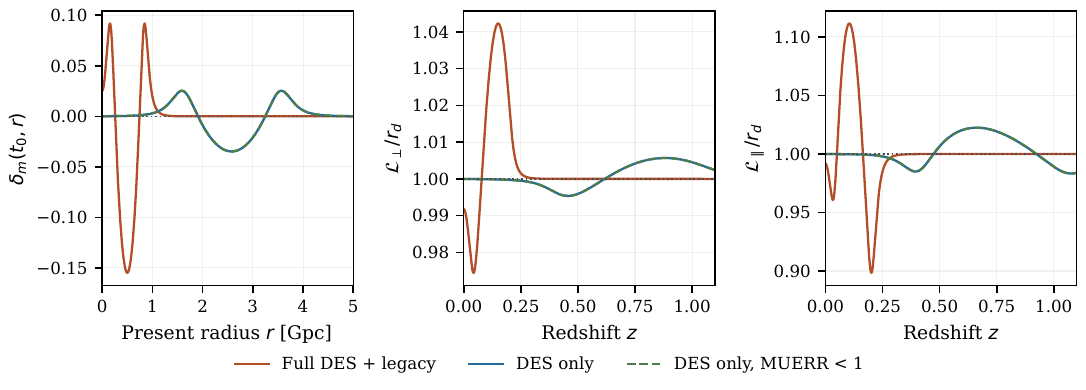}
\caption{Exact nonlinear best-fit density profiles and observed-comoving rulers relative to the single initial sound horizon, $\mathcal L_i/\rd$. All observers are at $r=0$; the displaced minima are underdense shells around them. The full DES+legacy fit is red, and the two DES-only curves nearly coincide. The dotted ruler baseline is FRW. This normalization exposes the localized oscillation of the nearby-shell solution: at $z=0.11$ its radial ruler is $1.1104\rd$, while at $z=0.65$ it has returned to approximately $\rd$. The endpoint ratio $0.9005$ is therefore not a $10\%$ deficit relative to $\rd$ at high redshift. Curves are fitted predictions, not confidence bands or direct ruler measurements.}\label{fig:des_only}
\end{figure*}

\begin{figure*}[tbp]
\centering
\includegraphics[width=\textwidth]{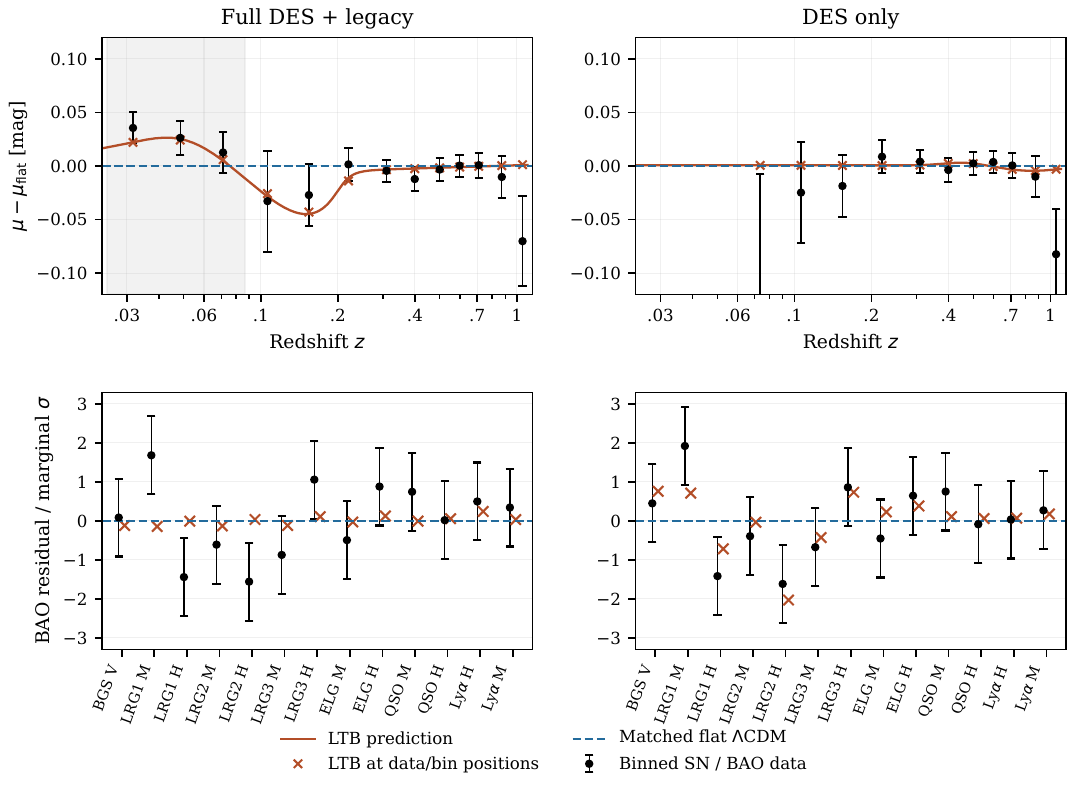}
\caption{Data and fitted predictions relative to the matched flat-$\Lambda$CDM solution for each SN selection. Top: SN distance-modulus residuals; black points are generalized least-squares bin averages with the full SN covariance, and the crosses apply the same binning operator to the LTB curve. Their errors are correlated and condition on the fitted reference model; these bins are only a visualization, while the fits use every retained SN. Gray shading marks the legacy-SN redshift range. Bottom: the thirteen DESI observations minus the flat prediction, divided by their marginal standard deviations; crosses show the LTB-minus-flat prediction in the same units. M, H, and V denote transverse, radial, and isotropic entries. The plotted one-entry errors are not independent: the likelihood uses the full covariance. Curves and crosses are best-fit predictions, not uncertainty bands.}\label{fig:data_residuals}
\end{figure*}

Figure~\ref{fig:data_residuals} displays the observations against the fitted models. For the SN visualization, with bin-assignment matrix $B$ and precision $P$, we use $W=(B^TPB)^{-1}B^TP$ and bin covariance $(B^TPB)^{-1}$; the same $W$ is applied to data residuals and model predictions. The full-sample nearby shell crosses the legacy-to-DES transition. The roughly $0.04$-mag offset discussed for the earlier DES5Y/Pantheon+ comparison~\cite{Efstathiou2025} is not an offset measured here in DES-Dovekie. In the DES-only fit, the density minimum is at $z=0.711$, with zero crossings at $0.496$ and $0.956$, bracketing the three LRG effective redshifts. The LRG2 radial marginal pull changes from $-1.62$ to $+0.41$; correlated residuals must be evaluated jointly, so its squared-pull change is not an additive decomposition of the total gain.

The fixed-geometry window check uses the same full covariance and the averaging prescription of main Appendix~B. Keeping the DES-only geometry fixed and re-profiling $\rd$ and the SN offset gives $T=6.97860$ (uniform) and $7.25030$ (triangular); with \texttt{MUERR<1} the values are $6.97697$ and $7.24873$. Increasing the quadrature from 48 to 192 nodes on each side of the effective redshift checks integration convergence. These values bound from below the improvement attainable by re-optimizing the geometry within each proxy. They are not new global fits, significance calibrations, or validations of the actual DESI survey response.

As Fig.~\ref{fig:des_only} shows, the DES-only selections change the preferred density shell and ruler history relative to the full DES+legacy fit, consistent with the redshift-cut result. A similar sensitivity to the SN selection was found in the $\Lambda$LTB Hubble-bubble analysis of Ref.~\cite{CamarenaBubble}. Earlier joint BAO--SN analyses~\cite{MarraPaakkonen}, unlike ours, used additional calibration information. The studies address different profile families and data combinations, so their inferred contrasts and model preferences cannot be compared directly with those in these tables.

The DESI-era analysis of Camarena et al.~\cite{CamarenaDESI2025} requires a more specific comparison. Its full $\Lambda$LTB fits use DES-Y5 or Pantheon+ SNe with compressed CMB information; its geometric consistency tests also include uncalibrated comparisons. Our DES-only selections use the DES objects in the published DES-Dovekie reanalysis, which updates the calibration and analysis of DES-SN5YR~\cite{Popovic}. We retain the released light-curve results, bias corrections, and covariance; we do not repeat that processing using DES data alone. Here $\rd$ and the SN magnitude offset are free and no CMB likelihood is imposed. Our density family also varies depth, width, and shell location independently, together with the exterior $\Omega_m$, instead of using their central-contrast and boundary-redshift parametrization. These differences change both the predictions and the allowed parameter space. Moreover, their profile-likelihood significance calculation and our homogeneous-mock search calibration are different procedures; our quoted mock result applies only to the primary full sample, not to these DES-only fits. Their reported $2.8\sigma$ (Pantheon+) and $5.2\sigma$ (DESY5) departures from FLRW are therefore not comparable to the full-sample lower-bound mock fraction quoted here. Thus the comparisons here neither reproduce nor exclude their preferred profiles, and fit improvements or significance levels cannot be compared directly, since they do not come from the same likelihood.

More generally, SNe without an external absolute-distance calibration retain information on the shape of the distance--redshift relation even when a constant BAO calibration is changed~\cite{PedrottiMiscalibration}. Our LTB application also allows redshift-dependent and direction-dependent ruler transport. The sound-horizon-calibrated inverse distance ladder of Ref.~\cite{NajeraDESOnly}, the effective potential-offset test of Ref.~\cite{BanikPotential}, and the Milgromian local-void scenario of Ref.~\cite{Haslbauer} are complementary tests based on different assumptions. The present uncalibrated fits do not independently infer absolute $H_0$ or decide whether a general local-void scenario resolves its tension.

\subsection{Searches and independent numerical checks}
For each free-location selection, the search starts from ten distinct initial profiles and uses differential evolution (seed 20261111, population 48) and local refinement. The global searches converge after 53 and 52 generations, respectively. Final evaluations use an exact shell table at the fitted $\Omega_m$, 2.5-Mpc radial spacing, and light-cone relative tolerance $2\times10^{-10}$. An independently implemented integration of physical radial shells reproduces the two total $\chi^2$ values within $4.3\times10^{-9}$. Covariance-whitened SN offset fits agree within $1.6\times10^{-12}$ in $\chi^2$, and the comparison of the LTB model without a void ($d=0$) with flat $\Lambda$CDM agrees within $2.3\times10^{-13}$. All sampled density, $q'$, $R'$, and metric-signature minima are positive. These numerical checks validate the implemented conditional likelihood; they do not validate the full DESI survey response or the Dovekie selection pipeline in LTB.

\subsection{Requiring the density minimum at the observer}
The observer is at $r=0$ in every fit, including those with $r_0>0$. A separate restriction sets the profile-location parameter itself to $r_0=0$, placing the density depression at the symmetry centre. We repeat both DES-only fits under this restriction, refitting $\Omega_m$, depth, width and the same two calibration parameters. These models have two additional shape parameters relative to flat $\Lambda$CDM, so $\Delta{\rm AIC}=4-T$, rather than $6-T$.

\begin{table*}[t]
\caption{Restricted $r_0=0$ fits, with the observer at the same symmetry centre as in every other fit. The flat reference is the corresponding row of Table~\ref{tab:des_totals}. These are best located fits within the stated profile family and bounds, not constraints on all possible central-void profiles.}\label{tab:des_central}
\begin{ruledtabular}
\begin{tabular}{lrrrrrr}
SN selection & $\Omega_m$ & $d$ & $r_w$ [Gpc] & $\chi^2_{r_0=0}$ & $T$ & $\Delta{\rm AIC}$\\
DES only & 0.299915 & 0.999000 & 0.274164 & 1426.216759 & 0.885153 & 3.114847 \\
DES only, \texttt{MUERR<1} & 0.299913 & 0.999000 & 0.274172 & 1425.881990 & 0.885385 & 3.114615 \\
\end{tabular}
\end{ruledtabular}
\end{table*}

Table~\ref{tab:des_central} shows that the restricted fits improve the matched flat likelihood by only $T\simeq0.885$ in either selection, at a cost of 6.553 and 6.551 in $\chi^2$ relative to the free-location shell fits. The best fits found in both restricted searches reach the imposed upper depth bound $d=0.999$. This boundary result is not a measurement of an almost empty central void: the likelihood gain is small and the AIC favors flat $\Lambda$CDM. The restricted comparison tests only this compensated HSW-shaped family and search range; it neither relocates the observer nor excludes central voids with different profiles or external calibration assumptions. We assign no detection significance to it.

Each restricted search uses eight distinct local starts, differential evolution (seed 20261112, population 36), and local refinement, and explicitly includes the homogeneous boundary. The same exact-exterior refinement, independent radial-shell integration, common-ruler profiling and sampled regularity checks are applied. The maximum coarse-to-fine change in total $\chi^2$ is $9.4\times10^{-6}$, and the independent integration agrees with the refined total within $1.3\times10^{-6}$. These errors are small compared with the quoted fit differences.

\section*{Data and code availability}
The executable numerical package is publicly available on Zenodo~\cite{NumericalArchive}, version 1.2.0, DOI \href{https://doi.org/10.5281/zenodo.24374395}{10.5281/zenodo.24374395}. It includes the baseline likelihood and numerical checks together with the DES-only fits and central-depression restriction of Sec.~\ref{sec:des_only}. The numerical addendum provides the fixed-geometry redshift-window tests, recorded results, and code for the ruler histories and data/model residual figures. Its README gives the reproduction commands. A checksum-verified downloader retrieves the pinned inputs from their original providers.

\section*{Computational and editorial assistance}
OpenAI Codex assisted with numerical code, checking scripts, derivations, and exposition under the authors\textquotesingle{} direction. GPT-6 was used for the final review; exact model-version identifiers for earlier sessions were not retained. The main paper describes the authors\textquotesingle{} direction of the work and the independent-integration, covariance, limiting-case and resolution checks. The authors retain responsibility for scientific verification and interpretation; automated comparisons are not external human peer review.